\documentclass[  amssymb, amsmath, aps,twocolumn,superscriptaddress, pra]{revtex4-2}
\usepackage{comment,amsmath,braket,tikz}
\usepackage{graphicx}
\usepackage{mathtools}

\usepackage{bbold}

\usepackage[colorlinks=true,
            linkcolor=blue,
            citecolor=blue,
            urlcolor=blue,
            bookmarks=true,
            pdfstartview=FitH,
            bookmarksopen=true,
            bookmarksdepth=subsubsection,
            pdfencoding=auto,
            bookmarksopen,
            bookmarksnumbered]{hyperref}

\newcommand{\tru}{\text{Tr}{\ddot{\text{u}}}\text{benbacher}}
\newcommand{\lam}{\text{L}{\ddot{\text{a}}}\text{mmerzahl}}
\newcommand{\schro}{\text{Schr}\ddot{\text{o}}\text{dinger}}
\newcommand{\inner}[2]{\braket{\braket{#1|#2}}_{\text{KG}} }
\newcommand{\innere}[2]{{\braket{#1|#2}}_{\text{KG}} }

\newcommand{\scaled}[1]{{#1}}

\newcommand{\elem}[3]{\bra{#1}#2\ket{#3}}
\newcommand{\poisson}[2]{\lceil #1 , #2\rfloor }

\newcommand{\deriv}[3]{\frac{\partial^{#3} #1}{\partial #2^{#3} }}
\newcommand{\derivb}[2]{\partial^{#2}_{#1}}
\newcommand{\derfunc}[2]{\frac{\delta #1}{\delta #2}}
\newcommand{\lad}[0]{\phi}

\newcommand{\bfr}[1]{{\bf #1}}

\newcommand{\pbdim}[2]{\{#1,#2\}_{*}}
\newcommand{\scale}[1]{\bar{#1}}
\newcommand{\scalex}[0]{\scale{X}}
\newcommand{\scalep}[0]{\scale{P}}
\newcommand{\scaleh}[0]{\scale{H}}

\newcommand{\moment}[0]{\mu}
\newcommand{\liep}[0]{\mathfrak{p}_{1+1}}

\newcommand{\Psilad}[0]{\Xi}
\newcommand{\psilad}[0]{\chi}
\newcommand{\psiladhalf}[0]{\psilad_{\frac{1}{2}}}
\newcommand{\psiladhalfb}[0]{\psilad_{\frac{1}{2}}^*}
\newcommand{\psiladb}[0]{{\psilad}^*}

\newcommand{\Phialpha}[0]{\Psilad_\alpha}
\newcommand{\phialpha}[0]{\psilad_\alpha}
\newcommand{\phialphadot}[0]{\dot{\psilad}_\alpha}

\newcommand{\hatom}[0]{\hat{\Omega}}

\newcommand{\bary}[0]{{\mathcal{X}}}

\newcommand{\mbaros}[0]{\rho_{\frac{1}{2}}}
\newcommand{\norm}[0]{\gamma}
\newcommand{\vecpp}[0]{v}
\newcommand{\Vecpp}[0]{V}

\DeclareMathAlphabet\mathbfcal{OMS}{cmsy}{b}{n}

\newcounter{note}
\renewcommand{\thenote}{\Roman{note}}

\makeatletter
\newcommand{\notefn}[1]{%
  \refstepcounter{note}%
  \hyperlink{note:\thenote}{\textsuperscript{\thenote}}%
  \insert\footins{%
    \reset@font\footnotesize
    \interlinepenalty\interfootnotelinepenalty
    \hypertarget{note:\thenote}{}%
    \hbox to 0.4em{{\hss\color{blue}\thenote}}%
    \vspace*{-\baselineskip}\setlength\belowdisplayshortskip{0pt}         
    \vspace{+0.05cm}
    #1\par
  }%
}
\makeatother

\begin{document}

\title{Algebra of quantum mechanics \textit{via} classical phonons. II: 
Klein-Gordon dynamics, the Heisenberg formalism, the Dirac canonical commutation rule 
and the Poincar\'e algebra through the continuous Poisson bracket formalism.  
}

\author{Emmanuel Giner}%
\email{emmanuel.giner@lct.jussieu.fr}
\affiliation{Laboratoire de Chimie Th\'eorique, Sorbonne Universit\'e and CNRS, F-75005 Paris, France}

\begin{abstract}
In the first part of this series we have shown how the $\schro$ equation for a single particle 
and the corresponding non relativistic 
quantum observables can be obtained from a purely classical phonon model through the Newtonian equations of motion. 
In this work we focus instead on how the classical Hamiltonian formalism 
applied to the same phonon system allows to recover the feature of relativistic quantum mechanics 
for a single spinless particle. 
Using the classical nature of the phonon model, we naturally define continuous Poisson brackets between classical observables, which  
allows to recover the dynamics of such observables, \textit{i.e.} the Ehrenfest relations associated to real-valued Klein-Gordon fields. 
The Poisson brackets also permits to obtain the generic form of constants of motions, thus generalizing the concept 
of inner products and momentum on Klein-Gordon fields. 
We then connect the formalism of real-valued classical functionals with that of hermitian operators 
and complex-valued wave functions. 
This is done through the introduction of a non-local complex-valued change of variables which 
allows to rewrite the real-valued Klein-Gordon equation in a form akin to the $\schro$ equation, 
and the classical observables as quantum expectation values. 
Then, we show how this change of variables allows to rewrite the classical Poisson brackets as commutators of hermitian operators.  
This points out the strict equivalence between the Heisenberg formalism and the formalism of classical Poisson bracket. 
Eventually, we illustrate how the Poisson brackets allows to recover the transformations of 
Poincar\'e group in 1+1 dimension together with its algebra. 
The latter makes the link between the Lorentz invariant inner product of Mostafazadeh and 
the Casimir invariant associated to the mass of particle. 
\end{abstract}

\maketitle

\section{Introduction}
Among the distinct features of quantum mechanics with respect to classical mechanics lies their fundamentally different mathematical formulations. 
The physical observables in quantum mechanics are obtained as expectation values 
of Hermitian operators over complex-valued wave functions, while they are simply functions of the 
real-valued position and momentum variables in classical mechanics. 
The dynamics of quantum observables is computed with the Heisenberg representation, which involves the commutator 
between the Hamiltonian of the system and the Hermitian operator representing the observable. 
It is noteworthy that the use of commutators is not restricted to the time evolution but is more general 
in both relativistic and non relativistic quantum mechanics, and 
through the formalism of Lie groups one can then implement fundamental continuous transformations 
such as translations and rotations. The precise structure of the commutators between operators generating 
these transformations (\textit{i.e.} the Lie algebra) leads, for instance, to either the Galilean or Poincar\'e symmetry groups. 
In that regard, the commutators are a fundamental aspect of the formalism of quantum mechanics, 
which resembles the Poisson brackets of classical Hamiltonian mechanics. 
Nevertheless, while commutators and Poisson brackets share some common mathematical features, such as the antisymmetry and the Jacobi identity, 
these two formalisms however appear as fundamentally distinct: 
the classical Poisson brackets are computed between functions of the phase 
space variables $(q,p)$, while commutators are computed between linear operators acting on an infinite dimensional 
Hilbert space. Also, even though the quantum observables result in real-valued functionals, 
the Hermitian operators and their algebra intrinsically involve imaginary numbers, which differ 
from classical mechanics. 
In that regard, the question of whether imaginary numbers are intrinsic in quantum mechanics 
or simply a mathematical tool particularly well suited for the theory was debated at 
the early stages of the development of the theory\cite{Schrodinger-26,Ehrenfest-ZP-32,Pauli-ZP-33,Karam-AJP-20},  
and remains\cite{Chen-JMP-89,McKagMosGis-PRL-09,RenTriWeiLeTavGisAci-nat-21,Wu-PRL-22,Li-PRL-22,HofWoo-arxiv-25,Makris-PO-25,HitTruKamEppBru-PRL-26} an open question. 

The link between classical and quantum mechanics is often made by means of the canonical quantization scheme 
initially proposed by Dirac\cite{Dirac-book-30}, which relates the Poisson bracket between two classical observables
and the commutator between the two associated quantum operators. 
While this approach typically works for many cases in non relativistic 
quantum mechanics and consists in a milestone of the quantum theory, 
it nevertheless have to be adapted when considering causality requirements in relativistic context, 
or external constraints. 

There exists nevertheless another way to connect the formalism of quantum mechanics for a single spinless particle 
with classical mechanics (\textit{i.e.} Newtonian mechanics), 
which is the linearized Frenkel-Kontorova model\cite{FreKon-ZETF-38}. 
The latter consists in a class of partial differential equations which are of second-order in both time and space variables, 
and which can be obtained 
from the continuous limit of the classical equation of motion of a transverse classical phonon model. 
When a specific type of interactions between particles is chosen (\textit{i.e.} harmonic springs), 
the Klein-Gordon equation can then be obtained in the continuous limit,  
which corresponds to an infinite number of coupled classical oscillators. 
In the introduction of his book on quantum field theory (QFT)\cite{QFTZee}, Zee sketched this model to illustrate how the 
Lagrangian associated to the Klein-Gordon equation can be obtained from this classical model that he refers to as a "mattress". 
Nevertheless, the classical nature of the model is quickly left aside as Zee uses then the Lagrangian to quantize the Klein-Gordon fields 
through the path integral formalism. 
Another recent work\cite{HeiMaaMakPom-PF-22} uses this classical phonon model to derive the Klein-Gordon equation 
from which the $\schro$ equation is obtained through a non relativistic limit. 
Nevertheless, in order to recover the $\schro$ equation in the $c \rightarrow \infty$ limit,   
the Klein-Gordon field is promoted to the complex plane,
which breaks the link with the classical (\textit{i.e.} Newtonian) formalism  
where the fields are intrinsically real-valued. 
In the first part of this series\cite{Giner-schro}, the present author bypassed the latter limitation by using a complex-valued change of variables 
mixing the real-valued displacements and velocity fields. 
This change of variable, borrowed from the work of Mostafazadeh on Klein-Gordon fields\cite{Mostafazadeh-CQG-02} 
and related to the two-components formalism of Feshback and Villars\cite{FesVil-RMP-58}, allows then to recover the $\schro$ equation in the non relativistic limit 
while remaining within a strict classical formalism. 
The present author also used the classical nature of the model to derive the usual functionals entering 
in the study of Klein-Gordon fields, such as the total energy, Lagrangian or momentum. 
Then, the complex-valued change of variables together with the $c\rightarrow \infty$ limit 
allows to recover the usual Hermitian operators of non relativistic quantum mechanics, 
thus establishing a direct link between the classical phonon model and the usual formalism of non relativistic 
quantum mechanics for a single spinless particle.   

The aim of the present work is to continue the derivation strictly based on classical mechanics of the algebra of 
quantum mechanics for a single spinless particle. 
As for the first part of this series\cite{Giner-schro}, we emphasize that our goal is not to re interpret quantum mechanics 
but rather to highlight the tight link between its formalism and that of classical mechanics. 
In the present work, we focus on the relativistic regime, the dynamics of observables, 
the link between quantum commutators and classical Poisson brackets, and eventually the Poincar\'e group. 
The main results of this work can be grouped into three broad parts that we briefly present here. 

The first part (\textit{i.e.} Sec. \ref{sec:pb_kg} and Sec. \ref{sec:dynamic_pb} and Sec. \ref{sec:ehrenfest_big}) 
focusses on how the classical phonon model together with the corresponding classical Poisson bracket formalism 
allows to recover the general structure of the algebra of relativistic quantum mechanics associated 
to the real-valued Klein-Gordon fields.  
Then the second part (\textit{i.e.} Sec. \ref{sec:connect}) focusses on establishing 
the equivalence between this classical formalism and that of the usual quantum mechanics, 
\textit{i.e.} Hermitian operators acting on complex-valued functions in $L^2$ Hilbert spaces. 
The last part (\textit{i.e.} Sec. \ref{sec:pb_transform}) builds up on the previous results and shows how 
the algebra of the Poincar\'e group can be recovered from the Poisson bracket structures, 
and how it is linked to the general framework introduced by Foldy\cite{Foldy-PR-56}. 
A discussion of the link between the latter and the usual QFT is also sketched out. 
We give now the organization of this paper. 

We begin in Sec. \ref{sec:pb_kg} by describing the classical phonon model at the heart of this work. 
In Sec. \ref{sec:kg_cl} we first describe its Hamiltonian and Lagrangian, which are then used to obtain, 
in the continuous limit, the corresponding equation of motion through the Hamiltonian formalism. 
We then show in Sec. \ref{sec:kg_fk} how these two-components Hamilton's equation of motion can then be combined to recover the Klein-Gordon equation. 
Having introduced the Hamilton's framework, we then use in Sec. \ref{sec:dynamic_pb} the corresponding Poisson brackets formalism  
to establish the general setup for flow transformation in phase space. 
Starting from the usual Poisson bracket formalism for the discrete model with a finite number of particles, we establish in Sec. \ref{sec:continuous_pb}  
the general Poisson brackets in the continuous limit. Then, in Sec. \ref{sec:two_comp}, 
by focussing on a specific class of classical physical quantities 
(such as the total momentum or energy), we obtain the general form of the 
$\{F,G \}$ Poisson bracket which will be used thoroughly in this work. 
The latter result is used in Sec. \ref{eq:cst_motion} to obtain the dynamics of observables from the Poisson bracket with the 
Hamiltonian (\textit{i.e.} $\{F,H\}$), and we derive the general structure of constants of motion, which can be grouped into two distinct categories. 
The first type of constants motions are strictly positive functionals which can be used to define time-independent inner products on real-valued 
Klein-Gordon fields, and we make the link with existing proposals such as that of Mostafazadeh\cite{Mostafazadeh-CQG-02}.  
The other type of constant of motions found are generalization of momentums, either linear or angular, which establishes 
a link with well known functionals of Klein-Gordon fields. 
We then use in Sec. \ref{sec:ehrenfest_big} the formalism of Poisson brackets to recover the Ehrenfest relations between observables.  
Two types of position estimates are first discussed in Sec. \ref{sec:pos}, and the associated time evolution are established 
through Poisson brackets in Sec. \ref{sec:ehrenfest}. 

We then switch to the second part of this paper where in Sec. \ref{sec:connect} we show how a complex-valued non local change of variables 
allows to rewrite the real-valued formalism of the classical phonon model as the complex-valued formalism 
of quantum mechanics. 
After having introduced in Sec. \ref{sec:change_var} the general class of change of variables which allows to rewrite 
the Klein-gordon equation as a $\schro$-like equation, we then give in Sec. \ref{sec:func_alpha} 
the conditions to rewrite the real-valued functionals as quantum expectation values. 
This allows in Sec. \ref{sec:flow} to translate the formalism of flow transformations in phase space 
to this new complex-valued variables, and we then show in Sec. \ref{sec:equiv_pb} how a special case 
of this change of variables allows to rewrite the Poisson brackets as quantum commutators. 
The non relativistic limit together with the link between Poisson brackets and the Heisenberg formulation 
are given in Sec. \ref{sec:nr_alpha} and Sec. \ref{sec:heisenberg}, respectively. 
Some links between the present derivations and the square-root 
formalism\cite{Schweber-QFT,Sucher-JMP-63,Castorina-84,Fiziev-85,Trubenbacher-ZFN-89,BriEngSus-ZFN-91,Lammerzahl-JMP-93,Namsrai-IJTP-98} are sketched out in Sec. \ref{sec:foldy_rep_funct}. 

We conclude this work by Sec. \ref{sec:pb_transform} where we show how some specific real-valued functionals together with the 
formalism of Poisson brackets allows to recover the transformation of the Poincar\'e group. 
We begin in Sec. \ref{sec:poincarre_pb} by using the classical Poisson brackets formalism to 
establish the algebra of the Poincar\'e group for a 1+1 dimensional.  
This allows us then in Sec. \ref{seq:algebra} to identify its Casimir invariant 
which happens to be the 
inner product introduced by Mostafazadeh\cite{Mostafazadeh-CQG-02}. 
We extend these results to the 3+1 Poincar\'e group in Sec. \ref{sec:full_poincarre}, 
and we then use the link between the Poisson brackets and commutators to recover the 
Foldy representation of the Poincar\'e group in Sec. \ref{sec:foldy_group} and its non relativistic limit. 
Eventually, some links with QFT are presented in Sec. \ref{sec:qft}. 
 
As the present derivations contain many intermediate results, 
we very often refer to the Appendix where calculations are reported in more details. 
Also, we acknowledge that this is a rather lengthy paper, and therefore we begin each section by a preambule 
contextualizing and summarizing the main results of that section.  
Eventually, a supplementary information file provides various important materials for the present derivations. 
More precisely, it contains a rather simple introduction to the use of Poisson bracket for flow transformations, 
the details of calculations of specific Poisson brackets, 
and eventually a derivation of the specific form of quantum operators corresponding to some important functionals. 

Also, as this work introduces various notations, we summarize them in Table \ref{table}. 
\begin{table*}[t]
\caption{\label{table}
        Summary of the different variables used in this work and their main features.
        We also recall that $a$ is the lattice parameter in the discrete model, $\lambda_c=\hbar/mc$, $\omega_0=mc^2/\hbar$, and $\hatom_v^\alpha=(1+2\hat{h}_v/mc^2)^{\frac{\alpha}{2}}$, where $\hat{h}_v$ is the usual non relativistic quantum hamiltonian with an external potential $v(x)$. 
        Here, "fast" means that the typical time variation is on the order of $(\omega_0)^{-1}$,
        while "slow" implies the typical time variations appears at a much larger time scale.
        }
\begin{tabular}{l|c|l|c|c|c}
 Variables       & definition & \,\,\, unit & discrete/continuous &real/complex & fast/slow  \\
\hline
$u_{n}(t)$            & transverse displacement   & $[L]$  & discrete  & real  & fast \\
$\psi_{n}(t)$         & $\psi_{n}(t)=\sqrt{a} \lambda_c u_{n}(t)$ &  $[L]^{-1/2}$  & discrete  & real  & fast \\
$\psi(x,t)$           & $\lim_{a\rightarrow 0}\psi_{n}(t)$ &  $[L]^{-1/2}$  & continuous & real  & fast \\
$\psilad_\alpha(x,t)$         
  & $\frac{1}{\sqrt{2}}\big(\hatom_v^{\alpha}\psi(x,t) + \frac{i}{\omega_0}\hatom_v^{1-\alpha}\dot{\psi}(x,t)\big)$ &  $[L]^{-1/2}$  & continuous & complex & fast \\
$\psilad(x,t)$         
  & $\frac{1}{\sqrt{2}}\big(\hatom_v^{\frac{1}{2}}\psi(x,t) + \frac{i}{\omega_0}\hatom_v^{-\frac{1}{2}}\dot{\psi}(x,t)\big)$ &  $[L]^{-1/2}$  & continuous & complex & fast \\
$\lad(x,t)$         & $e^{+i\omega_0 t}\psilad(x,t)$ &  $[L]^{-1/2}$  & continuous & complex & slow \\
\end{tabular}
\end{table*}

\section{The dynamics of real-valued Klein-Gordon fields from a classical phonon model}
\label{sec:pb_kg}
\subsection*{Summary and context}
In this section we briefly recall the classical phonon model at the heart of the present work, which is the 
linearized Frenkel-Kontorova\cite{FreKon-ZETF-38}, and how it allows to recover the real-valued solutions of the Klein-Gordon equation. 
More precisely, in Sec. \ref{sec:kg_cl} we give the description of this system through the classical Hamiltonian formalism, 
and then in Sec. \ref{sec:kg_fk} we show that a specific set of parameters allows to recover the 
dynamics of real-valued Klein-Gordon fields. 
\subsection{Hamilton equation of the linearised Frenkel-Kontorova model}
\label{sec:kg_cl}
\begin{figure}[t]
        \centering
        \includegraphics[width=\columnwidth]{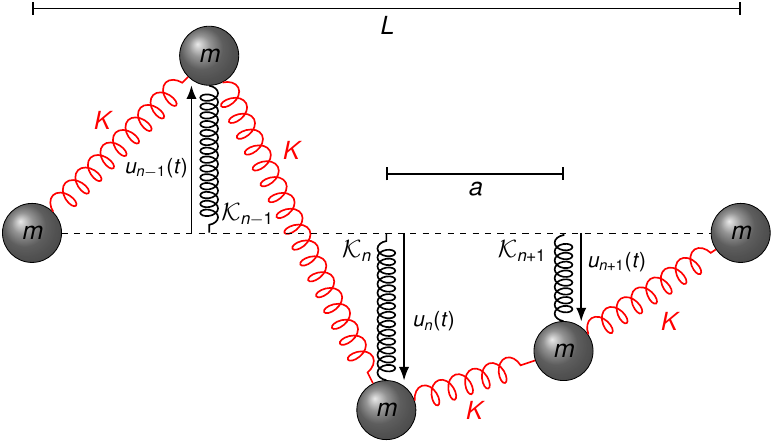}
        \caption{Pictorial representation of the Frenkel-Kontorova model used here: 
        $u_n(t)$ is the real number representing the transverse displacement, $\mathcal{K}_n$ the "on-site" harmonic force and $K$ the "between-site" harmonic force.  }
        \label{fig:FK_draw}
\end{figure}

Consider a one-dimensional regular lattice of periodicity $a$ and length $L$. 
At each vertex $\mathbf{X}_n=na \mathbf{e_x}, \,\, 1\le n \le N=L/a$ lies a mass $m$, attached to a spring, 
characterized by a constant $\mathcal{K}_n$ which can vary with $n$. 
All masses are identical. 
Each mass is allowed to move only in the transverse direction, and we 
label $u_n(t)\in\mathbb{R}$ the transverse displacement for the $n$-th mass at time $t$, 
and $\mathbf{u}(t)=\{ u_n(t),1\le n\le N \}$ is the set of all simultaneous displacement in the system. 
We impose that the displacements at the borders of the chain vanishes, \textit{i.e.} $u_1(t)=u_N(t)=0$. 
Also, each mass is connected to its two nearest neighbours, each one by a spring characterized by the constant $K$ 
and an equilibrium length of $a$.  
Therefore, the system is characterized by two types of forces: an "on-site" force, characterized by $\mathcal{K}_n$, 
and a "between-site" force, characterized by $K$. 
A pictorial representation of this model is shown in Fig. \ref{fig:FK_draw}.
This classical phonon model is the one-dimensional linearized Frenkel-Kontorova model\cite{FreKon-ZETF-38,BraKiv-Springer-04}, 
and we now describe its dynamics through the corresponding classical Hamilton's equation,  
which therefore requires the total Hamiltonian and Lagrangian of the system. 
As the system is allowed to move only in the transverse direction, we need only its 
transverse total kinetic and potential energies, and we will from now on drop the transverse specification 
for the sake of compaction of notation. 

The total kinetic energy of the system is simply 
\begin{equation}\label{eq:kinetic_u}
 T\big(\dot{\mathbf{u}}(t)\big) = \frac{m}{2}\sum_{n=1}^N  \dot{u}_n(t)^2,
\end{equation}
while its total potential energy, keeping only quadratic terms in the "between-site" terms, is given by 
\begin{equation}\label{eq:potential_u}
 \begin{aligned}
 V\big(\mathbf{u}(t)\big) & = \frac{1}{2}\bigg(\sum_{n=1}^{N}\mathcal{K}_n u_n(t)^2 + K(u_n(t) - u_{n+1}(t))^2\bigg), 
 \end{aligned}
\end{equation}
such that we can naturally define the total Lagrangian as 
\begin{equation}\label{eq:lagrang_0}
 \mathcal{L}
 \big(\mathbf{u}(t),\dot{\mathbf{u}}(t)\big) =  
 T\big(\dot{\mathbf{u}}(t)\big) - V\big(\mathbf{u}(t)\big),
\end{equation}
and the corresponding Hamiltonian as 
\begin{equation}\label{eq:hamilton_def_u}
 H\big(\mathbf{u}(t),\dot{\mathbf{u}}(t)\big) =  
 T\big(\dot{\mathbf{u}}(t)\big) + V\big(\mathbf{u}(t)\big).
\end{equation}
Having defined the Lagrangian of the system, we can then 
introduce the momentum $p_n(t)$ associated to the $n$-th mass as 
\begin{equation}\label{eq:def_PI}
 p_n(t) = \deriv{L\big(\mathbf{u}(t),\dot{\mathbf{u}}(t)\big)}{\dot{u}_n(t)}{} = m \dot{u}_n(t),  
\end{equation}
such that the corresponding Hamilton equations are then given by 
\begin{equation}\label{eq:hamilton_0}
 \begin{aligned}
&  \frac{d}{dt}u_n(t) = \deriv{H}{p_n(t)}{}, 
&  \frac{d}{dt}p_n(t) = - \deriv{H}{u_n(t)}{}.  
 \end{aligned}
\end{equation}
One can write the "between-site" spring constant as $K=\alpha/a^2$, such that the Hamilton's equation of 
Eq. \eqref{eq:hamilton_0} are then given by 
\begin{equation}\label{eq:hamilton_1}
 \begin{aligned}
&  \frac{d}{dt}u_n(t) = \frac{1}{m}p_n(t),\\
&  \frac{d}{dt}p_n(t) = \alpha \frac{u_{n+1}(t) + u_{n-1}(t) - 2 u_{n}(t)}{a^2} - \mathcal{K}_n u_n(t).
 \end{aligned}
\end{equation}
In order to smoothly perform the continuous limit, we introduce the rescaled variables $\psi_n(t)$ and $\pi_n(t)$ as follows 
\begin{equation}\label{eq:scaled_var}
u_n(t) = \lambda \sqrt{a} \psi_n(t), \quad
p_n(t) = \lambda \sqrt{a} \pi_n(t),
\end{equation}
where $\lambda$ is a positive real number which has the unit of a length. 
The introduction of these rescaled variables is equivalent to measuring $u_n(t)$ in units of $\lambda$ while 
introducing the variable $\psi_n(t)$ having the unit of $[L]^{-1/2}$. 
As the equations of motion of Eq. \eqref{eq:hamilton_1} are linear and homogeneous, they are also satisfied by the 
scaled variables $\psi_n(t)$ and $\pi_n(t)$. 

One can then consider the continuous limit for this classical system which is characterized by the following conditions 
\begin{equation}\label{eq:continuous_limit}
N\rightarrow \infty, \,a\rightarrow 0, \,Na =L, 
\end{equation}
where now 
\begin{equation}
 \begin{aligned}
 &\lim_{\substack{a \rightarrow 0 \\ N\rightarrow \infty}} {\psi}_n(t) = {\psi}(x,t), \,
  \lim_{\substack{a \rightarrow 0 \\ N\rightarrow \infty}} {\pi}_n(t) = {\pi}(x,t), \, \\
 &\lim_{\substack{a \rightarrow 0 \\ N\rightarrow \infty}} {\mathcal{K}}_n = \mathcal{K}(x), \\
 \end{aligned}
\end{equation}
\begin{equation}
 \begin{aligned}
  \lim_{\substack{a \rightarrow 0 \\ N\rightarrow \infty}} \frac{{\psi}_{n+1}(t) + {\psi}_{n-1}(t) - 2 {\psi}_n(t)}{a^2} = \deriv{{\psi}(x,t)}{x}{2},
 \end{aligned}
\end{equation}
such that the Hamilton equation of motion of Eq. \eqref{eq:hamilton_1} become then
\begin{equation}\label{eq:hamilton_3}
 \begin{aligned}
&  \frac{d}{dt}\psi(x,t) = \frac{1}{m}\pi(x,t),\\
&  \frac{d}{dt}\pi(x,t) = \alpha \deriv{{\psi}(x,t)}{x}{2} - \mathcal{K}(x) \psi(x,t). 
 \end{aligned}
\end{equation}

We can then rewrite the global energetic related quantities in the continuous limit. 
The kinetic and potential terms of Eq. \eqref{eq:kinetic_u} and Eq. \eqref{eq:potential_u} can then be written in 
terms of the rescaled variables $\big(\psi_n(t),\pi_n(t)\big)$ as follows 
\begin{equation}\label{eq:kinetic_pi}
 T\big(\mathbf{\Pi}(t)\big) = \frac{\lambda^2}{2m}\sum_{n=1}^N  a \,\,\pi_n(t)^2, 
\end{equation}
\begin{equation}\label{eq:potential_psi}
 \begin{aligned}
 V\big(\mathbf{\Psi}(t)\big) & = \frac{\lambda^2}{2}\sum_{n=1}^{N}a \bigg(\mathcal{K}_n \psi_n(t)^2 
 + \alpha\frac{\big(\psi_n(t) - \psi_{n+1}(t)\big)^2}{a^2}\bigg), 
 \end{aligned}
\end{equation}
where $\mathbf{\Pi}(t)=\{ \pi_n(t), 1\le n \le N\}$ and $\mathbf{\Psi}(t)=\{ \psi_n(t), 1\le n \le N\}$. 
Therefore, in the continuous limit the summations appearing in Eq. \eqref{eq:kinetic_pi} and \eqref{eq:potential_psi} properly become integrals, 
such that the functions of $\mathbf{\Pi}(t)$ and $\mathbf{\Psi}(t)$ 
become functionals of $\psi(x,t)$ and $\pi(x,t)$, \textit{i.e.} 
\begin{equation}\label{eq:kinetic_continous}
 \begin{aligned}
 T[\pi(t)] & \equiv \lim_{\substack{a \rightarrow 0 \\ N\rightarrow \infty}} 
  T\big(\mathbf{\Pi}(t)\big) \\
 & = \frac{\lambda^2}{2m} \int dx \,\,\pi(x,t)^2, \\
 V[\psi(t)] & \equiv \lim_{\substack{a \rightarrow 0 \\ N\rightarrow \infty}} V\big(\mathbf{\Psi}(t)\big) \\
 &= \frac{\lambda^2}{2} \int dx \bigg(\mathcal{K}(x) \psi(x,t)^2  + \alpha \big(\derivb{x}{} \psi(x,t)\big)^2\bigg).
 \end{aligned}
\end{equation}

The generalization to the two-dimensional case can be immediatly done by considering now a 
regular two-dimensional lattice of masses (say on the $(x,y)$ plane) which are allowed to move only in the transverse direction (the $z$ direction), 
which is similar to the "materas" model sketched by Zee in the introduction of his book on QFT\cite{QFTZee}. 
The transverse displacement $u_{n,m}(t)$ depends now on a couple of indices, 
and one can introduce the rescaled variables $u_{n,m}(t) = \lambda a\psi_{n,m}(t)$  
which become, in the continous limit, a function of time and the two-dimensional spacial coordinates, 
\textit{i.e.} $\psi(x,y,t)$. 
Then, we assume that the "on-site" spring is of strength $\mathcal{K}_{n,m}$ and 
that all masses are coupled through springs with only their nearest neighbours, such that the 
corresponding equation of motion become in the continuous limit
\begin{equation}\label{eq:hamilton_3d}
 \begin{aligned}
&  \frac{d}{dt}\psi(x,y,t) = \frac{1}{m}\pi(x,y,,t),\\
&  \frac{d}{dt}\pi(x,y,t) = - \mathcal{K}(x,y) \psi(x,y,t) \\
 & \qquad \qquad \qquad +\alpha \big(\deriv{{\psi}(x,y,t)}{x}{2} 
 + \deriv{{\psi}(x,y,t)}{y}{2}\big) .
 \end{aligned}
\end{equation}
When considering the global energetic quantities, the scaling units ensure that, in the continuous limit, 
the kinetic and total energies are expressed as integrals over $\mathbb{R}^2$, namely  
\begin{equation}\label{eq:kinetic_continous_2d}
 \begin{aligned}
 T[\pi(t)] & = \frac{\lambda^2}{2m} \int dx dy \,\,\pi(x,y,t)^2, \\
 V[\psi(t)] & = \frac{\lambda^2}{2} \int dx dy \bigg(\mathcal{K}(x,y) \psi(x,y,t)^2  \\ 
 & \qquad \qquad + \alpha \big(\derivb{x}{} \psi(x,y,t)\big)^2+ \alpha \big(\derivb{y}{} \psi(x,y,t)\big)^2\bigg).
 \end{aligned}
\end{equation}
The extension to the three-dimensional case is more conceptual as a transverse motion would 
require a fourth-dimensional spatial dimension, but one can simply assume that $u(x,y,z,t)$ 
represent the tension in an elastic medium whose unit corresponds to that of a length,
and whose equation of motion fulfills a classical equation of motion. 
For the sake of simplicity of notations, we will very often in this paper focus on the one-dimensional model unless 
stated explicitly.

\subsection{Link with the generalized Klein-Gordon equation}\label{sec:kg_fk}
As shown in the first part of this series\cite{Giner-schro}, we can connect the Frenkel-Kontorova model 
with the Klein-Gordon equation if we choose $\alpha = mc^2$ and $\mathcal{K}(x)$ as follows 
\begin{equation}\label{eq:k_x}
 \mathcal{K}(x) = m\omega_0^2\big(1+ 2 \frac{v(x)}{mc^2}\big),
\end{equation}
where $\omega_0=mc^2/\hbar$ is the pulsation associated to the rest mass energy. 
In Eq. \eqref{eq:k_x}, $v(x)$ has the units of an energy and can be thought as a potential energy 
which locally modifies the strength of the springs with respect to $m\omega_0^2$,  
and $v(x)$ will be frequently referred to as an "external potential". 
One can then combine the Hamilton's equation of Eq. \eqref{eq:hamilton_3} to obtain a generalization 
of the Klein-Gordon equation, \textit{i.e.} 
\begin{equation}
 \label{eq:kg_one_d}
 \begin{aligned}
  \ddot{\psi}(x,t) - c^2 \deriv{{\psi}(x,t)}{x}{2} +  \omega_0^2  (1+2\frac{v(x)}{mc^2}){\psi}(x,t) =0.
 \end{aligned}
\end{equation}
As discussed in the first part of this series\cite{Giner-schro} where the present author analyzed the dynamics of Klein-Gordon fields 
through the lens of this classical phonon system, $\omega_0^{-1}$ represents here the shortest time scale 
of the system, and the term $\omega_0^2\psi(x,t)$ in Eq. \eqref{eq:kg_one_d} generates 
very rapid "on-site" oscillations which induces inertia in the wave propagation.  
As the momentum $\pi(x,t)$  is simply proportional to $\dot{\psi}(x,t)$, from thereon we use $\dot{\psi}(x,t)$ 
as the conjugated variable to $\psi(x,t)$. Nevertheless, one should keep in mind that,  within the Hamiltonian formalism, 
the variable $\psi$ and its time derivative $\dot{\psi}$ are considered as two different variables 
coupled by the Hamiltonian equation of the form of Eq. \eqref{eq:hamilton_3}. 

Also, we choose the scaling parameter $\lambda$ to be the Compton wave length 
(\textit{i.e.} $\lambda_c= \hbar/mc$), which 
corresponds in measuring the displacement in the phonon field in units of $\lambda_c$. 
As a consequence, the total kinetic and potential energies of Eq. \eqref{eq:kinetic_continous} are then written as 
\begin{equation}\label{eq:kin_psin}
 T[\dot{\psi}(t)] = \frac{mc^2}{2}\int dx \bigg(\frac{\dot{\psi}(x,t)}{\omega_0}\bigg)^2,
\end{equation}
\begin{equation}\label{eq:pot_psi}
 V[\psi(t)] = \frac{1}{2}\int dx \bigg(\big(mc^2+2v(x)\big)\psi(x,t)^2 + \frac{\hbar^2}{m}\big( \derivb{x}{}\psi(x,t)\big)^2\bigg).
\end{equation}
We can then define the total energy associated to the phonon field $\psi(x,t)$ as 
\begin{equation}\label{eq:cl_htot}
 \begin{aligned}
 H_v[\psi(t),\dot{\psi}(t)] & = T[\dot{\psi}(t)] + V[\psi(t)]  \\
            & \equiv mc^2 \int dx \,\,\epsilon_v(\psi,\dot{\psi},x), 
 \end{aligned}
\end{equation}
where we introduced the energy density per units of $mc^2$ at a given point $x$ as follows 
\begin{equation}\label{eq:energy_rho}
 \begin{aligned}
 \epsilon_v(\psi,\dot{\psi},x) = 
 & \frac{1}{2} \bigg(\frac{\dot{\psi}(x,t)^2}{\omega_0^2} + \psi(x,t)^2\bigg) \\
 +& \frac{1}{2mc^2}\bigg(2v(x) \psi(x,t)^2 +  \frac{\hbar^2}{m} \big(\derivb{x}{}\psi(x,t)\big)^2\bigg).
 \end{aligned}
\end{equation}
We highlight here that we used $"v"$ as a lower script for both $H_v$ and $\epsilon_v$ to emphasize that 
these objects depends on the external potential $v(x)$. Also, we will use the notations $H$ and $\epsilon$ 
when $v(x)=0$, \textit{i.e.} $H\equiv H_{v=0}$ and $\epsilon \equiv \epsilon_{v=0}$. 

As the model uses only conservative forces (\textit{i.e.} harmonic springs), 
$H_v[\psi(t),\dot{\psi}(t)]$ is necessarily a constant of motion
\begin{equation}
 \frac{d}{dt}H_v[\psi(t),\dot{\psi}(t)] = 0.
\end{equation}
We can also naturally define the Lagrangian as
\begin{equation}
 \begin{aligned}
 L[\psi(t),\dot{\psi}(t)] & = T[\dot{\psi}(t)] - V[\psi(t)],\\
 \end{aligned}
\end{equation}
which, through the Euler-Lagrange equations, yields the generalized Klein-Gordon equation of Eq. \eqref{eq:kg_one_d}.  
Therefore, as long as the Klein-Gordon fields are not quantized, one can then consider the dynamics of these 
fields as the continuous limit of a classical system, \textit{i.e.} simply obeying Newton's or Hamilton's 
equation of motions. 

It is worth mentioning that 
the function $\epsilon(\psi,\dot{\psi},x)$ (\textit{i.e.} Eq. \eqref{eq:energy_rho} when $v=0$) 
is precisely the $T^{00}$ component of the energy-momentum tensor, 
which is recovered here naturally through the classical model. 
Also, we obtained without calculus that the corresponding integral of Eq. \eqref{eq:cl_htot} 
is a constant of motion due to the conservative nature of harmonic forces. 
These results are usually obtained through Noether's theorem applied to the Lorentz group. 

It is also interesting to connect the dynamics and total energy of this classical phonon system to the 
quantum non relativistic Hamiltonian $\hat{h}_v$ 
\begin{equation}\label{eq:nr_h}
 \hat{h}_v = -\frac{\hbar^2}{2m} \derivb{x}{2} + v(x),
\end{equation}
by introducing the following operator
\begin{equation}\label{eq:def_hatom_v}
 \hatom_v = \bigg(1 + 2 \frac{\hat{h}_v}{mc^2}\bigg)^{\frac{1}{2}}.
\end{equation}
Then, the Klein-Gordon equation of Eq. \eqref{eq:kg_one_d} is written in a compact way as follows
\begin{equation}\label{eq:kg_compact}
 \derivb{t}{2}\psi(x,t) + \omega_0^2 \hatom_v^2 \psi(x,t) = 0,
\end{equation}
and the total energy of Eq. \eqref{eq:cl_htot} is written as 
\begin{equation}\label{eq:total_h}
 H_v[\psi,\dot{\psi}] = \frac{1}{2}\int dx \bigg(mc^2\psi(x) \hatom_v^2 \psi(x) + \frac{\hbar}{\omega_0}\big(\dot{\psi}(x)\big)^2\bigg).
\end{equation}
As the case $v(x)=0$ is perticularly important for the relativistic regime, we introduce here the following notations 
\begin{equation}
 \hat{h} \equiv \hat{h}_{v=0} = -\frac{\hbar^2}{2m}\derivb{x}{2}, 
\end{equation}
\begin{equation}
 \hatom \equiv \hatom_{v=0} = \bigg(1 + 2 \frac{\hat{h}}{mc^2}\bigg)^{\frac{1}{2}}.
\end{equation}
As it will appear clearly along the paper, the operators $\hatom_v$ and $\hatom$ are central objects for the present work. 

\section{Dynamics and flow in phase space of Klein-Gordon fields through the continuous classical Poisson brackets}
\label{sec:dynamic_pb}
\subsection*{Summary and context}
Having established that the dynamics of real-valued Klein-Gordon fields can be obtained by 
the classical Hamiltonian formalism applied to the phonon model in the continuous limit, 
it is then natural to consider the corresponding classical Poisson bracket formalism, which is the aim of the present section.  
We begin by establishing in Sec. \ref{sec:continuous_pb} the general Poisson brackets formalism 
(\textit{i.e.} the generators of flow in phase space) in the continuous limit. 
We then introduce the two-components formulation of functionals and Poisson bracket in Sec. \ref{sec:two_comp}, 
which allows to compact notations. 
Then we study in Sec. \ref{eq:cst_motion} the constants of motion 
of the Klein-Gordon dynamics and show that two types of invariants can be obtained. 
The first category corresponds to linear and angular momentums, while the second category 
are positively defined constants, which can be used to define densities and time invariant inner products. 

We only review here the main results which are treated with more details in Appendix \ref{sec:pb_continous}.  
Also, one can refer to Sec. I of the supplementary materials for a brief overview of the Poisson bracket formalism 
for a single degree of freedom and its link with Galilean transformations. 

\subsection{Continuous Poisson brackets and flow transformations of Klein-Gordon fields}\label{sec:continuous_pb}
As the discrete phonon model is made of classical particles, it is natural to use the Poisson brackets formalism 
to generate the usual flow transformations in the classical phase space. 
More precisely, any function $G\big(\mathbf{u},\mathbf{p}\big)$ of the $2N$ variables 
$u_n$ and $p_n$ can be used to generate a transformation of the $2N$ phase space variables, 
which become $u_n(s)$ and $p_n(s)$ as they are now parametrized by a real number $s$ 
\notefn{As shown by Koopman\cite{Koopman-PNAS-31}, the classical functions $G\big(\mathbf{u},\mathbf{p}\big)$ of the $2N$ dimensional 
phase space can be associated with a Hilbert-space structure. We refer the reader to Ref. \cite{Barandes-EPJH-26} for 
a very pedagogical and insightful historical review 
of the main works on the Hilbert-space formulation of classical mechanics. }.
As an example, if 
$G\big(\mathbf{u},\mathbf{p}\big)=H_v\big(\mathbf{u},\mathbf{p}\big)$, 
the associated flow transformation simply generates 
the dynamics of the system, and therefore the parameter $s$ controlling the transformation is simply the time $t$, 
such that $u_n(t)$ and $p_n(t)$ are the time dependent coordinates and momentum of the system. 
The latter can nevertheless be generalized to perform transformations of the coordinates which are not 
directly related to the time evolution, such as translations, rotations or Galilean change of frame of reference 
(see Sec. I of the supplementary materials for an illustration with the Galilean transformations). 

To be more specific, a function $G\big(\mathbf{u},\mathbf{p}\big)$ 
generates a set of coupled flow equations which transform the variables $\big(\mathbf{u}(s),\mathbf{p}(s)\big)$ 
according to a continuous parameter $s$ as follows 
\begin{equation}\label{eq:pb_trans_0}
 \begin{aligned}
&  \frac{d}{ds}u_n(s) =  \! \deriv{G\big(\mathbf{u}(s),\mathbf{p}(s)\big)}{p_n(s)}{}, \,
  \frac{d}{ds}p_n(s) =\! - \deriv{G\big(\mathbf{u}(s),\mathbf{p}(s)\big)}{u_n(s)}{}.  
 \end{aligned}
\end{equation}
The form of Eq. \eqref{eq:pb_trans_0} can be seen as a generalization of the usual Hamilton's equations  
where we replaced $H_v$ by a generic function $G$, and $\big(\mathbf{u}(s),\mathbf{p}(s)\big)$ 
can be seen as a trajectory in phase space where now $s$ does not necessarily represent time. 
By changing the functional $G\big(\mathbf{u},\mathbf{p}\big)$, one changes the type of flow in phase space. 
Also, it is worth highlighting that the $s$ parameter has the units of $[A][G^{-1}]$, where $[A]$ is the unit 
of an action and $[G]$ is the unit of $G\big(\mathbf{u},\mathbf{p}\big)$, such that $s$ can be identified with a clear physical meaning. 
For instance, if $G=H$, the parameter $s$ has the unit of time. 

After some calculations detailed in Appendix \ref{sec:pb_continous}, 
when the scaled variables $\big(\psi_n,\pi_n\big)$ introduced in Eq. \eqref{eq:scaled_var} 
are used and after performing the continuous limit of Eq. \eqref{eq:continuous_limit}, the flow equations of Eq. \eqref{eq:pb_trans_0} are written 
as 
\begin{equation}\label{eq:flow_5_main}
 \begin{aligned}
 &\deriv{\psi(x,s)}{s}{} = \frac{\omega_0}{\hbar}\derfunc{G[\psi,\dot{\psi}]}{\dot{\psi}(x,s)} ,\\
 &\deriv{\dot{\psi}(x,s)}{s}{} = -\frac{\omega_0}{\hbar}\derfunc{G[\psi,\dot{\psi}]}{{\psi}(x,s)} ,\\
 \end{aligned}
\end{equation}
where the pre factor $\omega_0/\hbar$ comes from the fact that used scaled units, and where $G[\psi,\dot{\psi}]$ 
is the functional obtained from the continuous limit applied to the function $G\big(\mathbf{u},\mathbf{p}\big)$. 

Let us now consider another function $F\big(\mathbf{u},\mathbf{p}\big)$ of the variables $u_n$ and $p_n$, 
and establish how it is changed by a generic flow transformation of Eq. \eqref{eq:pb_trans_0}. 
This change is induced by the flow transformation generated by $G\big(\mathbf{u},\mathbf{p}\big)$ which 
implies that the variables $\big(\mathbf{u}(s),\mathbf{p}(s)\big)$ depend on $s$,  
and therefore the function $F\big(\mathbf{u}(s),\mathbf{p}(s)\big)$ itself acquires an indirect $s$ dependence. 
The derivative with respect to $s$ of the function $F$ can be obtained by using the chain rule 
and compactly written as 
\begin{equation}\label{eq:pb_transfo_discrete}
 \frac{d}{ds}F(s) = \big\{ F(s),G\big\}_N\quad ,
\end{equation}
where $\big\{ F, G\big\}_N$ is the Poisson bracket of two functions $F$ and $G$ depending on the $2N$ variables 
$(\mathbf{u},\mathbf{p})$
\begin{equation}\label{eq:pb_general_un}
 \begin{aligned}
 \big\{ F, G\big\}_N & \equiv 
 \sum_{n=1}^N \deriv{F(\mathbf{u},\mathbf{p})}{u_n}{} 
          \deriv{G(\mathbf{u},\mathbf{p})}{p_n}{} 
        - \deriv{F(\mathbf{u},\mathbf{p})}{p_n}{} 
          \deriv{G(\mathbf{u},\mathbf{p})}{u_n}{}, \\
 \end{aligned}
\end{equation}
and where we used the following notation $F(s) \equiv F\big(\mathbf{u}(s),\mathbf{p}(s)\big)$. 
Then, as shown in Annex \ref{sec:pb_continous}, if one expresses the Poisson bracket 
$\big\{ F, G\big\}_N$ in terms of the scaled variables $\big(\psi_n,\pi_n\big)$ 
previously introduced and then perform the continuous limit, it leads to 
\begin{equation}\label{eq:pb_limit}
 \begin{aligned}
 \lim_{\substack{a \rightarrow 0 \\ N\rightarrow \infty}} \big\{ F, G\big\}_N 
 &\equiv \big\{ F, G\big\}, \\
 \end{aligned}
\end{equation}
where the continuous Poisson bracket $\big\{ F, G\big\}$ is defined as 
\begin{equation}\label{eq:pb_limit}
 \begin{aligned}
\big\{ F, G\big\} &=  
 \frac{\omega_0}{\hbar} \int dx 
  \frac{\delta {F}[\psi,\dot{\psi}]}{\delta \psi(x)}  \frac{\delta {G}[\psi,\dot{\psi}]}{\delta \dot{\psi}(x)} 
- \frac{\delta {F}[\psi,\dot{\psi}]}{\delta \dot{\psi}(x)}  \frac{\delta {G}[\psi,\dot{\psi}]}{\delta {\psi}(x)},\\ 
 \end{aligned}
\end{equation}
and where the pre factor $\omega_0/\hbar$ again comes from the scaling units. 
Therefore in the continuous limit, when considering the $"s"$ dependence of a functional $F[\psi,\dot{\psi}]$ 
which is induced by the flow transformation generated by another functional $G[\psi,\dot{\psi}]$, 
it is simply given by  its continuous Poisson bracket, \textit{i.e.} 
\begin{equation}\label{eq:general_pb}
 \begin{aligned}
  \frac{d}{ds} F(s) = \big\{ F(s), G\big\}.
 \end{aligned}
\end{equation}

\subsection{The two-components formalism for functionals of real-valued Klein-Gordon fields} 
\label{sec:two_comp}
From here we will restrict our study to specific types of functionals which are relevant for 
the study of real-valued Klein-Gordon fields obeying a linear equation of motion, 
which leads to a two-components formulation of these functionals, allowing then to compact the notations. 
We then give the generic form of the Poisson bracket between any two of such functionals, 
which is the central object allowing us to compute flow transformations in phase space. 

We consider now a specific class of real-valued functionals which are written as  
\begin{equation}\label{eq:a_two-comp_0}
 \begin{aligned}
 G[\psi,\dot{\psi}] = & \int dx \psi(x) \hat{G}_{11} \psi(x) + \int dx \psi(x) \hat{G}_{12} \dot{\psi}(x) \\
                    + & \int dx \dot{\psi}(x) \hat{G}_{21} \psi(x) + \int dx \dot{\psi}(x) \hat{G}_{22} \dot{\psi}(x), \\
 \end{aligned}
\end{equation}
where $\hat{G}_{11}$, $\hat{G}_{21}$, $\hat{G}_{12}$ and $\hat{G}_{22}$ are linear operators 
with real-valued coefficients. 
The functionals of the type of Eq. \eqref{eq:a_two-comp_0} are therefore bilinear and/or at most quadratic 
in the variables $(\psi,\dot{\psi})$. 
The energetic quantities previously introduced fall in that category of functionals. 
For instance, the total energy of the classical phonon system of Eq. \eqref{eq:total_h} is explicitly written as in Eq. \eqref{eq:a_two-comp_0} 
with $\hat{H}_{11} = mc^2\hatom_v^2/2$, $\hat{H}_{12} = \hat{H}_{21} = 0$ and $\hat{H}_{22} = \hbar/2\omega_0$.
As the functions $\psi$ and $\dot{\psi}$ are real-valued, 
the operators $\hat{G}_{ij}$ need not to be hermitian to obtain a real-valued functional $G[\psi,\dot{\psi}]$. 
Nevertheless, because the diagonal element of an anti hermitian operator vanish for functions 
with Dirichlet boundary conditions, \textit{i.e.} 
\begin{equation}
 \begin{aligned}
 \int_a^b dx f(x) \hat{G}_{ah} f(x) = 0 &\text{ if  }\hat{G}_{ah}^\dagger = -\hat{G}_{ah} \\&\text{ and }f(b)=f(a)=0,
 \end{aligned}
\end{equation}
we can choose, without loss of generality, the operators $\hat{G}_{11}$ and $\hat{G}_{22}$ to be hermitian
\begin{equation}
 \hat{G}_{11} = \hat{G}_{11}^\dagger, \quad 
 \hat{G}_{22} = \hat{G}_{22}^\dagger.
\end{equation}
We can then introduce the two-components vector $\Psi(x)=\big(\psi(x),\dot{\psi}(x)\big)$ 
such that the functional of Eq. \eqref{eq:a_two-comp_0} can be written as a quadratic form 
\begin{equation}\label{eq:g_psi}
 \begin{aligned}
 G[\Psi] =  \int dx \Psi^\dagger(x) \,\hat{\mathcal{G}} \,\Psi(x),  
 \end{aligned}
\end{equation}
where the $2\times 2$ matrix $\hat{\mathcal{G}} $ is defined as 
\begin{equation}
 \hat{\mathcal{G}} = 
 \begin{pmatrix}
  \hat{G}_{11} &  \hat{G}_{12} \\
  \hat{G}_{21} &  \hat{G}_{22}
 \end{pmatrix}.
\end{equation}
The operators $\hat{G}_{11}$ and $\hat{G}_{22}$ will be often referred to as "diagonal operators", 
while $\hat{G}_{12}$ and $\hat{G}_{21}$ are referred to as "off-diagonal operators". 
We illustrate this notation in the case of the total energy, where Eq. \eqref{eq:total_h} is written in the two-components framework simply as 
\begin{equation}
 H_v[\Psi] = \frac{1}{2}\int dx \Psi(x)^\dagger  
\begin{pmatrix} 
 mc^2\hatom_v^2   & 0 \\
 0 & \frac{\hbar}{\omega_0} \\
\end{pmatrix} 
 \Psi(x). 
\end{equation}
Considering now another functional $F[\psi,\dot{\psi}]$ written as in Eq. \eqref{eq:g_psi}, 
the $\{F,G \}$ Poisson bracket as defined in Eq. \eqref{eq:pb_limit} can also be written as a functional 
of the type of Eq. \eqref{eq:g_psi}. 
As shown in Annex \ref{ann:pb_general}, the Poisson bracket $\{F,G \}$ between two functionals 
is then 
\begin{equation}\label{eq:poisson_final_main}
 \begin{aligned}
 \{F,G \}  & = \int dx \Psi^\dagger(x) \{\hat{\mathcal{FG}}\}
   \Psi(x) ,
 \end{aligned}
\end{equation}
where the matrix $\{\hat{\mathcal{FG}}\}$ reads 
\begin{equation}\label{eq:pb_matrix}
 \begin{aligned}
\{\hat{\mathcal{FG}}\}\! = \!
2\frac{\omega_0}{\hbar} 
    \begin{pmatrix}
 \hat{F}_{11} \hat{g}_{21}-\hat{G}_{11}\hat{f}_{21} & 2 \big(\hat{F}_{11}\hat{G}_{22}-\hat{G}_{11}\hat{F}_{22}\big)\\
\frac{1}{2}\big(\hat{f}_{12}^\dagger\hat{g}_{21}-\hat{g}_{12}^\dagger\hat{f}_{21}\big) 
 & \hat{f}_{12}^\dagger \hat{G}_{22} - \hat{g}_{12}^\dagger\hat{F}_{22}\\
    \end{pmatrix},
 \end{aligned}
\end{equation}
and where the operators $\hat{f}_{12}$ and $\hat{f}_{21}$ are defined as follows 
\begin{equation}\label{eq:def_alpha_main}
 \begin{aligned}
 \hat{f}_{12} &= \hat{F}_{12}+ \hat{F}_{21}^\dagger, \quad \hat{f}_{21} = \hat{F}_{21}+ \hat{F}_{12}^\dagger,
 \end{aligned}
\end{equation}
and similarly for the functional $G$ with the $\hat{g}_{12}$ and $\hat{g}_{21}$ operators. 
Having established the generic form of the Poisson bracket for two functionals, 
we can now use it to study the dynamics and generic flow transformations in phase space. 

\subsection{Constants of motion from the generic form of $\{F,H_v\}$: inner products and momentums }
\label{eq:cst_motion}
In view of studying the dynamics of the functionals of Klein-Gordon fields through the lens of classical dynamics 
(see Sec. \ref{sec:ehrenfest_big}), 
in this section we first study the structure of the $\{F,H_v\}$ Poisson bracket. 
Based on the result of Eq. \eqref{eq:poisson_final_main} and Eq. \eqref{eq:pb_matrix}, we give in Sec. \ref{sec:h_pb} the generic 
form of $\{F,H_v\}$ which allows to find 
the conditions for $F$ to be a constant of motion (\textit{i.e.} $\{F,H_v\}=0$). 
These conditions are then grouped into two distinct categories, 
which are studied in Sec. \ref{sec:inner} and Sec. \ref{sec:momentum}, respectively. 
From the first category one can introduce a class of time invariant inner products on Klein-Gordon fields, 
while the second category can be related to linear and angular momentum.  
The details of the calculations of this section can be found in the Appendix \ref{ann:der_cst}. 

\subsubsection{Dynamics and constants of motion for functionals of Klein-Gordon fields}\label{sec:h_pb}
Applying the general Poisson bracket formalism of Eq. \eqref{eq:general_pb} to the case where $G$ is 
the total energy of the system, we obtain the time dependence of any functional $F$ through 
\begin{equation}\label{eq:dynamic_def}
 \frac{d}{dt}F(t) = \{ F(t),H_v\}.
\end{equation}
We can then use the general form of the Poisson bracket of Eq. \eqref{eq:poisson_final_main}, 
which yields 
\begin{equation}\label{eq:h_pb_main_expl}
 \begin{aligned}
 \{F,H_v \} 
 = & \int dx  \dot{\psi}(x)\hat{f}_{12}^\dagger \dot{\psi}(x) 
- \frac{\omega_0}{\hbar}\int dx \psi(x) mc^2\hatom_v^2 \hat{f}_{21}\psi(x) \\
 +&\frac{2\omega_0}{\hbar} \int dx \psi(x) \big( \frac{\hbar}{\omega_0}\hat{F}_{11}- mc^2\hatom_v^2 \hat{F}_{22} \big) \dot{\psi}(x). 
 \end{aligned}
\end{equation}
Among all functionals $F$, the constants of motion are of particular importance and are characterized by $\{F,H_v \} = 0$. 
As detailed in Annex \ref{ann:der_cst}, the conditions on $F$ allowing $\{F,H_v \} = 0$  
can be grouped into two independent constraints on the couples of of diagonal operators $(\hat{F}_{11},\hat{F}_{22})$ and 
off-diagonal operators $(\hat{F}_{12},\hat{F}_{12})$. 
More precisely, if the diagonal operator are such that 
\begin{equation}\label{eq:cst_mot_diag}
 \hat{F}_{11} = \hatom_v^{2\zeta}  \text{ and } \hat{F}_{22} = \frac{1}{\omega_0^2}\hatom_v^{2(\zeta-1)}, \zeta \in \mathbb{R},\\
\end{equation}
and the off-diagonal operators are such that 
\begin{equation}\label{eq:cst_mot_odiag}
 \begin{aligned}
  \big[\hat{F}_{21} + \hat{F}_{12}^\dagger,\hat{h}_v \big]= 0, \text{ and } 
\big( \hat{F}_{21} + \hat{F}_{12}^\dagger\big)^\dagger = - \big(\hat{F}_{21} + \hat{F}_{12}^\dagger\big),
 \end{aligned}
\end{equation}
then $\{F,H_v\}=0$  and as a consequence, $F$ is a constant of motion. 
We now analyze how these two different constraints yield two different types of functionals.  

\subsubsection{Positive constants of motion and inner products}\label{sec:inner}
In the present section we focus on the conditions for the diagonal operators by analyzing Eq. \eqref{eq:cst_mot_diag}. 
We therefore assume in the present section that $\hat{F}_{12}=\hat{F}_{21}=0$. 
We then introduce the following class of functionals parametrized by the real number $\zeta$ 
\begin{equation}\label{eq:cst_scalar_main}
 \begin{aligned}
 \mathcal{C}_\zeta[\psi,\dot{\psi}] = 
 & \frac{1}{2}\int dx \psi(x,t) \hatom_v^{2\zeta}\psi(x,t) \\
 +& \frac{1}{2\omega_0^2} \int dx \dot{\psi}(x,t) \hatom_v^{2(\zeta-1)} \dot{\psi}(x,t), \quad \zeta\in \mathbb{R}, 
 \end{aligned}
\end{equation}
which are dimensionless constants of motion under the Klein-Gordon dynamics as they fulfill 
\begin{equation}
 \{ \mathcal{C}_\zeta, H_v\}=0,\quad\forall\,\,\zeta \in \mathbb{R}. 
\end{equation}
An important property of the functionals defined in Eq. \eqref{eq:cst_scalar_main} is that, 
because $\hatom_v=(1+2\hat{h}_v/mc^2)^\frac{1}{2}$, if the spectrum of the non relativistic Hamiltonian $\hat{h}_v$ 
is strictly bounded by $-mc^2/2$, then the functionals $ \mathcal{C}_\zeta[\psi,\dot{\psi}]$ 
are strictly positive whatever the value of $\zeta$, \textit{i.e.}  
\begin{equation}\label{eq:condition_positive}
 \begin{aligned}
 &\text{if } \hat{h}_v \phi_n(x) = e_n \phi_n(x), \quad e_n > -\frac{mc^2}{2} \\
 &\text{then }\mathcal{C}_\zeta[\psi,\dot{\psi}] > 0\quad  \forall \,\,\zeta\text{ and }\forall \,\, (\psi,\dot{\psi}).
 \end{aligned}
\end{equation}
We can then use the positivity of Eq. \eqref{eq:condition_positive} together with 
the quadratic structure of $\mathcal{C}_\zeta[\psi,\dot{\psi}]$ 
in terms of $\psi$ and $\dot{\psi}$ to introduce a class of inner products 
on the vector space of Klein-Gordon fields. More precisely, given a couple of 
two-dimensional vectors $\Psi_1=(\psi_1,\dot{\psi}_1)$ and $\Psi_2=(\psi_2,\dot{\psi}_2)$ of Klein-Gordon fields, 
we define their inner product, hereafter labelled $\inner{\Psi_1}{\Psi_2}^\zeta$, as follows 
\begin{equation}\label{eq:inn_prod}
 \begin{aligned}
 \inner{\Psi_1}{\Psi_2}^\zeta \equiv 
 & \frac{1}{2}\int dx \psi_1(x,t) \hatom_v^{2\zeta}\psi_2(x,t) \\
 +& \frac{1}{2\omega_0^2} \int dx \dot{\psi}_1(x,t) \hatom_v^{2(\zeta-1)} \dot{\psi}_2(x,t), \quad \zeta\in \mathbb{R}. 
 \end{aligned}
\end{equation}
It is easy to show that the definition of Eq. \eqref{eq:inn_prod} fulfills the axioms of inner products.  
The inner products of Eq. \eqref{eq:inn_prod} are parametrized by the real number $\zeta$ and 
are time invariant with respect to the Klein-Gordon dynamics. 
Therefore, the Klein-Gordon dynamic is unitary with respect to this class of inner product, \textit{i.e.} 
\begin{equation}
 \frac{d}{dt} \inner{\Psi_1(t)}{\Psi_2(t)}^\zeta = 0. 
\end{equation}
Among the set of positive constants of motion $\mathcal{C}_\zeta$ of the form of Eq. \eqref{eq:cst_scalar_main}, 
and the associated inner products, some are well known functionals. 
The case where $\zeta=1$ is simply proportional to the total energy, \textit{i.e.}  
\begin{equation}\label{eq:total_e_inner}
 H_v[\psi,\dot{\psi}] = mc^2\inner{\Psi}{\Psi}^{\zeta=1},
\end{equation}
which implies that the energy itself could be used as a time-independent inner product \notefn{This is actually the choice done in the first part of this series\cite{Giner-schro} which focusses on the non relativistic regime}. 
Another interesting case is when $\zeta=1/2$ where the inner product of Eq. \eqref{eq:inn_prod} then reads
\begin{equation}\label{eq:mosta_inner_v}
 \begin{aligned}
 \inner{\Psi_1}{\Psi_2}^{\zeta=\frac{1}{2}} =   
 & \frac{1}{2}\int dx \psi_1(x,t) \hatom_v \psi_2(x,t) \\
 +& \frac{1}{2\omega_0^2} \int dx \dot{\psi}_2(x,t)\hatom_v^{-1} \dot{\psi}_1(x,t), 
 \end{aligned}
\end{equation}
and in the specific case of free Klein-Gordon fields (\textit{i.e.} $v(x)=0$), 
the functional of Eq. \eqref{eq:mosta_inner_v} is explicitly written as 
\begin{equation}\label{eq:number_part}
 \begin{aligned}
 \inner{\Psi_1}{\Psi_2}^{\zeta=\frac{1}{2}}  =  &\frac{1}{2}\int dx \psi_1(x,t) \hatom \psi_2(x,t) \\
 + &\frac{1}{2\omega_0^2} \int dx \dot{\psi}_2(x,t) \hatom^{-1} \dot{\psi}_1(x,t),  
 \end{aligned}
\end{equation}
where we recall that $\hatom = \big( 1 -\lambda_c^2 \derivb{x}{2}\big)^{\frac{1}{2}}$ is the restriction of $\hatom_v$ when $v(x)=0$. 
The functional of Eq. \eqref{eq:number_part} is the Lorentz invariant inner-product 
on the space of real-valued Klein-Gordon fields which was introduced by Mostafazadeh 
in Ref. \onlinecite{Mostafazadeh-CQG-02}.    
Also, when considering the norm associated to Eq. \eqref{eq:number_part} (\textit{i.e.} when $\Psi_1=\Psi_2$), 
we obtain the following functional 
\begin{equation}\label{eq:norm_psi}
 \begin{aligned}
 \mathcal{N}[\psi,\dot{\psi}] = 
 & \frac{1}{2}\int dx \psi(x,t) \hatom \psi(x,t) \\
 + &\frac{1}{2\omega_0^2} \int dx \dot{\psi}(x,t)\hatom^{-1} \dot{\psi}(x,t), 
 \end{aligned}
\end{equation}
which coincides with the functional called "number of particles" in the work of Barros and Gomes
(see Eq. 21 of Ref. \onlinecite{BarGom-EPJC-21}). 
As it will be shown and discussed in more details in Sec. \ref{seq:algebra}, 
the functional $\mathcal{N}[\psi,\dot{\psi}]$ 
of Eq. \eqref{eq:norm_psi} is the Casimir invariant of the Poincar\'e group associated to the mass of the field. 
Therefore, such functional is not only time invariant but also invariant with respect to all other symmetry 
operations of the Poincar\'e group, \textit{i.e.} Lorentz boosts, rotations and spatial translations. 
For this reason, we use the form of Eq. \eqref{eq:mosta_inner_v} to define the inner product on Klein-Gordon fields, 
and therefore we introduce the following notation 
\begin{equation}\label{eq:def_innere}
 \innere{\Psi_1}{\Psi_2} \equiv \inner{\Psi_1}{\Psi_2}^{\zeta=\frac{1}{2}}.
\end{equation}
Also, as the Klein-Gordon equation is linear and homogeneous, 
its dynamic is stable by multiplication by any scalar, so its solution can be arbitrarily multiplied 
by a scaling factor without changing the dynamics. 
Nevertheless, functionals such as the total energy are not invariant with respect to an arbitrary scaling factor, 
which suggests to either normalized the vectors $\Psi=(\psi,\dot{\psi})$ or to normalize the functionals themselves, 
which is equivalent. 
We will therefore work with states $\big(\psi,\dot{\psi}\big)$ which are normalized to unity with respect to the 
norm associated to the scalar product of Eq. \eqref{eq:def_innere}, \textit{i.e.} 
\begin{equation}\label{eq:normalized_psi}
 \innere{\Psi}{\Psi}  = 1, 
\end{equation}
and as the inner product is time invariant  it implies that the norm of the vector is conserved in the dynamics. 

Eventually, it is noteworthy that, because $\lim_{c\rightarrow \infty} \hatom_v^\zeta = 1$ (see Appendix \ref{sec:hatom_taylor}), 
all these inner products are equal in the non relativistic limit, \textit{i.e.} 
\begin{equation}
 \lim_{c \rightarrow \infty} \inner{\Psi_1}{\Psi_2}^\zeta = \int dx 
 \big( \psi_1(x) \psi_2(x) + \frac{1}{\omega_0^2} \dot{\psi}_1(x) \dot{\psi}_2(x)\big).  
\end{equation}

\subsubsection{Momentum and its generalization}\label{sec:momentum}
Another class of constants of motion appears when considering the off-diagonal operators, 
such as the stationary condition $\{F,H_v\}=0$ of Eq. \eqref{eq:cst_mot_odiag} reads then 
\begin{equation}\label{eq:commut_f12}
 \big[\hat{F}_{21} + \hat{F}_{12}^\dagger,\hat{h}_v \big] = 0,
\end{equation}
where we recall that $\hat{h}_v$ is the usual non relativistic quantum Hamiltonian (see Eq. \eqref{eq:nr_h}). 
We treat first the case where $v(x)=0$, such that Eq. \eqref{eq:commut_f12} 
necessarily implies that both $\hat{F}_{21}$ and $\hat{F}_{12}$ are linear combination of 
differential operators, \textit{i.e.} 
\begin{equation}
 \big[\hat{F}_{21} + \hat{F}_{12}^\dagger,\derivb{x}{2} \big] = 0 \Leftrightarrow 
  \hat{F}_{21} = \sum_{n} a_n \derivb{x}{n}, \quad \hat{F}_{21} = \sum_{n} b_n \derivb{x}{m}. 
\end{equation}
The anti hermitian condition of Eq. \eqref{eq:cst_mot_odiag}  necessarily implies that 
both operators $\hat{F}_{21}$ and $\hat{F}_{12}$ are anti hermitian, such that the differential operators 
must be only of odd orders, \textit{i.e.} 
\begin{equation}
 \begin{aligned}
 &\big( \hat{F}_{21} + \hat{F}_{12}^\dagger\big)^\dagger = - \big(\hat{F}_{21} + \hat{F}_{12}^\dagger\big) \\
 \Leftrightarrow 
 & \hat{F}_{21} = \sum_{n} a_n \derivb{x}{2n+1} \text{ and } \hat{F}_{21} = \sum_{n} b_n \derivb{x}{2m+1}. 
 \end{aligned}
\end{equation}
Therefore, any functional of the form 
\begin{equation}
 \begin{aligned}
 \label{eq:momentum_general}
 F[\psi,\dot{\psi}] = & \sum_{n} a_n \int dx \dot{\psi(x)}\derivb{x}{2n+1} \psi(x) \\+ 
                      & \sum_{n} b_n \int dx  \psi(x)\derivb{x}{2n+1}\dot{\psi(x)},  
 \end{aligned}
\end{equation}
is a constant of motion, 
and through integration by parts, the non nullity of the functional $F[\psi,\dot{\psi}]$ of Eq. \eqref{eq:momentum_general} 
is guaranteed as long as $a_n \ne b_n$.
Among the infinite set of constants of motion of the type of Eq. \eqref{eq:momentum_general}, 
two important special cases are the following functionals 
\begin{equation}\label{eq:momentum_classic}
 {P}[\psi,\dot{\psi}] = -m\lambda_c^2 \int dx \dot{\psi}(x)\derivb{x}{} \psi(x) ,
\end{equation}
\begin{equation}\label{eq:momentum_relat}
 {\mathcal{P}}[\psi,\dot{\psi}] = -m\lambda_c^2 \int dx \dot{\psi}(x)\derivb{x}{} \hatom^{-1} \psi(x) ,
\end{equation}
where the $"-m\lambda_c^2"$ prefactor has been put for further strict identification with various definitions of momentums. 
As it will be shown in Sec. \ref{sec:ehrenfest}, the functionals of Eq. \eqref{eq:momentum_classic} 
and Eq. \eqref{eq:momentum_relat} correspond to the momentum associated with different definitions of 
the average position of the field.  
The two definitions of Eq. \eqref{eq:momentum_classic} and Eq. \eqref{eq:momentum_relat} are nevertheless 
both consistent with the fact that if the system is invariant by translations, 
they are constant in time, thus justifying their label of momentum. 

Another interesting case is when considering a three-dimensional system with a central potential 
(\textit{i.e.} $v(\bfr{r}) = v(|\bfr{r}|)$). 
In that case, we know from quantum mechanics textbooks that the operator $\hat{\bfr{l}}$ defined as 
\begin{equation}\label{eq:ang_mom} 
\hat{\bfr{l}} = \bfr{r} \times \nabla 
\end{equation}
commutes with $\hat{h}_v$. Therefore, because $\nabla$ is anti hermitian, 
we know that $\hat{\bfr{l}}$ as defined in Eq. \eqref{eq:ang_mom} 
is also anti hermitian, which therefore fulfills the condition of Eq. \eqref{eq:cst_mot_odiag} to be 
a constant of motion. 
As a consequence, we know that the following functional 
\begin{equation}\label{eq:ang_mom_tot}
 \bfr{L}[\psi,\dot{\psi}] = -m\lambda_c^2 \int d\bfr{r} \dot{\psi}(\bfr{r})\,\,\bfr{r} \times \nabla \psi(\bfr{r}),  
\end{equation}
is a constant of motion, where the prefactor $"\!\!-m\lambda_c^2"$ has been put for further identification 
with the angular momentum. 

We conclude this section by highlighting that the linear momentum of Eq. \eqref{eq:momentum_classic} corresponds to the energy-momentum tensor $T^{01}$ 
associated to a Lorentz boost, while the angular momentum of Eq. \eqref{eq:ang_mom_tot} corresponds to the conserved quantity associated 
to a rotation. 
We therefore recovered some results usually obtained from relativistic symmetry consideration 
using only the classical Poisson bracket formalism. 

\section{Applications of the Poisson bracket formalism: Ehrenfest relations}
\label{sec:ehrenfest_big}
\subsection*{Summary and context}
As an application of the formalism of classical Poisson brackets introduced previously,  
in the present section we describe the dynamics of the functionals of the Klein-Gordon fields 
which are associated to the position, momentum and forces. 
We begin in Sec. \ref{sec:pos} by introducing two different functionals to estimate the localization of the field. 
Then, in Sec. \ref{sec:ehrenfest} we use the Poisson bracket formalism to obtain the dynamics corresponding to 
the two different position functionals, which leads to different definitions of momentum functionals. 
\subsection{Different choices of localization estimates}\label{sec:pos}
As a first application of the classical Poisson brackets formalism, we propose here to derive the dynamics 
of the position and momentum functionals, 
\textit{i.e.} the analogue of the usual Ehrenfest relations of the non relativistic\cite{Ehrenfest-27} 
and relativistic\cite{ReiFer-PRL-91} spin zero quantum mechanics. 
The starting point of these derivations are the definition of the localization of the field, 
for which there is no unique choice in the case of Klein-Gordon fields because of the requirements of causality\cite{NewWig-RMP-49,Wightman-62,Hegerfeldt-74,Hegerfeldt-85,MosZam-AP-06,Moretti-23}. 
As the latter goes beyond this work and because our aim is to illustrate the Poisson brackets formalism, 
we will simply here consider two particular choices and show how it leads to different definitions of momentums. 

The position functionals that we consider here share the same conceptual content, 
which is that they are written as the barycentre of the position over a density, \textit{i.e.}
\begin{equation}\label{eq:def_pos_general}
 \tilde{O}_\rho[\psi,\dot{\psi}] = \frac{O_\rho[\psi,\dot{\psi}]}{ \int \, dx \, \, \rho(\psi,\dot{\psi},x)},
\end{equation}
where the unnormalized functional $O_\rho[\psi,\dot{\psi}]$ is defined as 
 \begin{equation}
 O_\rho[\psi,\dot{\psi}] = \int \, dx \, \, x \, \, \rho(\psi,\dot{\psi},x),
 \end{equation}
with $\rho(\psi,\dot{\psi},x)$ being the density (\textit{i.e.} a positively defined function). 
Therefore, by varying the density $\rho$ one changes the definition of the localization of the field. 
We propose here to consider to types of densities: one given by the energy density $\epsilon_v$ of Eq. \eqref{eq:energy_rho}
and one inspired by the class of positively defined constants of motion 
$\mathcal{C}_\zeta[\psi,\dot{\psi}]$ obtained in Sec. \ref{sec:inner}
(see Eq. \eqref{eq:cst_scalar_main}). 
Indeed, to each positively defined constants of motion $\mathcal{C}_\zeta[\psi,\dot{\psi}]$ we can associate the density $\rho_\zeta(\psi,\dot{\psi},x)$ which verifies
\begin{equation}
 \mathcal{C}_\zeta[\psi,\dot{\psi}] = \int dx \rho_\zeta(\psi,\dot{\psi},x),
\end{equation}
and, using the fact that $\hatom_v$ is hermitian, $\rho_\zeta$ can be explicitly written as 
\begin{equation}\label{eq:def_measure_zeta}
 \rho_\zeta(\psi,\dot{\psi},x)= \frac{1}{2}  \bigg(\hatom_v^{\zeta} \psi(x,t)\bigg)^2 
 + \frac{1}{2\omega_0^2} \bigg(\hatom_v^{\zeta-1} \dot{\psi}(x,t)\bigg)^2.
\end{equation}

Let us first consider the density associated to the inner product of Mostafazadeh (\textit{i.e.} $\zeta=1/2$ in Eq. \eqref{eq:cst_scalar_main}) which therefore verifies the following property
\begin{equation}\label{eq:measure_baros}
 \begin{aligned}
 \int dx \,\,\mbaros(\psi,\dot{\psi},x) = \innere{\Psi}{\Psi}. 
 \end{aligned}
\end{equation}
The explicit form of $\mbaros$ can be obtained by setting $\zeta=1/2$ in Eq. \eqref{eq:def_measure_zeta}, which then reads
\begin{equation}\label{eq:measure_baros_expl}
 \mbaros(\psi,\dot{\psi},x) = \frac{1}{2}\big(\hatom_v^{\frac{1}{2}}\psi(x)\big)^{2} + 
  \frac{1}{2\omega_0^2}\big(\hatom_v^{-\frac{1}{2}}\dot{\psi}(x)\big)^{2} .
\end{equation}
Using the density of Eq. \eqref{eq:measure_baros_expl} is rigorously equivalent to the position operator defined Barros and Gomes\cite{BarGom-EPJC-21} when $v(x)=0$. 
Also, as we work with vectors $\Psi=\big(\psi,\dot{\psi}\big)$ normalized to unity precisely according 
to the density $\innere{\Psi}{\Psi}$ (see Eq. \eqref{eq:normalized_psi}), the integral of the density appearing in the denominator 
of Eq. \eqref{eq:def_pos_general} is equal to one, such that the normalized position functional reduces then to
\begin{equation}\label{eq:def_x_baros}
 \tilde{\bary}[\psi,\dot{\psi}] = \int dx \,\, x\,\, \mbaros(\psi,\dot{\psi},x).  
\end{equation}
By using the fact that $\hatom_v$ is hermitian, the corresponding position functional can then be written in a two-components framework as follows 
\begin{equation}\label{eq:barybaros}
 \tilde{\bary}[\psi,\dot{\psi}] = \frac{1}{2} \int dx \Psi^\dagger(x) 
 \begin{pmatrix}
 \hatom_v^{\frac{1}{2}} x \hatom_v^{\frac{1}{2}} & 0 \\
 0 & \frac{1}{\omega_0^2}\hatom_v^{-\frac{1}{2}} x \hatom_v^{-\frac{1}{2}} \\
 \end{pmatrix}
 \Psi(x).
\end{equation}
This two-components form will be useful in Sec. \ref{sec:func_alpha} to rewrite $\tilde{\bary}$ 
as the usual position operator of quantum mechanics. 

We then proceed with the position functional corresponding to the energy density of Eq. \eqref{eq:energy_rho},  
such that the corresponding position of the field is defined as the energy barycentre, \textit{i.e.} 
\begin{equation}\label{eq:barycenter_x}
 \tilde{X}[\psi,\dot{\psi}] = \frac{X[\psi,\dot{\psi}] }{\norm[\psi,\dot{\psi}]},
\end{equation}
where 
\begin{equation}\label{eq:x_no_norm}
 X[\psi,\dot{\psi}] = \int dx \,\,x \,\,\epsilon_v(\psi,\dot{\psi},x),
\end{equation}
\begin{equation}\label{eq:def_gamma}
 \norm[\psi,\dot{\psi}] = \int dx \epsilon_v(\psi,\dot{\psi},x),
\end{equation}
and we recall that $\epsilon_v(\psi,\dot{\psi},x)$ is the positively defined energy density of Eq. \eqref{eq:energy_rho}. 
The definition of $\norm$ in Eq. \eqref{eq:def_gamma} allows to rewrite the total energy as 
\begin{equation}
 H_v[\psi,\dot{\psi}] = mc^2\norm[\psi,\dot{\psi}] ,
\end{equation}
and therefore $\norm[\psi,\dot{\psi}]$ bears a similar meaning than that of the Lorentz factor of special relativity 
(see Sec. \ref{sec:summary} for more detailed discussion). 

\subsection{Ehrenfest relations from classical Poisson brackets}
\label{sec:ehrenfest}
Having defined two types of position functionals, we can then determine their time dependence through the Poisson bracket formalism. 
The time evolution of any normalized functional $\tilde{O}_\rho(t)$ is then given by the Poisson bracket relation, \textit{i.e.} 
\begin{equation}
 \frac{d}{dt} \tilde{O}_\rho(t) = \big\{ \tilde{O}_\rho(t)  , H_v  \big\},
\end{equation}
and as the normalized functionals $\tilde{O}_\rho$ are obtained with a time invariant density, 
it is then given by 
\begin{equation}
 \frac{d}{dt} \tilde{O}_\rho(t) = \frac{1}{\int dx \rho(x) }\big\{ O_\rho(t)  , H_v  \big\}.
\end{equation}
We therefore only have to compute the Poisson bracket between the total energy functional $H_v$ and 
each of the two unnormalized position functionals ${O}_\rho$ previously defined 
(\textit{i.e.} either that of Eq. \eqref{eq:barybaros} or Eq. \eqref{eq:x_no_norm}). 
We will treat separately the dynamics generated by these two choices of position functionals. 
\subsubsection{The Ehrenfest relations based on the energy barycentre functional }
\label{sec:ehrenfest_bary}
We begin by considering the Ehrenfest relations based on the energy functional of Eq. \eqref{eq:barycenter_x}, 
and after some algebra provided in Sec. II-B of the supplementary materials, we obtain the following Poisson brackets 
\begin{equation}\label{eq:x_h_pb_main}
 \big\{ {X}  , H_v  \big\} = - \lambda_c^2\int dx \dot{\psi}(x,t) \derivb{x}{}\psi(x,t),
\end{equation}
which suggests to introduce the unnormalized momentum of Eq. \eqref{eq:momentum_classic} 
such that 
\begin{equation}\label{eq:d_x_dt}
 \frac{d}{dt} {X}(t) = \frac{{P}(t)}{m}.
\end{equation} 
Therefore, the normalized momentum is then obtained as
\begin{equation}\label{eq:normalized_mom}
 \tilde{P}[\psi,\dot{\psi}] = \frac{{P}[\psi,\dot{\psi}]}{\gamma[\psi,\dot{\psi}]},
\end{equation}
such that 
\begin{equation}\label{eq:d_x_dt}
 \frac{d}{dt} \tilde{X}(t) = \frac{\tilde{P}(t)}{m}.
\end{equation} 
One can then obtain the time derivative of the unnormalized total momentum ${P}$ by 
\begin{equation}
 \frac{d}{dt} {P}(t) = \{ {P}(t),H_v\},
\end{equation}
and, after some algebra (see Sec. II-B of the supplementary materials), we get 
\begin{equation}
 \{ {P},H_v\} =  -\int dx \psi^2(x) \derivb{x}{}v(x),
\end{equation}
which suggests to introduce the unnormalized total external forces as 
\begin{equation}\label{eq:ex_force}
 \mathcal{{F}}[\psi,\dot{\psi}] = -\int dx \psi(x,t)^2 \derivb{x}{}v(x),
\end{equation}
such that we can write 
\begin{equation}
 \label{eq:ehrenfest}
 \frac{d}{dt} {P}(t) = {\mathcal{F}}(t) .
\end{equation}
The normalized forces are then obtained as 
\begin{equation}
 \tilde{\mathcal{F}}[\psi,\dot{\psi}]= \frac{\mathcal{{F}}[\psi,\dot{\psi}]}{\gamma[\psi,\dot{\psi}]}, 
\end{equation}
such that we can write the Ehrenfest relations 
\begin{equation}
 \frac{d}{dt} \tilde{P}(t) = \tilde{\mathcal{F}}(t).
\end{equation}
If the external potential is constant (\textit{i.e.} $\derivb{x}{}v(x)=0$) we recover the usual conservation of momentum 
\begin{equation}
 \frac{d}{dt} \tilde{P}(t) =0,
\end{equation}
as in the case of the free Klein-Gordon field. 
Therefore, as long as the external potential does not break the translational invariance of the system, the momentum is conserved. 

Also, in a three-dimensional space, provided that the external potential is invariant by rotation around the origin 
(\textit{i.e.} $v(\bfr{r})=v(|\bfr{r}|)$), the angular momentum vector defined as in Eq. \eqref{eq:ang_mom_tot}
is a conserved quantity, as  
\begin{equation}
 \{ {\textbf{L}}, H_v\} = 0.
\end{equation}
We can then introduce the normalized angular momentum as 
\begin{equation}
 \tilde{\textbf{L}}[\psi,\dot{\psi}] = \frac{{\textbf{L}}[\psi,\dot{\psi}]}{\gamma[\psi,\dot{\psi}]},
\end{equation}
which can then be written as 
\begin{equation}
 \tilde{\textbf{L}}[\psi,\dot{\psi}] = \int d\bfr{r} \,\,\,\bfr{r} \times \tilde{\bfr{p}},
\end{equation}
where we introduced the local normalized momentum 
\begin{equation}
 \tilde{\bfr{p}} = -\frac{m \lambda_c^2}{\gamma[\psi,\dot{\psi}]}\dot{\psi}(\bfr{r},t) \nabla\psi(\bfr{r},t).
\end{equation}

We have therefore established the Ehrenfest relations of Ref. \onlinecite{ReiFer-PRL-91} associated to the functionals of 
Klein-Gordon fields using only the classical Poisson bracket formalism. 

\subsubsection{The Ehrenfest relations based on the $\bary$ functional }
\label{sec:ehrenfest_bary}
Turning now to the case of the $\bary$ functional proposed by Barros and Gomes (see Eq. \eqref{eq:barybaros}), 
we cannot give results in the case where $v(x)\ne 0$ as 
we were not able to compute explicitly operators such as $\hatom_v^{\frac{1}{2}} x \hatom_v^{\frac{1}{2}}$, 
so we impose $v(x)=0$. 
We therefore obtain the following Poisson bracket (see Sec. II-B of the supplementary materials) 
\begin{equation}\label{eq:bary_h_pb_main}
 \{ \bary, H\}= -\lambda_c^2 \int dx \dot{\psi}(x) \hatom^{-1}\derivb{x}{} \psi(x),
\end{equation}
where one recognizes a form similar to Eq. \eqref{eq:momentum_relat}, such that one can write 
\begin{equation}
 \frac{d}{dt}\bary = \frac{{\mathcal{P}}}{m}. 
\end{equation}
As mentioned earlier, as the density chosen here is the one corresponding to the inner product used for the Klein-Gordon space, 
the functionals obtained in this way are automatically normalized. 
One can then compute the time derivative of the momentum, but as the functional of Eq. \eqref{eq:momentum_relat} 
is a constant of motion, we know that 
\begin{equation}
 \frac{d}{dt}{\mathcal{P}}=0.
\end{equation}

\section{Connecting the classical and quantum formalisms}\label{sec:connect}
\subsection*{Summary and context}
Having shown how the dynamics of Klein-Gordon fields and their functionals can be obtained from 
the classical Poisson brackets formalism, we now propose to establish an equivalence with 
the usual formulation of quantum mechanics, which consists in the second part of this work.  
More precisely, we now make the connexion between the real-valued classical functionals and the quantum expectation values of hermitian operators acting on complex-valued wave functions living in $L^2$ Hilbert spaces. 
We begin by showing in Sec. \ref{sec:change_var} that there is a class of non local complex-valued change of variables 
allowing to rewrite the Klein-Gordon equation as a $\schro$-like equation, 
and which yields the actual $\schro$ equation in the non relativistic limit. 
Then, in Sec. \ref{sec:func_alpha} we give the conditions on the real-valued classical functionals 
of $(\psi,\dot{\psi})$ to be rewritten as quantum expectation values thanks to this complex-valued change of variables. 
We proceed in Sec. \ref{sec:flow} by translating the flow transformations in the complex-valued basis, 
and we show that they lead to uncoupled equations, unlike in the real-valued $(\psi,\dot{\psi})$ representation. 

Then, in Sec. \ref{sec:equiv_pb} we highlight that a specific case of this class of change of variable, 
coinciding with that originally introduced by Foldy\cite{Foldy-PR-56}, 
allows to establish an equivalence between the classical 
Poisson brackets formalism previously introduced and the algebra of hermitian operators. 
The Foldy representation then allows for a mapping between Poisson bracket and quantum commutators 
which takes a form very similar to the canonical quantization rule 
of Dirac, \textit{i.e.} $[\hat{F},\hat{G}] = i\hbar \{F,G\}$. 
We then summarize the main results of these two equivalent formulations in Sec. \ref{sec:summary}.  
We further show in Sec. \ref{sec:nr_alpha} how, in the $c\rightarrow \infty$ limit, one recovers the usual operator 
of non relativistic quantum mechanics, and how, by using the equivalence between quantum commutators 
and classical Poisson brackets to recover the Heisenberg representation in Sec. \ref{sec:heisenberg}. 
We conclude this section by discussing in Sec. \ref{sec:foldy_rep_funct} the link between the present derivations 
and previous works both regarding the two-components formalism of Klein-Gordon equation and the square-root formalism. 

\subsection{A continuous class of non local change of variables connecting the Klein-Gordon and $\schro$ equations}
\label{sec:change_var}
For any couple of Klein-Gordon fields $\big(\psi(x),\dot{\psi}(x)\big)$, 
we introduce the following class of complex-valued non local change of variables  
\begin{equation}\label{eq:phialpha_main}
 \phialpha(x) = \frac{1}{\sqrt{2}}\big(\hatom_v^{1-\alpha} \psi(x) + \frac{i}{\omega_0} \hatom_v^{-\alpha} \dot{\psi}(x)\big),
\end{equation}
which is parametrised by the number $\alpha\in \mathbb{R}$, and 
where we recall that the operator $\hatom_v$ is defined in Eq. \eqref{eq:def_hatom_v}.
This change of variables can be seen as a generalization of that originally proposed 
by Foldy\cite{Foldy-PR-56} which corresponds within our notations to $\alpha=1/2$ and $v(x)=0$. 
This change of variables is also related to the two-components formalism \cite{FesVil-RMP-58,Mostafazadeh-CQG-02}, 
and we give in Sec. \ref{sec:foldy_rep_funct} a more detailed discussion of the link of Eq. \eqref{eq:phialpha_main} 
with pre existing works.  
We also acknowledge that the change of variables of Foldy was used in the context of the study of existence of solutions 
of non linear Klein-Gordon equations\cite{SimTaf-CMP-93,BerGreRiv-AIHP-22}. 

After some algebra, the change of variable of Eq. \eqref{eq:phialpha_main} allows to rewrite 
the Klein-Gordon equation of Eq. \eqref{eq:kg_compact} as a $\schro$-like equation 
\begin{equation}\label{eq:kg_3_main}
 \begin{aligned}
 -i \phialphadot(x,t) + \omega_0 \hatom_v \phialpha(x,t) = 0, \quad \forall \alpha.
 \end{aligned}
\end{equation}
Nevertheless this results holds provided that the operator $\hatom_v$ is invertible, 
which implies that the spectrum of the non relativistic Hamiltonian $\hat{h}_v$ is strictly bounded by $-mc^2/2$. 

To actually recover the $\schro$ equation, one has to first expand in powers of $c^{-2}$ the operator $\hatom_v$ as follows 
(see Appendix \ref{sec:hatom_taylor}) 
\begin{equation}
 \begin{aligned}
  \hatom_v = 1 + \frac{1}{mc^2}\hat{h}_v + o(c^{-4}), 
 \end{aligned}
\end{equation}
such that inserted into Eq. \eqref{eq:kg_3_main} it yields 
\begin{equation}\label{eq:kg_4}
 \begin{aligned}
 -i \phialphadot(x,t) + \big( \omega_0+ \frac{1}{\hbar}\hat{h}_v \big)\phialpha(x,t) = 0.
 \end{aligned}
\end{equation}
Then, the fast oscillations at frequency $\omega_0$ in Eq. \eqref{eq:kg_4} can be absorbed by introducing the envelope $\lad_{\alpha}(x,t)$ 
as follows
\begin{equation}\label{eq:lad}
 \phialpha(x,t) = e^{-i\omega_0t} \lad_{\alpha}(x,t),
\end{equation}
such that inserted into Eq. \eqref{eq:kg_4} and multiplied by $\hbar$, it reduces simply the usual Schrodinger equation 
\begin{equation}
 i\hbar \dot{\lad}_{\alpha}(x,t) = \hat{h}_v \lad_{\alpha}(x,t).
\end{equation}
Therefore, the Schrodinger wave function $\lad_{\alpha}(x,t)$ can be seen as the slowly varying amplitude 
of the non local complex-valued mixing of the transverse displacement and velocities of the phonon field. 
It is noteworthy that the local change of variables used in the first paper of this series 
can be obtained in the non relativistic limit in Eq. \eqref{eq:phialpha_main}
\begin{equation}\label{eq:phiold}
 \lim_{c\rightarrow \infty}\psilad_{\alpha}(x,t) = \frac{1}{\sqrt{2}}\big(\psi(x,t) + \frac{i}{\omega_0} \dot{\psi}(x,t)\big).
\end{equation}
We refer to the reader to the first part of this series\cite{Giner-schro} for the physical analysis of the content of the $\schro$
equation under the light of this classical analogy. 

\subsection{Rewriting of functionals in the $\phialpha$ basis}
\label{sec:func_alpha}
Just as the change of variables $(\phialpha,\phialpha^*)$ of Eq. \eqref{eq:phialpha_main} allows to compact 
the writing of the Klein-Gordon equation, it also compacts the writing of the functionals $G[\psi,\dot{\psi}]$ of the real-valued Klein-Gordon fields 
$(\psi,\dot{\psi})$.  
More precisely, there is a whole class of functionals which,
once written in the $(\phialpha,\phialpha^*)$ basis, are expressed as usual quantum expectation values, 
and we discuss this result in the present section. 
We begin in Sec. \ref{sec:func_cst_motion} by giving the general conditions for such a property 
and show that the constants of motion fall 
in that category. Then in Sec. \ref{sec:func_specific} we provide  the explicit expression for 
important functionals by emphasizing the role of the external potential $v(x)$. 
We summarize here the main results, the details of the corresponding calculations can be found 
in the Appendix \ref{sec:func_alpha_annex} and the section III of the supplementary material.

\subsubsection{General conditions and application to constants of motions}
\label{sec:func_cst_motion}
If a real-valued functional $G$ is written in the $(\psi,\dot{\psi})$ basis as follows 
\begin{equation}\label{eq:general_func}
 G[\Psi] = \int dx \Psi^\dagger(x) 
 \begin{pmatrix}
  \hat{G}_{11} & \hat{G}_{12} \\
 -\hat{G}_{12} & \hat{G}_{22} \\
 \end{pmatrix}
 \Psi(x),
\end{equation}
and that the operators $\hat{G}_{11}$, $\hat{G}_{22}$ and $\hat{G}_{12}$ fulfill the following conditions 
\begin{equation}\label{eq:general_op}
 \begin{aligned}
& \hat{G}_{22} = \frac{1}{\omega_0^2}\hatom_v^{-1}\hat{G}_{11}\hatom_v^{-1}, 
 \quad \big(\hat{G}_{11}\big)^\dagger = \hat{G}_{11},\\& \big(\hat{G}_{12}\big)^\dagger = -\hat{G}_{12} , 
\quad [\hat{G}_{12},\hat{h}_v]=0,
 \end{aligned}
\end{equation}
then, once written in the $(\phialpha,\phialpha^*)$ basis, the same functional $G$ is written as a quantum expectation value, \textit{i.e.} 
\begin{equation}\label{eq:func_target_main}
 G[\phialpha,\phialpha^*] = \int dx \phialpha^*(x) \hat{G}_\alpha \phialpha(x), \quad \hat{G}_\alpha = \hat{G}_\alpha^\dagger, 
\end{equation}
with the hermitian operator $\hat{G}_\alpha$ being then obtained from $\hat{G}_{11}$ and $\hat{G}_{12}$ as follows
\begin{equation}\label{eq:f_qm_general}
 \hat{G}_\alpha = 2 \big(\hatom_v^{\alpha-1}\hat{G}_{11} \hatom_v^{\alpha-1} - i\omega_0\hatom_v^{2\alpha-1} \hat{G}_{12}\big).
\end{equation}
An interesting consequence of the form of Eq. \eqref{eq:func_target_main} is that 
the functional derivatives in the $(\phialpha,\phialpha^*)$ basis reads then simply
\begin{equation}\label{eq:derfunc_alpha}
  \derfunc{G[\phialpha,\phialpha^*]}{\phialpha^*(x)} = \hat{G}_\alpha\phialpha(x), 
\end{equation}
which will be useful for when considering flow transformations in Sec. \ref{sec:flow}. 

An important application of this  result concerns the constants of motion. 
More precisely, one can notice that the diagonal operators corresponding to the constants of motions 
(\textit{i.e.} Eq. \eqref{eq:cst_mot_diag}) are characterized by $\hat{G}_{11}=\hatom_v^{2\zeta}$
and $\hat{G}_{22} = \omega_0^{-2}\hatom_v^{-1}\hat{G}_{11}\hatom_v^{-1}$ and therefore  
fulfill Eq. \eqref{eq:general_op}, while the off-diagonal operators are anti Hermitian and fulfill  
$[\hat{G}_{12},\hat{h}_v]=0$ (see Eq. \eqref{eq:cst_mot_odiag}), which coincides with  Eq. \eqref{eq:general_op}. 
As a consequence, the conditions given in Eq. \eqref{eq:general_op} match the conditions for the constants of motion 
given in Sec. \ref{eq:cst_motion}, such that we obtain the following interesting results 
\begin{equation}\label{eq:cst_motion_qm}
  \begin{aligned}
&\text{Let }G[\psi,\dot{\psi}] \in \mathbb{R} \text{ be a constant of motion.}\\
&\text{Then, in the } \big(\phialpha,\phialpha^*\big) \text{ basis, it is written as }  \\ &G[\phialpha,\phialpha^*] = \int dx \phialpha^* \hat{G}_\alpha \phialpha(x) 
\text{ with } \hat{G}_\alpha^\dagger=\hat{G}_\alpha.
  \end{aligned}
\end{equation}
We emphasize however that the results of Eq. \eqref{eq:cst_motion_qm} is an implication and not an equivalence. 
More precisely, there are some functionals which are not constants of motion and which can nevertheless 
be written as expectation values, 
as it will be illustrated in mode details in Sec. \ref{sec:func_specific}. 

\subsubsection{Applications to important functionals and role of $v(x)$}
\label{sec:func_specific}
Before giving the explicit expression in the $\big(\phialpha,\phialpha^*\big)$ basis of the various 
functionals previously introduced, 
it should be emphasized that the external potential $v(x)$ plays here a rather subtle role. 
More precisely, while a functional  might not explicitly depend on the 
external potential (\textit{e.g.} the total linear momentum $P$ of Eq. \eqref{eq:momentum_classic}), 
the conditions of Eq. \eqref{eq:general_func} and Eq. \eqref{eq:general_op}, which allow to rewrite 
a functional as a quantum expectation value, depend explicitly on the external potential $v(x)$ 
through $\hat{h}_v$ and $\hatom_v$. 
As a consequence, the same functional expressed in the $(\psi,\dot{\psi})$ basis can be expressed or not 
as a quantum expectation value in the $\big(\phialpha,\phialpha^*\big)$ basis according to its properties 
with respect to $v(x)$, even though $v(x)$ does not explicitly appear and its definition. 
For instance, the unnormalized momentum $P\propto \int \dot{\psi}\derivb{x}{}\psi$ of Eq. \eqref{eq:momentum_classic} is independent 
of $v(x)$, but nevertheless it does not fulfill the commuting property $[\hat{G}_{12},v(x)]$ 
of Eq. \eqref{eq:general_op} 
in the case where of a generic $v(x)\ne cst$, and therefore can be written as a quantum expectation value 
in the $\big(\phialpha,\phialpha^*\big)$ basis only when $v(x)=cst$. 

We give now the explicit form of functionals in the $\big(\phialpha,\phialpha^*\big)$ basis, 
the details of the calculation can be found in Sec. III of the Supplementary information. 
We begin by the positive definite functionals $\mathcal{C}_\zeta[\psi,\dot{\psi}]$ of Eq. \eqref{eq:cst_scalar_main} 
whose expression in the $\big(\phialpha,\phialpha^*\big)$ basis reads 
\begin{equation}\label{eq:pos_const_alpha}
 \begin{aligned}
 \mathcal{C}_\zeta[\phialpha,\phialpha^*] & =  \int dx \phialpha^*(x)  \hatom_v^{2(\alpha+\zeta-1)}\phialpha(x).
 \end{aligned}
\end{equation}
Among the positive constants $\mathcal{C}_\zeta$, the case where $\zeta=1/2$ 
is the functional $\mathcal{N}[\psi,\dot{\psi}] $ of Eq. \eqref{eq:norm_psi} 
associated to the inner product, which then reads 
\begin{equation}\label{eq:n_phialpha}
 \mathcal{N}[\phialpha,\phialpha^*] = \int dx \phialpha^*(x)\hatom_v^{2\alpha-1}\phialpha(x),
\end{equation}
while the total energy as written in Eq. \eqref{eq:total_e_inner} reads 
\begin{equation}\label{eq:h_phialpha}
 \begin{aligned}
 H_v[\phialpha,\phialpha^*] &= mc^2 \int dx \phialpha^*(x)\hatom_v^{2\alpha}\phialpha(x).
 \end{aligned}
\end{equation}
We continue with the functionals being expressed as the average of a function $f(x)$ over unnormalized time 
invariant density $\rho_\zeta$ of Eq. \eqref{eq:def_measure_zeta}, \textit{i.e.} 
\begin{equation}\label{eq:f_general}
 f_\zeta[\psi,\dot{\psi}] = \int dx f(x)\rho_\zeta(\psi,\dot{\psi},x).
\end{equation}
These functionals are then written in the $(\phialpha,\phialpha^*)$ basis as follows 
\begin{equation}
 f_\zeta[\phialpha,\phialpha^*] = \int dx \phialpha^*(x) \hatom_v^{\alpha+\zeta-1} f(x) \hatom_v^{\alpha+\zeta-1} 
 \phialpha(x). 
\end{equation}
As an example of the latter formula, the position estimate of Barros introduced in Sec. \ref{sec:pos} 
(see Eq. \eqref{eq:def_x_baros}) falls in the category of Eq. \eqref{eq:f_general} with $\zeta=1/2$, 
and is therefore written as 
\begin{equation}\label{eq:x_alpha}
 \bary[\phialpha,\phialpha^*]=\int dx \phialpha^*(x)\hatom_v^{\alpha-\frac{1}{2}} x\hatom_v^{\alpha-\frac{1}{2}}\phialpha(x).
\end{equation}
The case of the energy barycentre given in Eq. \eqref{eq:x_no_norm} must be taken with care 
because it can be written as in Eq. \eqref{eq:f_general} only in the case when $v(x)=cst$. 
This is due to the fact that $[x,\hatom_{v}] = \lambda_c^2\derivb{x}{}\hatom_v^{-1}$ \textit{only when $v(x)=cst$}. 
Therefore, in the case where $v(x)=cst$, the un normalized energy barycentre of Eq. \eqref{eq:x_no_norm} 
is expressed as follows 
\begin{equation}\label{eq:x_no_norm_alpha}
 X[\phialpha,\phialpha^*]=\int dx \phialpha^*(x) 
 \frac{1}{2}\big(\hatom^{2\alpha} x + x\hatom^{2\alpha} \big)\phialpha(x),
\end{equation}
where we used some commutators relations to obtain the explicit form of Eq. \eqref{eq:x_no_norm_alpha} 
(see Sec. III-B of the supplementary materials for more details). 

Switching now to the class of momentum functionals introduced in Sec. \ref{sec:momentum}, 
the linear momentum are written as an expectation value only in the case where $v(x)=cst$. 
As an example, the momentum as defined in Eq. \eqref{eq:momentum_classic} becomes then 
\begin{equation}\label{eq:p_phialpha}
 P[\phialpha,\phialpha^*] = -i\hbar\int dx \phialpha^*(x) \hatom^{2\alpha-1}\derivb{x}{} \phialpha(x), 
\end{equation}
while that defined as Eq. \eqref{eq:momentum_relat} becomes then 
\begin{equation}\label{eq:p_relatphialpha}
 \mathcal{P}[\phialpha,\phialpha^*] = -i\hbar\int dx \phialpha^*(x) \hatom^{2\alpha-2}\derivb{x}{} \phialpha(x). 
\end{equation}
In the case of a central potential (\textit{i.e.} $v(\bfr{r})=v(|\bfr{r}|)$), 
the angular momentum of Eq. \eqref{eq:ang_mom_tot} being a constant of motion, 
it can then be written as an expectation value, whose explicit form is then 
\begin{equation}\label{eq:l_phialpha}
 \bfr{L}[\phialpha,\phialpha^*] = -i\hbar\int d\bfr{r} \phialpha^*(\bfr{r}) \big(\bfr{r} \times \hatom_v^{2\alpha-1}\nabla\big) \phialpha(\bfr{r}). 
\end{equation}

\subsection{Flow transformation and Poisson brackets in the $(\phialpha,\phialpha^*)$ representations}
\label{sec:flow}
As we introduced a non local change of variables allowing to rewrite functionals in a form akin to the usual 
quantum expectation values, we can now make the connexion with the classical Poisson brackets 
and the continuous transformations in phase space. 
We provide here the main results, which are detailed in Appendix \ref{sec:pb_link}. 

Considering a generic functional $G[\psi,\dot{\psi}]$, we know that it 
generates a transformation in phase space $(\psi(x,s),\dot{\psi}(x,s))$ as follows
\begin{equation}\label{eq:flow_general}
 \begin{aligned}
 \deriv{}{s}{}\psi(x,s) & = \frac{\omega_0}{\hbar} \derfunc{G[\psi,\dot{\psi}]}{\dot{\psi}(x,s)},\\
 \deriv{}{s}{}\dot{\psi}(x,s) & = -\frac{\omega_0}{\hbar} \derfunc{G[\psi,\dot{\psi}]}{{\psi}(x,s)}, 
 \end{aligned}
\end{equation}
and if now one imposes that such functional $G$ satisfies the condition to be written as a quantum expectation value 
in the $(\phialpha,\phialpha^*)$ basis (\textit{i.e.} Eq. \eqref{eq:general_func} and Eq. \eqref{eq:general_op}),
the functional derivatives are given explicitly by
\begin{equation}\label{eq:func_der_psi}
 \begin{aligned}
 & \derfunc{G[\psi,\dot{\psi}]}{{\psi}(x)} = 2 \big(\hat{G}_{11} \psi(x)- \hat{G}_{12} \dot{\psi}(x)\big), \\
 & \derfunc{G[\psi,\dot{\psi}]}{\dot{\psi}(x)} =  2\big(\frac{1}{\omega_0^2}\hatom_v^{-1}\hat{G}_{11}\hatom_v^{-1} \dot{\psi}(x)
 - \hat{G}_{12} {\psi}(x)\big) .
 \end{aligned}
\end{equation}
We can then perform the change of variables $(\psi(s),\dot{\psi}(s))\rightarrow(\phialpha(s),\phialpha^*(s))$ 
to get how the flow generated by $G$ acts on the complex-valued variables and, 
after some calculations provided in Appendix \ref{sec:flow_link}, 
it yields 
\begin{equation}\label{eq:flow_phi_general_1_main}
 \begin{aligned}
 \deriv{}{s}{}\phialpha(x,s) & = -\frac{i}{\hbar} \hatom_v^{1-2\alpha} \hat{G}_\alpha\phialpha(x,s), \\
 \end{aligned}
\end{equation}
where we recall that $\hat{G}_\alpha$ is the operator associated to the functional $G$ expressed 
in the $(\phialpha,\phialpha^*)$ basis (see Eq. \eqref{eq:f_qm_general}). 
We notice from Eq. \eqref{eq:flow_phi_general_1_main} that the flow transformation does not couple $\phialpha$ 
and $\phialpha^*$, 
in opposition with Eq. \eqref{eq:func_der_psi} where both $\psi$ and $\dot{\psi}$ are coupled.

One can then compute how a generic functional $F$ is changed by 
the action of the flow transformation generated by the functional $G$. 
The latter can be simply obtained by the chain rules 
using the $\big(\phialpha(s),\phialpha^*(s)\big)$ variables, \textit{i.e}
\begin{equation}\label{eq:fg_alpha}
 \begin{aligned}
 \frac{d}{ds}F[\phialpha(s),\phialpha^*(s)]\! =\! 
 \int\! dx \bigg(&\derfunc{F[\phialpha(s),\phialpha^*(s)]}{\phialpha(x,s)}\deriv{}{s}{}{\phialpha(x,s)}\\
 + &\derfunc{F[\phialpha(s),\phialpha^*(s)]}{\phialpha^*(x,s)}\deriv{}{s}{}{\phialpha^*(x,s)}\bigg).
 \end{aligned}
\end{equation}
Using now Eq. \eqref{eq:flow_phi_general_1_main} and Eq. \eqref{eq:derfunc_alpha}, together with the fact 
that the operators $\hat{F}_\alpha$, $\hatom_v$ and $\hat{G}_\alpha$ are  hermitian, 
one can rewrite the Eq. \eqref{eq:fg_alpha} as follows 
\begin{equation}\label{eq:flow_fg_main}
 \begin{aligned}
& \frac{d}{ds}F[\phialpha(s),\phialpha^*(s)]\! \\
 &=  -\frac{i}{\hbar}
 \int dx  \phialpha^*(x,s)\big(
 \hat{F}_\alpha\hatom_v^{1-2\alpha} \hat{G}_\alpha- \hat{G}_\alpha \hatom_v^{1-2\alpha}\hat{F}_\alpha 
 \big)\phialpha(x,s).
\end{aligned}
\end{equation}
We notice here that the right-hand side of Eq. \eqref{eq:flow_fg_main} is written as a quantum expectation value, 
and therefore we can connect the general flow transformations formalism in phase space a quantum-like formalism. 
We therefore established how the flow transformations in phase space can be translated from 
the usual real-valued $\big(\psi(s),\dot{\psi}(s)\big)$ basis to the complex-valued basis $\big(\phialpha(s),\phialpha^*(s)\big)$, whatever the value of $\alpha$. 

\subsection{Equivalence between classical and quantum formulation through the Foldy representation: the Dirac canonical commutation rule}
\label{sec:equiv_pb}
In Sec. \ref{sec:flow} we established how the flow transformation formalism 
is translated within the complex-valued representation for the change of variable $\big(\phialpha,\phialpha^*\big)$ with a generic $\alpha$. 
We show here that in the specific case of $\alpha=1/2$ (\textit{i.e.} the Foldy representation), 
all the expressions are simplified and coincide with the algebra of quantum mechanics. 

Setting $\alpha=1/2$ in the flow transformations of Eq. \eqref{eq:flow_phi_general_1_main} yields 
\begin{equation}\label{eq:flow_phi_half_1_main}
 \begin{aligned}
 \deriv{}{s}{}\psiladhalf(x,s) &= -\frac{i}{\hbar}\hat{G}_{\frac{1}{2}}\psiladhalf(x,s), \\
 \end{aligned}
\end{equation}
where $\hat{G}_{\frac{1}{2}}$ reads 
\begin{equation}\label{eq:f_qm_half}
 \hat{G}_{\frac{1}{2}} = 2 \big(\hatom_v^{-\frac{1}{2}}\hat{G}_{11} \hatom_v^{-\frac{1}{2}} - i\omega_0 \hat{G}_{12}\big),
\end{equation}
which is obtained by setting $\alpha=1/2$ in Eq. \eqref{eq:f_qm_general}. 
We therefore see from Eq. \eqref{eq:flow_phi_half_1_main} and Eq. \eqref{eq:derfunc_alpha} that 
\begin{equation}\label{eq:flow_phi_half_2}
 \begin{aligned}
 \deriv{}{s}{}\psiladhalf(x,s) &= -\frac{i}{\hbar}\derfunc{G[\psiladhalf,\psiladhalf^*]}{\psiladhalfb(x,s)},
 \end{aligned}
\end{equation}
such that the flow transformation in the Foldy basis translates very simply. 
Also, using $\alpha=1/2$ in Eq. \eqref{eq:flow_fg_main} yields 
\begin{equation}\label{eq:flow_fg_half_3_main}
\begin{aligned}
 \frac{d}{ds}F[\psiladhalf(s),\psiladhalf^*(s)] &= -\frac{i}{\hbar}
 \int dx  \psiladhalf^*(x,s)\big[ \hat{F}_{\frac{1}{2}} , \hat{G}_{\frac{1}{2}} \big] \psiladhalf(x,s). 
\end{aligned}
\end{equation}
Therefore, Eq. \eqref{eq:flow_fg_half_3_main} establishes the link between a flow transformation in phase space 
of the classical phonon model and the corresponding commutator akin to quantum mechanics. 

We can go further if we then define the new Poisson bracket in the $(\phialpha,\phialpha^*)$ basis, labelled here $\poisson{F}{G}_\alpha$, 
as follows 
\begin{equation}\label{eq:poisson_bis_main_1}
 \begin{aligned}
 \poisson{F}{G}_\alpha \equiv \int dx \bigg(& \derfunc{F[\phialpha,\phialpha^*]}{\phialpha^*}\derfunc{G[\phialpha,\phialpha^*]}{\phialpha} \\ 
 - &\derfunc{F[\phialpha,\phialpha^*]}{\phialpha}\derfunc{G[\phialpha,\phialpha^*]}{\phialpha^*}\bigg),
 \end{aligned}
\end{equation} 
and we notice that, only in the case where $\alpha=1/2$, the latter coincide with the expectation value of the 
usual commutators, \textit{i.e.}  
\begin{equation}
 \poisson{F}{G}_{\frac{1}{2}} = \elem{\psiladhalf}{\big[\hat{F}_{\frac{1}{2}},\hat{G}_{\frac{1}{2}}\big]}{\psiladhalf}, 
\end{equation}
such that Eq. \eqref{eq:flow_fg_half_3_main} can be rewritten as 
\begin{equation}\label{eq:flow_fg_half_4}
\begin{aligned}
 \frac{d}{ds}F(s) = -\frac{i}{\hbar}\poisson{F(s)}{G(s)}_{\frac{1}{2}}.
\end{aligned}
\end{equation}
We therefore conclude that we have the following identities
\begin{equation}
 \begin{aligned}
 \{F,G\} & = -\frac{i}{\hbar} \poisson{F}{G}_{\frac{1}{2}} \\
         & = -\frac{i}{\hbar} \elem{\psiladhalf}{\big[\hat{F}_{\frac{1}{2}},\hat{G}_{\frac{1}{2}}\big]}{\psiladhalf},
 \end{aligned}
\end{equation}
or formally as 
\begin{equation}\label{eq:dirac_q}
 \big[ \hat{F}_{\frac{1}{2}} , \hat{G}_{\frac{1}{2}} \big] = i\hbar\{F,G\},
\end{equation}
which is similar to the Dirac canonical quantization scheme. 
As a consequence, the change of variable of Foldy (\textit{i.e.} $\alpha=1/2$ in the complex-valued change of variables) 
allows to establish a direct link between the classical 
Poisson brackets, the quantum commutators, the quantum Poisson brackets defined in Eq. \eqref{eq:poisson_bis_main_1}, 
and the Dirac canonical quantization scheme. 
We will therefore use the value $\alpha=1/2$ for the rest of this paper, unless explicitly stated.

\subsection{Summary of the equivalence between classical and quantum formalisms}
\label{sec:summary}
Based on the results established in Sec. \ref{sec:change_var}, \ref{sec:func_alpha}, \ref{sec:flow} and \ref{sec:equiv_pb}, 
we can then conclude that the real-valued classical functionals and their associated flow in phase space 
are represented either by the real-valued $\big(\psi,\dot{\psi}\big)$ basis together with the usual Poisson brackets, 
or by the complex-valued change of variable of Foldy together with the usual algebra of hermitian operators 
of quantum mechanics. We emphasize that these two formalisms are rigorously equivalent as they describe 
the same classical system of an infinite number of coupled oscillators (\textit{i.e.} the Frenkel-Kontorova model of Sec. \ref{sec:kg_cl}).  
We summarize here the key features of this two equivalent formalisms for a three-dimensional system. 
\begin{enumerate}
 \item The dynamics of the classical phonon model coincides with the generalized Klein-Gordon equation written as 
  \begin{equation}
  \derivb{t}{2}\psi(\bfr{r},t) + \omega_0^2 \hatom_v^2 = 0,
  \end{equation}
 and we recall that $\hatom_v=\big(1+2\hat{h}_v/mc^2\big)^{\frac{1}{2}}$.
 Provided that the spectrum of the corresponding non relativistic Hamiltonian $\hat{h}_v$ is strictly bounded by $-mc^2/2$, 
 the same dynamics can be written as a $\schro$-like equation 
\begin{equation}
 \begin{aligned}
 -i \dot{\psilad}(\bfr{r},t) + \omega_0 \hatom_v \psilad(\bfr{r},t) = 0, 
 \end{aligned}
\end{equation}
when expressed into the Foldy representation $(\psilad,\psilad^*)$ defined as 
\begin{equation}\label{eq:phifoldy}
 \psilad(\bfr{r}) = \frac{1}{\sqrt{2}}\big(\hatom_v^{\frac{1}{2}} \psi(\bfr{r}) + \frac{i}{\omega_0} \hatom_v^{-\frac{1}{2}} \dot{\psi}(\bfr{r})\big),
\end{equation}
 which is the special case $\alpha=1/2$ of the change of variable of Eq. \eqref{eq:phialpha_main}. 
\notefn{The present author does not see yet a straightforward connection between 
the non-local complex-valued change of variables used here and  
the concept of complex-valued "classical wave functions" $\Theta_n$ in phase space introduced by Sch{\ifmmode\ddot{o}\else\"{o}\fi}nberg\cite{Schonberg-NC-52,Schonberg-NC-53}.} 
 \item If a real-valued functional $G$ is expressed in the $\big(\psi,\dot{\psi}\big)$ basis as 
 \begin{equation}\label{eq:general_func_2}
  G[\Psi] = \int d\bfr{r} \Psi^\dagger(\bfr{r}) \,\hat{\mathcal{G}} \,
  \Psi(\bfr{r}),
 \end{equation}
where the two-components matrix $\hat{\mathcal{G}}$ being of the form 
\begin{equation}\label{eq:g_matrix}
 \hat{\mathcal{G}} = 
  \begin{pmatrix}
    \hat{G}_{11} & \hat{G}_{12} \\
  -\hat{G}_{12} & \hat{G}_{22}\\
  \end{pmatrix},
\end{equation}
 and with the operators fulfilling the following conditions
 \begin{equation}\label{eq:general_func_3}
 \begin{aligned}
 &\hat{G}_{11}^\dagger = \hat{G}_{11}, \quad 
   \hat{G}_{22} = \frac{1}{\omega_0^2}\hatom_v^{-\frac{1}{2}}\hat{G}_{11}\hatom_v^{-\frac{1}{2}}, \\
 &\hat{G}_{12}^\dagger = - \hat{G}_{12}, \quad [\hat{G}_{12},\hat{h}_v]=0,
 \end{aligned}
 \end{equation}
 then the same functional can be re expressed as a quantum expectation value using the Foldy $(\psilad,\psilad^*)$ basis, \textit{i.e.} 
\begin{equation}\label{eq:func_target_main_bis}
 \begin{aligned}
 G[\psilad,\psilad^*] & = \int d\bfr{r} \psilad^*(\bfr{r}) \hat{G} \psilad(\bfr{r}),\\
 \end{aligned}
\end{equation}
with $ \hat{G}$ being the following hermitian operator 
\begin{equation}\label{eq:f_qm_general_bis}
 \hat{G} = 2 \big(\hatom_v^{-\frac{1}{2}}\hat{G}_{11} \hatom_v^{-\frac{1}{2}} - i\omega_0 \hat{G}_{12}\big) 
 , \quad \hat{G} = \hat{G}^\dagger.
\end{equation}

\item The constants of motion of the system are functionals of the type of Eq. \eqref{eq:general_func_2} 
whose operators satisfy the following conditions 
\begin{equation}
 \begin{aligned}
& \hat{G}_{11} = \hatom_v^{2\zeta}, \quad \hat{G}_{22} = \hatom_v^{2(\zeta-1)}, \\
& \hat{G}_{12}^\dagger = -\hat{G}_{12}, \quad [\hat{G}_{12},\hat{h}_v] = 0.
 \end{aligned}
\end{equation}
As a consequence, the constants of motion are expressed as quantum expectation values in the Foldy representation. 

\item We give here a list of important functionals expressed in the Foldy representation. 
 We begin by giving the results corresponding to a generic non vanishing potential $v(\bfr{r})$ fulfilling 
 the positivity condition of Eq. \eqref{eq:condition_positive}, 
 and then focus on the case of more specific external potentials (\textit{i.e.} $v(\bfr{r})=cst$ 
 and $v(\bfr{r})=v(|\bfr{r}|)$) which are particularly relevant for some applications. 

 The positive constant $\mathcal{C}_\zeta$ of Eq. \eqref{eq:cst_scalar_main} corresponding 
 to the density $\rho_\zeta$ are written in the Foldy representation $(\psilad,\psilad^*)$ as follows 
\begin{equation}\label{eq:c_zeta_phihalf}
 \mathcal{C}_\zeta[\psilad,\psilad^*] = \int d\bfr{r} \psilad^*(\bfr{r}) 
 \,\, \hatom_v^{2\zeta-1}\, \psilad(\bfr{r}). 
\end{equation}
The case $\zeta=1/2$ is particularly interesting as 
it corresponds to the inner product of Mostafazadeh\cite{Mostafazadeh-CQG-02}
on the space of Klein-Gordon fields, 
and applied to Eq. \eqref{eq:c_zeta_phihalf} it yields 
\begin{equation}
 \mathcal{C}_{\frac{1}{2}}[\psilad,\psilad^*] = \int d\bfr{r} |\psilad(\bfr{r})|^2, 
\end{equation}
such that the inner product associated to $\mathcal{C}_{\frac{1}{2}}$ (\textit{i.e.} Eq. \eqref{eq:def_innere}) is then 
\begin{equation}\label{eq:n_phihalf}
 \begin{aligned}
 \innere{\Psi_1}{\Psi_2} & = \int d\bfr{r} \psilad_1^*(\bfr{r})\psilad_2(\bfr{r}) \\
                         & = \braket{\psilad_1|\psilad_2}.
 \end{aligned}
\end{equation}
Therefore the Foldy representation $(\psilad,\psilad^*)$ allows to recover the usual $L^2$ Hilbert space 
formalism of quantum mechanics, 
but we emphasize that it is rigorously equivalent to the inner product on the space of the 
real-valued classical phonon fields $\Psi=(\psi,\dot{\psi})$. 
Another application of Eq. \eqref{eq:c_zeta_phihalf} is the total energy of the system which is simply proportional to the case where $\zeta=1$ (see Eq. \eqref{eq:total_e_inner}), which then reads 
\begin{equation}
 H_v[\psilad,\psilad^*]=mc^2\int d\bfr{r} \psilad^*(\bfr{r}) \hatom_v \psilad(\bfr{r}).
\end{equation}

Another type of functionals are the average of a function $f$ over the density $\rho_\zeta$ 
(\textit{i.e.} the functions $f_\zeta$ of Eq. \eqref{eq:f_general}), which are then written in the Foldy basis as 
\begin{equation}\label{eq:f_zeta_foldy}
 f_\zeta[\psilad,\psilad^*] = \int dx \psilad^*(x) \,\, \hatom_v^{\zeta-\frac{1}{2}} f(x) \hatom_v^{\zeta-\frac{1}{2}} 
 \, \phialpha(x). 
\end{equation}
Also, setting $\zeta=1/2$ in Eq. \eqref{eq:f_zeta_foldy} leads to 
\begin{equation}\label{eq:f_half_foldy}
 f_{\frac{1}{2}}[\psilad,\psilad^*] = \int d\bfr{r} |\psilad(\bfr{r})|^2f(\bfr{r}) ,
\end{equation}
and therefore the average of a function $f$ over the density $\rho_{\frac{1}{2}}$ associated to 
the inner product of Mostafazadeh becomes, in the Foldy representation $(\psilad,\psilad^*)$, 
simply the usual average of quantum mechanics. As further applications of Eq. \eqref{eq:f_half_foldy} 
we consider the position functional of Baros and Gomes of Eq. \eqref{eq:def_x_baros}, 
which corresponds to $f(\bfr{r})=\bfr{r}$ and then reads simply 
\begin{equation}\label{eq:x_half}
 \mathcal{\mathbfcal{R}}[\psilad,\psilad^*]=\int d\bfr{r} |\psilad(\bfr{r})|^2\bfr{r},
\end{equation}
which is the usual position expectation value of quantum mechanics. 
We emphasize however that, even if this expression is local in the $(\psilad,\psilad^*)$ basis, 
it is non local in the $(\psi,\dot{\psi})$ representation. 


We switch now to the case where the external potential vanishes, which corresponds to free space and which is useful for relativistic 
kinematic. 
Then, we notice that the operator $\hatom$ can be written as 
\begin{equation}
 \hatom = \bigg(1 + \frac{\hat{\bfr{P}}^2}{m^2c^2} \bigg)^\frac{1}{2},
\end{equation}
where $\hat{\bfr{P}}=-i\hbar \nabla$ is the usual momentum operator of non relativistic quantum mechanics. 
Therefore the operator $\hatom$ bears an interpretation similar to the Lorentz factor $\gamma=\sqrt{1+(p/mc)^2}$,  
and the total energy of the phonon field is then given by an expression similar to relativistic mechanics 
\begin{equation}\label{eq:h_phihalf}
 \begin{aligned}
 H[\psilad,\psilad^*] &= mc^2 \int d\bfr{r} \psilad^*(\bfr{r})\hatom\psilad(\bfr{r}) \\
                      &= mc^2 \elem{\psilad}{\bigg(1 + \frac{\hat{\bfr{P}}^2}{m^2c^2} \bigg)^\frac{1}{2}}{\psilad}.
 \end{aligned}
\end{equation}
On the other hand, the unnormalized momentum functional of Eq. \eqref{eq:momentum_classic} is then written as in quantum mechanics, \textit{i.e.} 
\begin{equation}\label{eq:p_phihalf}
 \bfr{P}[\psilad,\psilad^*] = \int d\bfr{r} \psilad^*(\bfr{r}) \hat{\bfr{P}}\psilad(\bfr{r}),
\end{equation}
and as a consequence, the normalized momentum functional $\tilde{\bfr{P}}$ of Eq. \eqref{eq:normalized_mom} 
is written as 
\begin{equation}
 \begin{aligned}
 \tilde{\bfr{P}}[\psilad,\psilad^*] & = 
 \frac{\elem{\psilad}{\hat{\bfr{P}}}{\psilad}}{\elem{\psilad}{\hatom}{\psilad}},\\
 \end{aligned}
\end{equation}
such that $\tilde{\bfr{P}}[\psilad,\psilad^*]$ can never exceed the limit of $mc$, in absolute value. 
Exploiting the known commutator $[\hatom,\bfr{r}]$ (see Ref. \onlinecite{Trubenbacher-ZFN-89}), 
we can then rewrite the unnormalized energy barycentre in the case of $v(\bfr{r})=0$ as 
\begin{equation}\label{eq:x_no_norm_half}
 \bfr{R}[\psilad,\psilad^*]=\int d\bfr{r} \psilad^*(\bfr{r}) 
 \frac{1}{2}\big(\hatom \bfr{r} + \bfr{r}\hatom \big)\psilad(\bfr{r}),
\end{equation}
and its normalized counterpart reads then 
\begin{equation}
 \begin{aligned}
 \tilde{\bfr{R}}[\psilad,\psilad^*] & = 
 \frac{\elem{\psilad}{\frac{1}{2}\big(\hatom \bfr{r} + \bfr{r}\hatom \big)}{\psilad}}{\elem{\psilad}{\hatom}{\psilad}},\\
 \end{aligned}
\end{equation}
which has the meaning of a true energy barycentre. 
The momentum associated to the position functional of Barros and Gomes (see Eq. \eqref{eq:x_half}) can be written in the case $v(\bfr{r})=0$ 
as follows 
\begin{equation}\label{eq:momentum_relat_phihalf}
 \mathcal{\mathbfcal{P}}[\psilad,\psilad^*] = \int dx \psilad^*(x) \hat{\bfr{P}}\hatom^{-1} \psilad(x).
\end{equation}
We see from Eq. \eqref{eq:momentum_relat_phihalf} that 
the operator corresponding to $\mathcal{\mathbfcal{P}}[\psilad,\psilad^*]$ coincides with a relativistic definition 
of the momentum, and that the associated functional can never exceed $mc$, in absolute value. 

Eventually, in the case where $v(\bfr{r})=v(|\bfr{r}|)$, the classical angular momentum of Eq. \eqref{eq:ang_mom_tot} is then obtained as 
\begin{equation}\label{eq:l_phihalf}
 \bfr{L}[\psilad,\psilad^*] = -i\hbar\int dx \psilad^*(\bfr{r}) \big(\bfr{r} \times \nabla \big)\psilad(\bfr{r}).
\end{equation}
 \item Any functional $G$ can generate a flow in phase space parametrized by the real number $s$,   
 which changes the basis $\big(\psi(s),\dot{\psi}(s)\big)$ according to the following equations 
\begin{equation}\label{eq:flow_general_bis}
 \begin{aligned}
 \deriv{}{s}{}\psi(x,s) & = \frac{\omega_0}{\hbar} \derfunc{G[\psi,\dot{\psi}]}{\dot{\psi}(x,s)},\\
 \deriv{}{s}{}\dot{\psi}(x,s) & = -\frac{\omega_0}{\hbar} \derfunc{G[\psi,\dot{\psi}]}{{\psi}(x,s)}, 
 \end{aligned}
\end{equation}
where the functional derivatives are given by 
\begin{equation}\label{eq:func_der_psi_bis}
 \begin{aligned}
 & \derfunc{G[\psi,\dot{\psi}]}{{\psi}(x)} = 2 \big(\hat{G}_{11} \psi(x)- \hat{G}_{12} \dot{\psi}(x)\big), \\
 & \derfunc{G[\psi,\dot{\psi}]}{\dot{\psi}(x)} =  2\big(\frac{1}{\omega_0^2}\hatom_v^{-1}\hat{G}_{11}\hatom_v^{-1} \dot{\psi}(x)
 - \hat{G}_{12} {\psi}(x)\big) .
 \end{aligned}
\end{equation}
The same flow in phase space can be represented in the Foldy basis $\big(\psilad(s),\psilad^*(s)\big)$ as 
\begin{equation}\label{eq:flow_phi_half_1}
 \begin{aligned}
 \deriv{}{s}{}\psilad(x,s) &= -\frac{i}{\hbar}\derfunc{G[\psilad,\psilad^*]}{\psiladb(x,s)},
 \end{aligned}
\end{equation}
where the functional derivative is then simply given by 
\begin{equation}
 \derfunc{G[\psilad,\psilad^*]}{\psiladb(x,s)} = \hat{G}\psilad(x,s). 
\end{equation}
As a consequence, the flow transformation does not couple $\psilad(s)$ and $\psilad^*(s)$. 
 \item To each couple of functionals $F$ and $G$ one can associate a classical Poisson bracket $\{F,G\}$ given by 
\begin{equation}\label{eq:pb_def}
 \begin{aligned}
 \big\{ F, G\big\} = \frac{\omega_0}{\hbar} \int dx 
  \frac{\delta {F}[\psi,\dot{\psi}]}{\delta \psi(x)}  \frac{\delta {G}[\psi,\dot{\psi}]}{\delta \dot{\psi}(x)} 
- \frac{\delta {F}[\psi,\dot{\psi}]}{\delta \dot{\psi}(x)}  \frac{\delta {G}[\psi,\dot{\psi}]}{\delta {\psi}(x)}, 
 \end{aligned}
\end{equation}
which is a functional of $\Psi=(\psi,\dot{\psi})$ 
 \begin{equation}\label{eq:fg_func}
  \big\{ F, G\big\}[\Psi] = \int d\bfr{r} \Psi^\dagger(\bfr{r}) \,\hat{\mathcal{F}\mathcal{G}} \,
  \Psi(\bfr{r}). 
 \end{equation}
Because we assumed that the functionals $F$ and $G$ are of the form of Eq. \eqref{eq:general_func_2},\eqref{eq:g_matrix}, and \eqref{eq:general_func_3}, 
the matrix $\hat{\mathcal{F}\mathcal{G}}$ in Eq. \eqref{eq:fg_func} is given by 
\begin{equation}
 \begin{aligned}
 & \hat{\mathcal{FG}}  = -4 \frac{\hbar}{\omega_0}
  \begin{pmatrix}
    \hat{F}_{11} \hat{G}_{12} - \hat{G}_{11} \hat{F}_{12}\,\,& 
  \,\,\hat{G}_{11} \hat{F}_{22}
   - \hat{F}_{11} \hat{G}_{22}\\
  \hat{F}_{12}\hat{G}_{12} - \hat{G}_{12}\hat{F}_{12} \,\, &  
 \,\, \hat{F}_{12}\hat{G}_{22} -  \hat{G}_{12}\hat{F}_{22}\\
  \end{pmatrix}.
 \end{aligned}
\end{equation}
This Poisson bracket allows to obtain the dependence on a functional $F$ on the flow parameter $s$ 
\begin{equation}
 \deriv{F(s)}{s}{} = \{ F(s),G\}.
\end{equation}
One can then define the Poisson bracket in the $\big(\psilad,\psilad^*\big)$ basis as 
\begin{equation}\label{eq:poisson_bis_main}
 \poisson{F}{G} \equiv \int dx \derfunc{F[\psilad,\psilad^*]}{\psilad^*(x)}\derfunc{G[\psilad,\psilad^*]}{\psilad(x)} - \derfunc{F[\psilad,\psilad^*]}{\psilad(x)}\derfunc{G[\psilad,\psilad^*]}{\psilad^*(x)},
\end{equation} 
which, because the functionals are bilinear in the $\big(\psilad,\psilad^*\big)$ basis 
(see Eq. \eqref{eq:func_target_main}), turns out to be simply the expectation value of the commutator 
 \begin{equation}
  \poisson{F}{G} = \braket{\psilad|\big[\hat{F},\hat{G}\big]|\psilad}. 
 \end{equation}
The link between the two types of Poisson brackets is given simply by
 \begin{equation}\label{eq:link_pb_commut}
 \begin{aligned}
  \{F,G\}  &=  \frac{1}{i\hbar}\braket{\psilad|\big[\hat{F},\hat{G}\big]|\psilad}. 
 \end{aligned}
 \end{equation}

\end{enumerate}
We can therefore conclude that the formalism of a relativistic single spinless particle in quantum mechanics 
is formally identical to the classical phonon model by the complex-valued non local change of variable 
originally proposed by Foldy (\textit{i.e.} $\alpha=1/2$) which mixes positions and velocities. 



\subsection{Non relativistic limit in the general case $v(x)\ne 0$}
\label{sec:nr_alpha}
As mentioned in Sec. \ref{sec:func_alpha}, the ability to rewrite a classical real-valued functional of $(\psi,\dot{\psi})$ in a form akin 
to a quantum expectation value over the complex-valued functions $(\psilad,\psilad^*)$ depends 
implicitly on the external potential $v(x)$ appearing in the problem, 
even if the functional itself does not explicitly depend on the potential $v(x)$. 
This might seem odd as for instance the momentum of a particle in non relativistic quantum mechanics is defined by the expectation value 
of $\hat{P}=-i\hbar \nabla$, whatever the external potential acting on the quantum particle. 
We propose here to discuss how the general case of a non vanishing potential can nevertheless be treated in the non relativistic limit. 
We give in Sec. \ref{sec:nr_ideas} the main arguments allowing to recover the form of non relativistic quantum mechanics, 
and in Sec.\ref{sec:nr_alpha} the explicit form of the functionals. 
We refer the reader to Appendix  \ref{sec:nr_v_x} and Appendix \ref{sec:foldy_rep_funct} for a more detailed discussion.  

\subsubsection{General framework to obtain the non relativistic limit of functionals}\label{sec:nr_ideas}
In the case where $v(x)\ne0$, because of the non commuting relation between $v(x)$ 
and $\derivb{x}{}$, the non-local change of variable of Foldy 
does not allow in general to rewrite the real-valued functionals in a form akin to quantum mechanics, 
\textit{i.e.} as a bilinear form in terms of $\big(\psilad,\psilad^*\big)$.  
We emphasize nonetheless that there are exceptions which are i) the positive constants of motion of the type $\mathcal{C}_\zeta$ of 
Eq. \eqref{eq:cst_scalar_main} (such as the energy or the inner product), 
and ii) the averages $f_\zeta$ of Eq. \eqref{eq:f_general}. 
On the other hand, when considering a generic potential $v(x)$, 
functionals such as the momentum (either linear or angular) or the average forces $\mathcal{F}$ 
of Eq. \eqref{eq:ex_force} cannot strictly be written as a quantum expectation value as they contain 
quadratic forms of the type $\psilad\hat{F}\psilad$ and $\psilad^*\hat{F}\psilad^*$. 
Nevertheless, in the non relativistic limit these additional terms vanish, leading thus to the usual quantum expectation value. 
The reason for this nice feature can be summarized as follows: 
i) in the non relativistic limit, the relevant variable is no longer $\psilad$ but the slow-motion amplitude $\lad=e^{+i\omega_0t}\psilad$, 
ii) once written in the $(\lad,\lad^*)$ variables, the additional terms $\lad\hat{F}\lad$ and $\lad^*\hat{F}\lad^*$ 
are then proportional to a fast oscillating factor $e^{\pm 2i\omega_0t}$, 
iii) these terms vanish in the $c\rightarrow \infty$ limit when performing a small time average, 
provided that the potential is smooth enough (\textit{i.e.} fulfills the condition of Eq. \eqref{eq:condition_positive}). 

\subsubsection{Explicit form of functionals in the non relativistic limit}
To obtain the non relativistic limit of a functional, we first need 
to perform a large $c$ expansion of the $\hatom_v^\alpha$ operator, 
which reads
\begin{equation}\label{eq:hatom_alpha}
 \hatom_v^\alpha = 1 + \alpha \frac{\hat{h}_v}{mc^2} + o(c^{-4}), 
\end{equation}
and then replace the variables $(\psilad,\psilad^*)$ of Eq. \eqref{eq:phifoldy} by the slowly varying amplitude variables 
$(\lad,\lad^*)$ defined as 
\begin{equation}\label{eq:lad_foldy}
 \lad(\bfr{r},t) = e^{+i\omega_0t} \psilad(\bfr{r},t).
\end{equation}
Then, one notices that, because $\psilad$ and $\lad$ are related by a phase factor, the following identity holds
\begin{equation}
 \psilad^*(\bfr{r}) \hat{O} \psilad(\bfr{r}) = \lad^*(\bfr{r}) \hat{O} \lad(\bfr{r}), 
\end{equation}
provided that $\hat{O}$ is not an operator acting on the time variable, or that its time dependence is slow compared to $\omega_0^{-1}$.  
Following this procedure, we give here the non relativistic expression of the functionals 
of Appendix. \ref{sec:foldy_rep_funct}. 

Let us start with the positive constants functionals $\mathcal{C}_\zeta$ of Eq. \eqref{eq:cst_scalar_main} 
which are written in $(\psilad,\psilad^*)$ basis as in Eq. \eqref{eq:c_zeta_phihalf}, and therefore,  
using Eq. \eqref{eq:hatom_alpha}, the large $c$ expansion reads 
\begin{equation}
 \begin{aligned}
 &\lim_{c\rightarrow \infty}  \mathcal{C}_\zeta[\lad,\lad^*]  \\
 &=\lim_{c\rightarrow \infty} \int d\bfr{r} \lad^*(\bfr{r}) 
 \bigg(1 + (2\zeta-1) \frac{\hat{h}_v}{mc^2} + o(c^{-4})\bigg) \lad(\bfr{r}). 
 \end{aligned}
\end{equation}
Therefore, provided that $\lim_{c\rightarrow \infty}\hat{h}_v/mc^2=0$, 
the non relativistic limit of the positive constants of motion $\mathcal{C}_\zeta$ are then simply the usual $L^2$ norm 
on the basis of functions $(\lad,\lad^*)$ which satisfy the $\schro$ equation. 
As a consequence, the density associated to the Mostafazadeh inner product becomes the usual $L^2$ norm 
when written in the $(\lad,\lad^*)$ basis, 
\begin{equation}\label{eq:n_phialpha_bis}
 \lim_{c\rightarrow \infty} \mathcal{N}[\lad,\lad^*] = \int d\bfr{r} |\lad(\bfr{r})|^2 , 
\end{equation}
just as the density associated to the energy density 
\begin{equation}
 \lim_{c\rightarrow \infty} \elem{\psilad}{\hatom}{\psilad} = \int d\bfr{r} |\lad(\bfr{r})|^2. 
\end{equation}
Therefore, in the non relativistic limit, the precise choice of the density does not matter provided that it is 
of the form of $\mathcal{C}_\zeta$, whatever the value of $\zeta$.  

Following similar arguments, we can then show that the unnormalized energy barycentre of Eq. \eqref{eq:x_alpha} 
coincides with the usual non relativistic expression of the position operator
\begin{equation}\label{eq:x_phialpha_bis}
 \lim_{c\rightarrow \infty} \bfr{R}[\lad,\lad^*]=\int d\bfr{r} |\lad(\bfr{r})|^2 \bfr{r} , 
\end{equation}
just as the position functional associated to the $\bary$ functional of Eq. \eqref{eq:x_no_norm_alpha}
\begin{equation}
 \mathcal{\mathbfcal{R}}[\lad,\lad^*]=\int d\bfr{r} |\lad(\bfr{r})|^2 \bfr{r}. 
\end{equation}
Coming now to the momentum functionals, that associated to the $\bary$ position and that associated to the energy coincide in the non 
relativistic limit, \textit{i.e.} 
\begin{equation}\label{eq:p_phialpha_bis}
 \lim_{c\rightarrow \infty} \bfr{P}[\lad,\lad^*] = -i\hbar\int d\bfr{r} \lad^*(\bfr{r}) \nabla \lad(\bfr{r}), 
\end{equation}
\begin{equation}\label{eq:p_bary}
 \lim_{c\rightarrow \infty} \mathcal{\mathbfcal{P}}[\lad,\lad^*] = -i\hbar\int d\bfr{r} \lad^*(\bfr{r}) \nabla \lad(\bfr{r}). 
\end{equation}
The definition of the unnormalized angular momentum also coincides with that of the usual non relativistic angular momentum
\begin{equation}\label{eq:l_phialpha_bis}
 \lim_{c\rightarrow \infty} {\bfr{L}}[\lad,\lad^*] = -i\hbar\int d\bfr{r} \lad^*(\bfr{r}) \bfr{r} \times \nabla\lad(\bfr{r}). 
\end{equation}
One must take a special care in the case of the total energy of the system, as the naive non relativistic limit would read 
\begin{equation}
 \lim_{c \rightarrow \infty} H_v[\lad,\lad^*] = mc^2 \int |\lad(\bfr{r})|^2  
  + \int d\bfr{r} \lad^*(\bfr{r}) \hat{h}_v \lad(\bfr{r}).
\end{equation}
which therefore contains the diverging rest mass energy. 
Nevertheless, if one subtract the diverging rest mass energy by introducing the following functional 
\begin{equation}
 \mathcal{E}[\lad,\lad^*] = H_v[\lad,\lad^*]-mc^2  \int |\lad(\bfr{r})|^2,
\end{equation}
in the $c\rightarrow \infty$ limit, $\mathcal{E}[\lad,\lad^*]$ yields the expectation value 
of the non relativistic Hamiltonian 
\begin{equation}\label{eq:nr_h_tot}
 \lim_{c\rightarrow \infty} \mathcal{E}[\lad,\lad^*] = \int d\bfr{r} \lad^*(\bfr{r}) \hat{h}_v \lad(\bfr{r}).
\end{equation}

\subsection{Illustrative example: the Heisenberg formalism as Poisson brackets}
\label{sec:heisenberg}
An important application of the equivalence between classical Poisson brackets and quantum commutator consists 
in the case where one considers the Poisson bracket with the Hamiltonian. 
Therefore, using Eq. \eqref{eq:link_pb_commut}, one knows that  
 \begin{equation}\label{eq:link_pb_h}
 \begin{aligned}
  \{F,H_v\}  &=  \frac{1}{i\hbar}\braket{\psilad|\big[\hat{F},mc^2\hatom_v\big]|\psilad}. 
 \end{aligned}
 \end{equation}
As a consequence, the time dependence of a functional $F$, which is given in the $(\psi,\dot{\psi})$ basis by 
\begin{equation}\label{eq:time_pb}
 \frac{d}{dt}F(t) = \{F(t),H_v\}, 
\end{equation}
is then given, using Eq. \eqref{eq:link_pb_h}, as 
 \begin{equation}\label{eq:link_pb_h_2}
 \begin{aligned}
 \frac{d}{dt}F(t)   &=  \frac{\omega_0}{i}\braket{\psilad|\big[\hat{F},\hatom_v\big]|\psilad}, 
 \end{aligned}
 \end{equation}
which is of course the Heisenberg representation (we used that $\omega_0=mc^2/\hbar)$. 
Therefore, just as we established in Sec. \ref{sec:ehrenfest} the equation of motion for the functionals 
of the classical phonon field using the $(\psi,\dot{\psi})$ basis, the same can be done 
in the $(\psilad,\psilad^*)$ basis using the equivalence between classical Poisson brackets and commutators.  
We provide here the example of the position and momentum operators of Baros and Gomes in the case where $v(x)=0$. 

As the position functional of Baros and Gomes is written 
in the $(\psilad,\psilad^*)$ basis simply as the usual position operator of quantum mechanics 
(see Eq. \eqref{eq:x_half}), once inserted into Eq. \eqref{eq:link_pb_h} we obtain 
\begin{equation}\label{eq:d_dt_x}
 \frac{d}{dt}\bary(t) = -i\omega_0 \elem{\psilad}{\big[x,\hatom\big]}{\psilad}.
\end{equation}
The commutator between $x$ and $\hatom$ has been given in Ref. \onlinecite{Trubenbacher-ZFN-89} as follows 
\begin{equation}
 \big[x,\hatom\big] =  \lambda_c^2 \derivb{x}{} \hatom^{-1},
\end{equation}
such that inserted into Eq. \eqref{eq:d_dt_x} we obtain simply 
\begin{equation}\label{eq:d_dt_x_2}
 \frac{d}{dt}\bary(t) = -i\frac{\hbar}{m} \elem{\psilad}{\derivb{x}{} \hatom^{-1}}{\psilad},
\end{equation}
where we recognize the definition of the momentum of Eq. \eqref{eq:momentum_relat_phihalf} and therefore we can write 
\begin{equation}\label{eq:d_dt_x_3}
 \frac{d}{dt}\bary(t) = \frac{1}{m} \elem{\psilad}{\hat{\mathcal{\mathbfcal{P}}}}{\psilad},
\end{equation}
where the momentum operator is $\hat{\mathcal{\mathbfcal{P}}}=-i\derivb{x}{}\hatom^{-1} $. 
As $\big[\hat{\mathcal{\mathbfcal{P}}},\hatom\big]=0$, we then obtain that 
\begin{equation}
 \frac{d}{dt}\mathcal{\mathbfcal{P}}=0,
\end{equation}
and we recover the Ehrenfest relations of Sec. \ref{sec:ehrenfest_bary} but obtained within the $(\psilad,\psilad^*)$ 
basis. 

We conclude this section by noticing that the usual Heisenberg formalism of non relativistic quantum mechanics 
can be obtained if one chooses the non relativistic energy of Eq. \eqref{eq:nr_h_tot} as the functional generating the evolution. 

\subsection{Link with the two-components and square-root formalism}
\label{sec:foldy_rep_funct}
We now highlight some connections between the classical functionals expressed in the $\big(\psilad,\psilad^*\big)$ 
basis and previous works related to the free Klein-Gordon equation\cite{Foldy-PR-56,Trubenbacher-ZFN-89,Lammerzahl-JMP-93,MosZam-AP-06}. 

In the seminal work of Feshbach and Villars\cite{FesVil-RMP-58}, 
which was latter developed in a pseudo-hermitian framework by Mostafazadeh\cite{Mostafazadeh-CQG-02,Mostafazadeh-AP-04}, 
the authors introduced a two-components formalism of the Klein-Gordon fields 
through a complex-valued change of variable mixing both 
$\psi$ and $\dot{\psi}$. The difference between the latter formalism and the change of variables 
$(\phialpha,\phialpha^*)$ is that Feshbach and Villars mix  
$\psi$ and $\dot{\psi}$ with constants coefficients, and not with pseudo differential operators 
as in Eq. \eqref{eq:phialpha_main}.  
With respect to the constant mixing, the use of non-local differential operator as is Eq. \eqref{eq:phialpha_main} allows 
to decouple the dynamics of the two components, as it was initially noticed by Foldy\cite{Foldy-PR-56}, 
where he introduced a 
change of variables which is a special case corresponding to $\alpha=1/2$ within our notations.  
The Foldy representation, corresponding to the variables $(\psilad,\psilad^*)$ of Eq. \eqref{eq:phifoldy} in 
our notations, was latter used by Mostafazadeh in Ref. \onlinecite{MosZam-AP-06} in his seminal work 
on the Hilbert-space formulation of Klein-Gordon fields. 
Nevertheless, with respect to these work, we have shown that other classical functionals beside 
the total energy (such as the position, energy barycentre and momentums) are written as quantum expectation values 
once written in the complex valued representation $(\phialpha,\phialpha^*)$, 
whatever the value of $\alpha$. 

We turn now our attention to the square-root formalism of the Klein-Gordon equation,  
which was intensively studied to solve some problems 
associated with the Klein-Gordon equation\cite{Schweber-QFT,Sucher-JMP-63,Fiziev-85,Trubenbacher-ZFN-89,BriEngSus-ZFN-91,Lammerzahl-JMP-93,Namsrai-IJTP-98} and also used 
to obtain a $\schro$-like formalism phenomenological description of composite relativistic particles, 
such as the various flavours of quarkonium 
(see for instance early references \onlinecite{HenKelMoo-PLB-64,Refcharmmodel-78,KarMesSyd-PRL-80,MoxRos-PRD-83,Castorina-84} and more recent 
works such as Refs. \onlinecite{Rosner-JPCS-07,SegOrtEnt-PRD-16}). 
The main idea of this approach is the introduction of the square-root operator of the Klein-Gordon dynamics  
(corresponding to $\hatom$ in our framework), which is then used \textit{a la} $\schro$ to generate a dynamical equation 
(\textit{i.e.} Eq. \eqref{eq:kg_3_main} in our framework).  
In Ref. \onlinecite{Trubenbacher-ZFN-89}, $\tru$ emphasized that the square-root formalism 
differs from the Klein-Gordon formalism and identified the former as a proper quantum mechanical version 
of the Klein-Gordon equation due to its unitary first-order in time formulation. 
This quantum mechanical treatment is done by postulating the $\schro$-like dynamics 
generated by $\hatom$, and 
then introducing the following position operator 
$\braket{X} = \int \phi^*(x) x \phi(x)$, from which the relativistic momentum 
$\braket{P}=\int dx \phi^*(x) \hatom^{-1}\hat{p}\phi(x)$ (where $\hat{p}$ is the usual non relativistic momentum operator) 
is recovered from the Heisenberg formalism (\textit{i.e.} from $[\hatom,x]$).  
In a separated work\cite{Lammerzahl-JMP-93}, $\lam$ obtained the same conclusion using the language of pseudo-differential operators. 
As shown in Sec. \ref{sec:change_var}, it is worth noticing that the dynamical $\schro$-like equation 
associated to the square-root equation is nothing but a rewriting of the original real-valued Klein-Gordon 
using the change of variable $(\phialpha,\phialpha^*)$. 
Therefore, as originally shown by Foldy\cite{Foldy-PR-56} with its own non local change, 
there is no need for an explicit square-root formalism as the latter is equivalent to the original 
real-valued Klein-Gordon equation. 
This aspect can be further supported by the fact that 
the expression of $\braket{X}$ and $\braket{P}$ obtained in the context of the square-root formalism by both 
$\tru$\cite{Trubenbacher-ZFN-89} and $\lam$\cite{Lammerzahl-JMP-93} coincides precisely with the 
definition of 
the position and momentum functionals proposed by $\bary$ functional of Barros and Gomes 
(\textit{i.e.} Eq. \eqref{eq:x_half} and Eq. \eqref{eq:momentum_relat_phihalf}) expressed in the 
Foldy representation. 
Also, the Heisenberg equation used by $\tru$\cite{Trubenbacher-ZFN-89} and $\lam$\cite{Lammerzahl-JMP-93} 
to obtain the time derivative coincide precisely with the Poisson bracket computed in the Foldy representation, as 
shown in Sec. \ref{sec:heisenberg}.
Therefore, the change of variable of Foldy allows to understand the work of $\tru$ and $\lam$ 
as simply a rewriting of the Klein-Gordon equation and a specific choice of the density of the position of the associated fields.

\section{Using the Poisson brackets formalism to generate the transformations of the Poincar\'e group}
\label{sec:pb_transform}
\subsection*{Summary and context}
As another application of the formalism of continuous classical Poisson bracket, we consider there the case of 
free Klein-Gordon fields (\textit{i.e.} $v(x)=0$), and show how the functionals previously introduced 
together with the Poisson brackets can be used to generate the transformations of the Poincar\'e group. 
In Sec. \ref{sec:poincarre_pb} we focus on 
the unnormalized functionals $H$, $X$ and $P$ and show that they reproduce the Lie algebra 
of the 1+1 Poincar\'e group, which allows us in Sec.\ref{seq:algebra} to identify the inner product of Mostafazadeh 
as the Casimir invariant associated to the mass. 
We then show in Sec. \ref{sec:full_poincarre} that when considering the three-spatial dimension case, 
the functionals and Poisson brackets reproduce the algebra of the 3+1 Poincar\'e group.  
We then show in Sec. \ref{sec:foldy_group} how the Foldy representation allows to connect 
the Poisson bracket formalism with the usual representation of of the Poincar\'e group in quantum mechanics. 
Eventually, we show how some connections with QFT in Sec. \ref{sec:qft}. 

\subsection{The 1+1 Poincar\'e transformations through the continuous Poisson brackets}\label{sec:poincarre_pb}
Considering the transformation of phase space $(\psi(x,s),\dot{\psi}(x,s))$ generated by a functional 
$G[\psi,\dot{\psi}]$ (see Eq. \eqref{eq:flow_5_main}), we know from Eq. \eqref{eq:general_pb} that 
the $s$ dependence of a generic functional $F(s)$ is related to the continuous Poisson bracket 
$\big\{ F(s),G \big\}$.  
Therefore, the study of the $s$ dependence of $F(s)$ induced by the flow transformation generated by $G$  
is reduced to solving the following first-order differential equation 
\begin{equation}\label{eq:flow_eq_gen}
 \begin{aligned}
 \left\{ 
 \begin{array}{l}
  \frac{d}{ds}F(s) = \big\{ F(s),G \big\} \\
  F(0) = F[\psi_0,\dot{\psi}_0]  \\ 
  G(0) = G[\psi_0,\dot{\psi}_0]  
 \end{array}
 \right.,
 \end{aligned}
\end{equation}
where the initial conditions $F_0$ and $G_0$ are given through the initial conditions on the fields themselves 
\begin{equation}
 \big(\psi_0,\dot{\psi}_0\big) \equiv \big(\psi(s=0),\dot{\psi}(s=0)\big) .
\end{equation}
In the present case, we will consider the functionals $H$, $X$ and $P$ previously introduced, 
with the initial conditions given, without loss of generality, by
\begin{equation}\label{eq:init_cond}
 X(0) = X_0,\quad P(0) = P_0, \quad H(0) = mc^2 \gamma_0,
\end{equation}
where 
\begin{equation}
 \begin{aligned}
&  X_0 = \int dx \,\,x\epsilon(\psi_0,\dot{\psi}_0,x), \quad P_0 = -\int dx \dot{\psi}_0(x) \derivb{x}{} \psi_0(x),\\ 
& \gamma_0=\int dx \epsilon(\psi_0,\dot{\psi}_0,x).
 \end{aligned}
\end{equation}
We are therefore going to solve the following type of coupled differential equations 
\begin{equation}\label{eq:general_pb_qm}
 \left\{ 
 \begin{array}{l}
\frac{dH(s)}{ds} = \{H(s),G\}                \\
\frac{dX(s)}{ds} = \{X(s),G\}          \\
\frac{dP(s)}{ds} = \{P(s),G\}          \\
 \end{array}
 \right.,
\end{equation}
where by varying the functional $G[\psi,\dot{\psi}]$ we are changing the type of flow transformations. 
Therefore, the structure of the continuous Poisson brackets between $H$, $X$ and $P$ drives the flow equations of 
Eq. \eqref{eq:general_pb_qm}.  
As shown in Sec. II-B of the supplementary materials, these Poisson brackets are given by 
\begin{equation}\label{eq:pb_poincarre_11}
 \{X,P\} = \frac{H}{mc^2}, \quad \{X,H\} = \frac{P}{m}, \quad \{P,H\} = 0.
\end{equation}
Also, we recall that the parameter $s$ 
has the units of $[A][g^{-1}]$, where $[A]$ is the unit of an action and $[g]$ 
has the units of $G[\psi,\dot{\psi}]$. 
As a consequence, $s$ can be identified with a clear physical meaning such as time, length or momentum. 

Let us begin by choosing the energy functional $H$ as the generator of
the transformation (\textit{i.e.} $G=H$ in Eq. \eqref{eq:general_pb_qm}). 
We therefore use $s\equiv t$ because $s$ has the unit of time.  
In that case, Eq. \eqref{eq:general_pb_qm} is written as 
\begin{equation}\label{eq:h_qm}
 \left\{ 
 \begin{array}{l}
\frac{dH(t)}{dt} = 0                      \\
\frac{dX(t)}{dt} = \frac{P}{m}\Rightarrow \\
\frac{dP(t)}{dt} = 0                      \\
 \end{array}
 \right.
 \left\{ 
 \begin{array}{l}
  H(t) = \gamma_0mc^2              \\
  X(t) = X_0 + \frac{P_0}{m} t       \\
  P(t) = P_0        \\
 \end{array}
 \right.
\end{equation}
which, as expected, implies that $H$ generates the time evolution where the particle travels at the speed $P_0/m$. 
We proceed now by considering the momentum functional as the generator of the transformation (\textit{i.e.} $G=P$ in Eq. \eqref{eq:general_pb_qm}), 
which yields the following equations  
\begin{equation}\label{eq:momentum_qm}
 \left\{ 
 \begin{array}{l}
\frac{dH(l)}{dl} = 0                      \\
\frac{dX(l)}{dl} = \frac{H}{mc^2} \Rightarrow \\
\frac{dP(l)}{dl} = 0                      \\
 \end{array}
 \right.
 \left\{ 
 \begin{array}{l}
  H(l) = \gamma_0mc^2              \\
  X(l) = X_0 + \gamma_0l       \\
  P(l) = P_0        \\
 \end{array}
 \right. ,
\end{equation}
where we used $s\equiv l$ has its units are that of a length. 
As a consequence, 
we find that the total momentum functional $P$ generates the translations of the barycentre 
of the fields. Nevertheless, we notice that because $\{X,P\}=H/mc^2$, the translation is scaled by 
$\gamma_0=H(0)/mc^2$, although such scaling factor would reduce to $1$ when considered the 
normalized functional $\tilde{X}$ of Eq. \eqref{eq:barycenter_x}.

Let us now use $X$ as the generator of a continuous transformation, where now $s$ has the unit of a momentum, 
hereafter labelled by $\moment$. 
Remembering the structure of Poisson brackets of Eq. \eqref{eq:pb_poincarre_11}, it 
yields the following coupled differential equations
\begin{equation}\label{eq:x_qm_1}
 \left\{ 
 \begin{array}{l}
 \frac{dH(\moment)}{d\moment} =   -\frac{P(\moment)}{m}\\
\frac{dX(\moment)}{dl} = 0 \\
 \frac{dP(\moment)}{d\moment} =  -H(\moment)\\
 \end{array}
 \right. .
\end{equation}
From Eq. \eqref{eq:x_qm_1} we conclude that $X(\moment) = X_0$, such that we can focus on the two coupled components 
(\textit{i.e.} $H(\moment)$ and $P(\moment)$). The two coupled equations 
can be explicitly solved as 
\begin{equation}\label{eq:dp_s_3}
 \begin{aligned}
 \begin{pmatrix}
 H(\moment)/c \\
 P(\moment)\\
 \end{pmatrix}
  & = 
 \begin{pmatrix}
 \cosh\big(\tilde{\moment}\big)  &-\sinh\big(\tilde{\moment}\big) \\
 -\sinh\big(\tilde{\moment}\big) & \cosh\big(\tilde{\moment}\big)\\
 \end{pmatrix}
 \begin{pmatrix}
 \gamma_0mc \\
 P_0\\
 \end{pmatrix}
 \end{aligned},
\end{equation}
which then reads 
\begin{equation}\label{eq:dp_s_4}
 \begin{aligned}
 H(\moment)&= \cosh\big(\tilde{\moment}\big)\gamma_0mc^2-\sinh\big(\tilde{\moment}\big)P_0c \\
 P(\moment)&= \cosh\big(\tilde{\moment}\big)P_0         -\sinh\big(\tilde{\moment}\big)\gamma_0mc,
 \end{aligned}
\end{equation}
where $\tilde{\moment} = \moment/mc$ is the rapidity of the transformation. 
Eq. \eqref{eq:dp_s_4} has the form of the transformation of the energy-moment vector by a Lorentz boost in the $x$ direction, 
such that we deduce that the $X$ functional is a generator of such relativistic transformation. 

We can then conclude that the Poisson bracket relations of Eq. \eqref{eq:pb_poincarre_11} 
generates the Poincar\'e transformations in 1+1 dimension: the energy functional generates time translations, 
the total momentum $P$ spatial translations, while the energy barycentre $X$ generates Lorentz boosts. 
A more detailed investigation can be found in Appendix \ref{eq:annex_group} where we show that the 
derivations of the present section is linked to the adjoint representation of the Lie algebra of the Poincar\'e group. 

\subsection{Poisson brackets, Poincar\'e algebra and Casimir element}\label{seq:algebra}
An important aspect of the Lie groups are the so-called quadratic Casimir invariants, which are elements which are invariant with respect to all 
transformations of the group. Such invariant quantities are therefore characterized by the fact that they commute with all generators, 
or more precisely that their Lie brackets with all generators vanish. 
In our case, the generators are the set of three functionals $(H,X,P)$ and the Lie bracket is the continuous Poisson bracket. 
Therefore, the corresponding quadratic Casimir invariant, hereafter labelled $\mathcal{N}$ for reasons which will appear clearer in a few lines, must verify 
\begin{equation}
 \{\mathcal{N}, H\} = \{\mathcal{N}, X\} = \{\mathcal{N}, P\} = 0.
\end{equation}
As shown in Sec. II-D of the supplementary materials, there is only one functional fulfilling such property for the Poincar\'e group in 1+1 dimension, 
which is given by 
the Lorentz invariant inner product of Mostafazadeh (see Eq. \eqref{eq:number_part}). 
We will make the connexion with the mass of the field in Sec. \ref{sec:foldy_group} using the non local change of variables 
of Foldy. 
As there are no other Casimir quadratic element in the Poincar\'e group in 1+1  dimension, 
we can conclude that the definition of the functionals $(H,X,P)$ together with the continuous Poisson bracket 
allows to recover all properties of the Poincar\'e group in 1+1  dimension. 

\subsection{Extension to the Poincar\'e group in 3+1 dimensions}\label{sec:full_poincarre}
Having shown in details how the $X$, $P$ and $H$ functionals together with the continuous Poisson brackets 
provide the algebra of the Poincar\'e in 1+1 dimension, 
we give here a brief summary of the extension of these relations in 3+1 dimensions. 
In the three dimensional case, the functionals used to generate the Poincar\'e group are 
the extension of that defined in Sec. \ref{sec:ehrenfest}, which consists in the total energy of the system 
\begin{equation}
 H[\psi,\dot{\psi}] = mc^2 \int d\bfr{r}  \,\,\epsilon(\psi,\dot{\psi},\bfr{r}), 
\end{equation}
where the local energy density is defined as 
\begin{equation}
 \epsilon(\psi,\dot{\psi},\bfr{r}) = \frac{1}{2}\bigg( \psi(\bfr{r})^2 + \frac{\dot{\psi}(\bfr{r})^2}{\omega_0^2} 
 + \lambda_c^2\big(\nabla \psi(\bfr{r})\big)^2\bigg),
\end{equation}
the three components $R_{m}$ of the unnormalized energy barycentre functional defined as follows 
\begin{equation}\label{eq:def_bar_3d}
 \mathbf{R}[\psi,\dot{\psi}] = \int d\bfr{r}  \,\,\epsilon(\psi,\dot{\psi},\bfr{r})\,\,\bfr{r},
\end{equation}
the three components $P_m$ of the unnormalized total momentum functional defined as follows 
\begin{equation}
 \mathbf{P}[\psi,\dot{\psi}] = -m\lambda_c^2 \int d\bfr{r}  \dot{\psi}(\bfr{r}) \nabla \psi(\bfr{r}),
\end{equation}
and the three components $L_m$ of the total unnormalized angular momentum defined as 
\begin{equation}
 \mathbf{L}[\psi,\dot{\psi}] = -m\lambda_c^2 \int d\bfr{r}  \,\,\dot{\psi}(\bfr{r})\,\,\bfr{r}\times \nabla \psi(\bfr{r}).
\end{equation}
Then, by pedestrian calculus detailed in Sec. II-C of the supplementary materials, 
one can compute the various Poisson brackets between these ten functionals, which can be summarized as follows  
\begin{equation}\label{eq:pb_pc}
 \begin{aligned}
&\{ R_i, R_j \} = -\epsilon_{ijk} \frac{L_k}{m^2c^2}, \quad 
 \{ L_i, L_j \} = \epsilon_{ijk} L_k, \\
&\{ L_i, P_j \} = \epsilon_{ijk} P_k \quad
 \{ L_i, R_j \} = \epsilon_{ijk} R_k, 
\{ P_i, P_j \} = 0, \\
& \{ P_i, H   \} = 0, \quad 
 \{ R_i, P_j \} = \delta_{ij} \frac{H}{mc^2}, \quad 
 \{ R_i, H   \} = -\frac{P_i}{m},
 \end{aligned}
\end{equation}
where $\delta_{ij}$ is the Kronecker symbol and $\epsilon_{ijk}$ the Levi-Citta symbol. 
Therefore, these Poisson brackets relations precisely reproduce the Lie algebra of the generators 
of the Poincar\'e group. We can then conclude that the classical continuous Poisson brackets 
together with the ten functionals previously mentioned form the generators of the Poincar\'e group. 
It is therefore interesting to notice that, while we started with purely classical interpretation 
of the Klein-Gordon dynamics (\textit{i.e.} non quantum, non relativistic), 
we nevertheless obtain a typical relativistic behaviour when considering the global functionals of the fields. 

\subsection{Representation of the Poincar\'e group in the Foldy representation and non relativistic limit}
\label{sec:foldy_group}
We continue this section on the Poincar\'e group by illustrating how 
the use the variables $(\psilad,\psilad^*)$ allows to recover the algebra of operators 
introduced by Foldy\cite{Foldy-PR-56}, to associate the Casimir element with the mass of the field, 
and to recover the appropriate Galilean group in the non relativistic limit. 

By using the equivalence between the classical objects 
(\textit{i.e.} functionals of $(\psi,\dot{\psi})$ and Poisson bracket) and 
the quantum objects (\textit{i.e.} Hermitian operators and their commutators) 
summarized in Sec. \ref{sec:summary}, one can then rewrite the Lie algebra of Poincar\'e of Eq. \eqref{eq:pb_pc} 
directly in terms of the commutators between the hermitian operators, which then yields 
\begin{equation}\label{eq:pb_pc_half}
 \begin{aligned}
&\big[ \hat{R_i}, \hat{R_j} \big] = -\epsilon_{ijk} \hbar\frac{\hat{L}_k}{m^2c^2}, \quad 
 \big[ \hat{L_i}, \hat{L_j} \big] = i\hbar \epsilon_{ijk} \hat{L}_k, \\
&\big[ \hat{L_i}, \hat{P_j} \big] = i\hbar \epsilon_{ijk} P_k \quad
 \big[ \hat{L_i}, \hat{R_j} \big] = i\hbar \epsilon_{ijk} R_k, \quad 
\big[ \hat{P_i}, \hat{P_j} \big] = 0, \quad \\
 &\big[ \hat{P_i}, \hat{\hatom} \big] = 0, \quad 
 \big[ \hat{R_i}, \hat{P_j} \big] = i\hbar \delta_{ij} \hatom, \quad 
 \big[ \hat{R_i}, mc^2\hatom \big] = \frac{\hat{P}_i}{m},
 \end{aligned}
\end{equation}
where the operators are defined in Sec. \ref{sec:foldy_rep_funct}. 
Therefore, thanks to the link between quantum and classical objects, there is an equivalence between 
the Poincar\'e algebra expressed in terms of 
the usual classical Poisson brackets (\textit{i.e.} Eq. \eqref{eq:pb_pc}) and that represented in terms of non commuting 
hermitian quantum operators as in Eq. \eqref{eq:pb_pc_half}. 
It is also noteworthy that the form of these hermitian operators coincide with that 
given by Foldy in his seminal work on the relativistic quantum equations\cite{Foldy-PR-56}.  
Therefore, the Foldy representation is simply a more convenient way to manipulate the classical phonon fields (
\textit{i.e.} the real-valued Klein-Gordon fields), 
but these two representations are strictly equivalent. 

We also emphasize that the Casimir invariant of the Poincar\'e group associated with the mass,   
whose expression in terms of the generators of the group is 
\begin{equation}\label{eq:casimir}
 P^\mu P_\mu = P_0^2 - P_1^2 - P_2^2 - P_3^2,
\end{equation}
can be straightforwardly obtained in the Foldy representation. 
To see this, we use the fact that the time translation $P_0$ is the energy, 
whose hermitian operator in the $(\psilad,\psilad^*)$ representation is, with appropriate units, given by 
\begin{equation}
 \hat{P}_0 \equiv mc \hatom, 
\end{equation}
while the spatial translations are the momentum operators 
\begin{equation}
 \hat{P}_\mu \equiv -i \hbar \derivb{\mu}{}, \quad \mu=1,3.
\end{equation}
Therefore, Eq. \eqref{eq:casimir} is written as 
\begin{equation}
  P^\mu P_\mu = m^2c^2 \hatom^2 + \hbar^2 \Delta 
\end{equation}
but as $m^2c^2 \hatom^2 = m^2c^2 \mathbb{1}-\hbar^2 \Delta$, one can then rewrite the Casimir element as 
\begin{equation}
 P^\mu P_\mu  = m^2c^2 \mathbb{1},
\end{equation}
which is simply proportional to the identity operator. 
As the identity operator commutes with all generators of the Poincar\'e Lie algebra, 
it therefore satisfies the defining property of a Casimir invariant. 
Although this expression of the Casimir invariant seems trivial,  
it nevertheless allows us to establish a connection with Mostafazadeh's inner product. 
Indeed, in the $(\psilad,\psilad^*)$ Foldy representation 
the latter reduces to the identity operator (see Eq. \eqref{eq:n_phihalf}). 
Consequently, the $L^2$-norm of the wave function associated 
to the Klein-Gordon field (\textit{i.e.} the $(\psilad,\psilad^*)$ representation) 
is directly related to the mass of the particle. 

We conclude this section by noticing that, in the non relativistic limit, $\hatom$ tends to the identity operator, 
such that 
\begin{equation}
 \begin{aligned}
 \lim_{c\rightarrow \infty}
\big[ \hat{R_i}, \hat{R_j} \big] = \lim_{c\rightarrow \infty}\epsilon_{ijk} \hbar\frac{\hat{L}_k}{m^2c^2} =0, \\
 \lim_{c\rightarrow \infty}
 \big[ \hat{R_i}, \hat{P_j} \big] = i\hbar \delta_{ij} \lim_{c\rightarrow \infty}\hatom = i\hbar \delta_{ij} \mathbb{1},
 \end{aligned}
\end{equation}
and remembering that the unnormalized energy barycentre $\hat{\bfr{R}}$ become, in the non relativistic limit, 
the usual position operator, we recover the usual commutator of Galilean transformations 
of non relativistic quantum mechanics, \textit{i.e.} 
\begin{equation}\label{eq:b_pc_half}
 \begin{aligned}
&\big[ \hat{R_i}, \hat{R_j} \big] = 0, \quad 
 \big[ \hat{L_i}, \hat{L_j} \big] = i\hbar \epsilon_{ijk} \hat{L}_k, \quad 
 \big[ \hat{L_i}, \hat{P_j} \big] = i\hbar \epsilon_{ijk} P_k \\
&\big[ \hat{L_i}, \hat{R_j} \big] = i\hbar \epsilon_{ijk} R_k, \quad
 \big[ \hat{P_i}, \hat{P_j} \big] = 0, \quad 
 \big[ \hat{P_i}, \hat{\hatom} \big] = 0, \\
&\big[ \hat{R_i}, \hat{P_j} \big] = i\hbar \delta_{ij} \mathbb{1}, \quad 
 \big[ \hat{R_i}, \hat{P}_0 \big] = \hat{P}_i.
 \end{aligned}
\end{equation}
We can also notice that the $\big[ \hat{R_i}, \hat{P_j} \big] = i\hbar \delta_{ij} \mathbb{1}$ 
canonical commutation relation between the position operator $\hat{\bfr{R}}$ and the momentum operator $\hat{\bfr{P}}$  
emerges as a non relativistic limit of the corresponding commutator between the unnormalized 
energy barycentre and momentum operators. 

\subsection{Connection with QFT}\label{sec:qft}
In this section we briefly highlight some connexions between the complex-valued non-local change of variables 
of Foldy and some key ingredients involved in the steps of quantization of the free Klein-Gordon fields. 

As the eigenvectors of $\hatom$ are plane waves, the free Klein-Gordon equation 
is diagonal in the Fourier representation, \textit{i.e.}  
\begin{equation}\label{eq:fourier}
 \derivb{t}{2} \psi(k,t) + \omega_k^2 \psi(k,t) = 0,
\end{equation}
where $\psi(k,t)$ is the Fourier transform of the field $\psi(x,t)$, and where $\omega_k^2 = \omega_0^2 + k^2c^2 $. 
The dynamics of Eq. \eqref{eq:fourier} is that of a classical harmonic oscillator of pulsation $\omega_k$, 
and one can then introduce the following change of variable 
\begin{equation}\label{eq:a_k}
 a_k(t) = \frac{1}{\sqrt{2}}\big(\omega_k^{\frac{1}{2}} \psi(k,t) +i \omega_k^{-\frac{1}{2}}  \dot{\psi}(k,t) \big),
\end{equation}
such that the equation of motion of Eq. \eqref{eq:fourier} is written as 
\begin{equation}\label{eq:a_k_time}
 -i\dot{a}_k(t) + \omega_k a_k(t) = 0.
\end{equation}
We can notice that the definition of $a_k(t)$ as in Eq. \eqref{eq:a_k} 
is precisely the Fourier representation of the Foldy representation of Eq. \eqref{eq:phifoldy}, \textit{i.e.}  
\begin{equation}
 \psilad(k,t) = a_k(t),
\end{equation}
and that the equation of motion of Eq. \eqref{eq:a_k_time} satisfied by $a_k(t)$ is precisely the Fourier representation 
of the equation of motion satisfied by $\psilad(x,t)$ (\textit{i.e.} Eq. \eqref{eq:kg_3_main} in the case where $v(x)=0$).
Therefore, from thereon we will use $\psilad(k,t)$ instead of $a_k(t)$. 
The solution of Eq. \eqref{eq:a_k_time} reads then 
\begin{equation}
 \begin{aligned}
 &\psilad(k,t) = \psilad(k,0) e^{-i\omega_k t}, \\
 \end{aligned}
\end{equation}
where $\big(\psilad(k,0),\psilad^*(k,0)\big)$ are the two complex-valued initial conditions 
obtained from the two real-valued initial conditions $\big(\psi(k,t=0),\dot{\psi}(k,t=0)\big)$ 
\begin{equation}
 \begin{aligned}
 &\psilad(k,0) = \frac{1}{\sqrt{2}}\big(\omega_k^{\frac{1}{2}}\psi(k,t=0) + i\omega_k^{-\frac{1}{2}} \dot{\psi}(k,t=0)\big). 
 \end{aligned}
\end{equation}

One can then come back to the Fourier transform of the fields, \textit{i.e.} $\big(\psi(k,t),\dot{\psi}(k,t)\big)$,
by simply inverting the change of variables of Eq. \eqref{eq:a_k}
\begin{equation}
 \begin{aligned}
 \psi(k,t) & = \sqrt{\frac{\omega_k}{2}} \big( \psilad(k,t) + \psilad^*(k,t)(t) \big), \\
 \dot{\psi}(k,t) & = -i \sqrt{\frac{1}{2\omega_k}} \big(\psilad(k,t)(t) - \psilad^*(k,t)(t) \big), \\
 \end{aligned}
\end{equation}
and because both $\psi(x,t)$ and $\dot{\psi}(x,t)$ are real-valued, $\psilad^*(k,t) = \psilad^*(-k,t)$ such that 
such that the fields are then written in real space as 
\begin{equation}
 \begin{aligned}
 \psi(x,t)  = \int \frac{dk}{\sqrt{2\pi}} \sqrt{\frac{\omega_k}{2}} 
 \big( \phantom{+}&\psilad(k,0) e^{-i\omega_kt+ikx} \\ +& \psilad^*(k,0) e^{+i\omega_kt-ikx} \big) \\
 \end{aligned}
\end{equation}
\begin{equation}
 \begin{aligned}
 \dot{\psi}(x,t)  = -i\int \frac{dk}{\sqrt{2\pi}} \sqrt{\frac{1}{2\omega_k}} 
 \big(\phantom{-}& \psilad(k,0) e^{-i\omega_kt+ikx} \\ 
     - &\psilad^*(k,0) e^{+i\omega_kt-ikx} \big),\\
 \end{aligned}
\end{equation}
which is the usual representation of the classical field pre quantization in QFT. 
One can then express the functionals previously introduced in the $(\psilad,\psilad^*)$ basis.  
Exploiting the fact that $\hatom$ is diagonal on the Fourier basis and that these are constants of motion, 
such functional reads then 
\begin{equation}\label{eq:norm_k}
 \mathcal{N}[\psilad,\psiladb] = \int dk \psilad^*(k,0) \psilad(k,0) ,
\end{equation}
\begin{equation}
 H[\psilad,\psiladb] = mc^2 \int dk \psilad^*(k,0) \psilad(k,0) \omega_k,
\end{equation}
\begin{equation}
 P[\psilad,\psiladb] = -i\hbar \int dk \psilad^*(k,0) \psilad(k,0) k.
\end{equation}

We emphasize that so far, we have only rewritten the dynamics of the classical phonon field in the 
Fourier representation.  
When quantizing the field, one then promotes to operators the coefficients $a_k(0)$ and $a_k^*(0)$ and imposes 
the canonical commutation relation, which, in our language is equivalent to 
\begin{equation}\label{eq:qft}
 \big(\psilad(k,0), \psilad^*(k,0)\big) 
\xrightarrow[\text{of fields}]{\text{quantization}}
   \begin{cases}
  \big(\hat{\psilad}_{\frac{1}{2}}(k), \hat{\psilad}^\dagger_{\frac{1}{2}}(k)\big) \text{ and }\\
 \big[\hat{\psilad}_{\frac{1}{2}}(k),\hat{\psilad}_{\frac{1}{2}}^\dagger(k')\big] = \delta_{k,k'}, \\
   \end{cases}
\end{equation}
which is equivalent to quantize the initial conditions of the classical field equation.  
As done for instance in Ref. \onlinecite{BarGom-EPJC-21}, using the commutation relation 
and choosing an appropriate reordering (\textit{i.e.} the normal ordering) 
one can then rewrite the functionals as operators acting on a vacuum 
\begin{equation}
 \hat{\mathcal{N}} = \int dk 
 \hat{\psilad}_{\frac{1}{2}}^\dagger(k)\hat{\psilad}_{\frac{1}{2}}(k) ,
\end{equation}
\begin{equation}
 \hat{H}_0 = mc^2 \int dk \omega_k \hat{\psilad}_{\frac{1}{2}}^\dagger(k)\hat{\psilad}_{\frac{1}{2}}(k),
\end{equation}
\begin{equation}
 \hat{P}= -i\hbar \int dk  k \hat{\psilad}_{\frac{1}{2}}^\dagger(k)\hat{\psilad}_{\frac{1}{2}}(k).
\end{equation}
Therefore, the notion of particles and the associated vacuum,
 which emerge from the quantization step of Eq. \eqref{eq:qft}, appear here to be only step which is not directly 
connected to the classical model of phonons. 
Also, we notice that the $L^2$ norm of the Klein-Gordon wave function $\psilad$ (\textit{i.e.} Eq. \eqref{eq:norm_k}) 
which is the Casimir invariant associated to the mass of the 1+1 Poincar\'e group 
(see Sec. \ref{sec:foldy_group}), becomes the operator $\hat{\mathcal{N}}$ after quantization. 
Therefore, quantizing the norm implies to have integer eigenvalues for $\hat{\mathcal{N}}$, 
which coincides with integer multiples of the mass of the particles. 


\section{Conclusions}
In the second part of this series, we established a direct connexion between the formalism of relativist 
quantum mechanics for a single spinless particle and a classical phonon system, 
with an emphasize on the Hamiltonian description of such classical system. 
This study contains three broad parts. The first part 
(\textit{i.e.} Sec.\ref{sec:pb_kg}, Sec. \ref{sec:dynamic_pb} and Sec. \ref{sec:ehrenfest_big}) 
studies the dynamics of real-valued Klein-Gordon fields and their functionals 
through the lens of classical Poisson brackets. 
The second part (\textit{i.e.} Sec. \ref{sec:connect}) establishes the equivalence between 
the classical real-valued formalism and the usual complex-valued formalism of quantum mechanics, 
\textit{i.e.} Hermitian operators acting on $L^2$ Hilbert spaces. 
The last part consists in applying the formalisms developed in the two first parts 
to the Poincar\'e group and the link between classical Poisson brackets and quantum commutators. 
We review now the most important points addressed in this work. 

The cornerstone of this work is the Frenkel-Kontorova model\cite{FreKon-ZETF-38}, 
depicted in Fig. \ref{fig:FK_draw}, consisting in a regular lattice of classical massive particles 
coupled by springs and which, in the limit of an infinite number of particles, describes a continuous elastic medium whose Newtonian 
equation of motion yields the Klein-Gordon equation. 
The real-valued solutions $\psi(\bfr{r},t)$ of the Klein-Gordon equation can therefore be associated 
to the local transverse displacement of this phonon field, and $\dot{\psi}(\bfr{r},t)$ as the corresponding 
transverse velocity. 
The classical nature of this model naturally leads us to use the corresponding Hamiltonian formalism 
to study its dynamics, and more generally the flow transformation in phase space. 
We then naturally recover the main functionals of the real-valued Klein-Gordon fields as 
classical physical quantities, such as the total Hamiltonian of the system, its momentum or its average position. 
Starting from the usual Poisson brackets formalism applied to the phonon model with a finite number of classical particles, 
we established, in the continuous limit, the corresponding continuous Poisson bracket formalism expressed in terms 
of functional derivatives. 
By studying the structure of the Poisson bracket involving the total Hamiltonian of the system, 
we established the general conditions for a functional to be a constant of motion. 
Two types of constants of motion appears then, which can be categorized as either momentums (linear or angular) 
or a class of positive functionals. 
These positively defined functionals are parametrized by a real number $\zeta$, 
and consists in a class of time invariant densities and inner products on the space 
of Klein-Gordon fields. 
We show that the Hamiltonian of the system and the Lorentz invariant inner product 
of Mostafazadeh\cite{Mostafazadeh-CQG-02} are special cases of these class of functionals. 
We then use these two functionals to define the position of the fields as barycentre over a density, 
and recover the corresponding dynamics through the Poisson bracket formalism.  
This allows us to establish the Ehrenfest relations associated to real-valued Klein-Gordon fields 
as in Ref. \onlinecite{ReiFer-PRL-91}. We therefore establish most of the important aspects related to 
the dynamics of real-valued Klein-Gordon fields starting simply from the Hamiltonian description of the classical phonon model.  

We then focus our attention in Sec. \ref{sec:connect} on the connection between this classical formalism and the usual 
complex-valued formalism of quantum mechanics. 
This is done by introducing a class of non-local change of variable which mixes the real-valued fields 
$\psi(\bfr{r},t)$ and $\dot{\psi}(\bfr{r},t)$ with complex-valued coefficients. 
This complex-valued representation generalizes the one introduced by Foldy\cite{Foldy-PR-56} 
and allows to rewrite the Klein-Gordon equation as a $\schro$-like equation, 
which in the $c\rightarrow \infty$ limit yields the actual $\schro$ equation. 
We then rewrite the classical functional in the complex-valued representation and show that, 
under certain conditions, they are written as quantum expectation values. 
We then show that the Foldy representation, which is a specific case of our class of non-local 
complex-valued change of variable, establishes an interesting connection between the classical formalism 
and the quantum formalism. 
Among the distinctive features of the Foldy representation are that the total classical momentum 
of the phonon system is written as the usual quantum expectation value of the $-i\hbar \nabla$ operator. 
Also, the Foldy representation allows to rewrite the inner product of Mostafazadeh as the usual 
$L^2$ inner product of quantum mechanics. 
This highlights the fact that the wave-function interpretation of real-valued Klein-Gordon fields can be established 
properly by the Foldy representation. 
Moreover, the classical Poisson brackets are written as quantum expectation values in the Foldy representation, 
and the connection between these two formalisms appears akin to the Dirac canonical commutation rule. 
As an application of this mapping between classical and quantum formalisms, 
we show that the Heisenberg representation is nothing but the usual Poisson bracket formalism written in the 
Foldy representation. 
We also show how, in the $c\rightarrow \infty$ limit, the whole formalism of non 
relativistic quantum mechanics is recovered. 
This section is concluded with some connections with pre existing works, such as the two-components formalisms of Feshback and Villars\cite{FesVil-RMP-58} together with various works on the square-root formalism\cite{Trubenbacher-ZFN-89,Lammerzahl-JMP-93}. 

The last section proposes to apply the classical Poisson brackets formalism to the study of phase space transformations 
and the link with the relativistic symmetry group of Poincar\'e. 
We begin by studying the Klein-Gordon fields in the 1+1 dimensions, and we show that the Poisson brackets 
between the Hamiltonian, momentum and position functionals reproduces the 
algebra of the 1+1 Poincar\'e group. Then, we use the Poisson brackets to show that the only Casimir element 
associated to this group is nothing but the inner product of Mostafazadeh. 
We then generalize these results to the case in 3+1 dimensions, and then show that the use of the Foldy 
representation allows to rewrite the structure of the classical Poisson brackets between functionals as the 
algebra of non commuting operators, as usually done in relativistic quantum mechanics. 
We also highlight that the Casimir invariant associated to the mass 
of the field is actually the identity operator once written in the Foldy representation. 
Therefore, the $L^2$ inner product of quantum mechanics appears as deeply linked to the mass of the particle. 
We then conclude this work by establishing some connections with QFT 
and show how the Mostafazadeh inner product, which is the Casimir invariant associated to the mass, 
appears after quantization of the fields as the number operator. 
We therefore show that the integer eigenvalues of the number operators corresponds to integer values of the mass 
of the particle, which is a valid definition of the number of particles. 

As concluding remarks, we can therefore safely say that the usual formalism of quantum mechanics for a single spinless 
particle can be fully recovered by the classical phonon model. 
The Newtonian equations of motions yields the Klein-Gordon equation, and the $\schro$ equation 
is recovered in the $c\rightarrow \infty$ limit. 
On the other hand, the Hamiltonian or Poisson brackets formalisms leads to the Heisenberg formalism.  
As a consequence, left aside the probabilistic aspect of quantum mechanics, 
its algebra can be fully recovered by a classical 
framework, provided that one accepts that a single particle is actually represented by the oscillations 
of an elastic medium obeying a classical equation motion. 
As perspectives of this work, we hope to report soon on the application of the classical formalism 
to the Dirac equation, based on its link with the Klein-Gordon equation. 
Also, a classical procedure to recover the usual canonical quantization of real-valued Klein-Gordon fields 
is under study, and we hope to present it in future publications. 

\appendix
\section{Poisson brackets: from the point particle formulation to the continuous limit}\label{sec:pb_continous}
We consider the discrete one-dimensional Frenkel model of Sec. \ref{sec:kg_cl} made of the $N$ particles described 
by the set of $2N$ variables made of their displacement $\mathbf{u}=\{u_n,1\le n\le N \}$ and 
momentum $\mathbf{p} = \{ p_n=m\dot{u}_n,1\le n\le N \}$. 

Considering now a generic functions ${G}\big(\mathbf{u},\mathbf{p}\big)$ of the displacement $\mathbf{u}$ 
and momentum $\mathbf{p}$ variables, it can be used to generate a flow transformation 
of the variables $\big(\mathbf{u}(s),\mathbf{p}(s)\big)$ parametrized by the real number $s$  
as follows 
\begin{equation}\label{eq:flow_5}
 \begin{aligned}
 &\deriv{u_n(s)}{s}{} =  \deriv{{G}[\mathbf{u}(s),\mathbf{p}(s)]}{p_n}{} ,\\
 &\deriv{p_n(s)}{s}{} = - \deriv{{G}[\mathbf{u}(s),\mathbf{p}(s)]}{{u}_n}{} .
 \end{aligned}
\end{equation}
As the variables $\big(\mathbf{u}(s),\mathbf{p}(s)\big)$ are changed the flow transformation, 
any function ${F}\big(\mathbf{u},\mathbf{p}\big)$ is also changed 
by the transformation of Eq. \eqref{eq:flow_5}, and its rate of change with respect to $s$ is simply given 
by its Poisson bracket with ${G}$, \textit{i.e.} 
\begin{equation}
 \deriv{{F}(s)}{s}{} = \{ {F}(s) ,{G} \}_N, 
\end{equation}
where we labelled ${F}(s)\equiv {F}\big(\mathbf{u}(s),\mathbf{p}(s)\big)$, and 
where the Poisson bracket with $2N$ degrees of freedom is given by 
\begin{equation}\label{eq:pb_general_un}
 \begin{aligned}
 \big\{ {F}, {G}\big\}_N & \equiv 
 \sum_{n} \deriv{{F}(\mathbf{u},\mathbf{p})}{u_n}{} 
          \deriv{{G}(\mathbf{u},\mathbf{p})}{p_n}{} 
        - \deriv{{F}(\mathbf{u},\mathbf{p})}{p_n}{} 
          \deriv{{G}(\mathbf{u},\mathbf{p})}{u_n}{}. \\
 \end{aligned}
\end{equation}
If we restrict to functions ${F}$ and ${G}$ which are bilinear and/or at most quadratic forms in 
$\big(\mathbf{u},\mathbf{p}\big)$, 
they can be written in terms of matrices ${F}^{11}_{nm}$, ${F}^{22}_{nm}$, ${F}^{21}_{nm}$ and ${F}^{12}_{nm}$ as follows 
\begin{equation}\label{eq:F_mat}
 \begin{aligned}
 {F}(\mathbf{u},\mathbf{p}) = \sum_{n,m} \big(&u_n {F}^{11}_{nm}u_m + u_n {F}^{12}_{nm}p_m \\
  + &p_n {F}^{21}_{nm}u_m + p_n {F}^{22}_{nm}p_m \big).
 \end{aligned}
\end{equation}
For instance, the total energy of the phonon system of Eq. \eqref{eq:hamilton_def_u} is written as 
\begin{equation}
 \begin{aligned}
 {H}_v\big(\mathbf{u},\mathbf{p}\big) = 
  & \sum_{n=1}^{N}\bigg(\frac{m}{2}\omega_0^2 (1+ 2\frac{v_n}{mc^2}) u_n^2 
   + \frac{mc^2}{2} \frac{(u_n - u_{n+1})^2}{a^2}\bigg) \\
+& \sum_{n=1}^N  \frac{p_n^2}{2m},
 \end{aligned}
\end{equation}
and because $u_1=u_N=0$, the four matrices $H_v$ can be defined as 
\begin{equation}
 \begin{aligned}
& {H}^{11}_{nm} = \frac{m}{2}\omega_0^2 (1+ 2\frac{v_n}{mc^2})\delta_{nm}
  + \frac{mc^2}{2a^2}(2\delta_{nm} - \delta_{m,n+1} - \delta_{m+1,n}) \\
& {H}^{22}_{nm} = \frac{1}{2m} \delta_{nm},\quad  {H}^{12}_{nm} ={H}^{21}_{nm} = 0.
 \end{aligned}
\end{equation}
By introducing the following notation for the matrix vector product 
\begin{equation}
 \mathbf{v}^\dagger {F} \mathbf{u}=\sum_{m} v_n {F}_{nm}u_m ,
\end{equation}
the function of Eq. \eqref{eq:F_mat} can be written as 
\begin{equation}
 \begin{aligned}
 {F}(\mathbf{u},\mathbf{p}) & = \mathbf{u}^\dagger {F}^{11} \mathbf{u} + \mathbf{p}^\dagger {F}^{22}\mathbf{p} 
  + \mathbf{p}^\dagger {F}^{21}\mathbf{u} + \mathbf{u}^\dagger {F}^{12}\mathbf{p} .
 \end{aligned}
\end{equation}
We can then express these functions in terms of the $(\psi_n,\dot{\psi}_n)$ variables, 
and because $u_n = \sqrt{a}\lambda_c \psi_n$ and that the functions 
considered have necessarily quadratic forms, 
the Poisson bracket is written as 
\begin{equation}
 \begin{aligned}
\big\{ F, G\big\}_N  = \frac{mc^2 }{\hbar^2}
 \sum_{n} a \bigg(
 &\deriv{\scaled{F}(\psi,\dot{\psi})}{\psi_n}{} 
          \deriv{\scaled{G}(\psi,\dot{\psi})}{\dot{\psi}_n}{} \\
 &       - \deriv{\scaled{F}(\psi,\dot{\psi})}{\dot{\psi}_n}{} 
          \deriv{\scaled{G}(\psi,\dot{\psi})}{\psi_n}{} \bigg)
 \end{aligned}
\end{equation}
In the continuous limit, we have then 
\begin{equation}
 \begin{aligned}
 & \lim_{\substack{a \rightarrow 0 \\ N\rightarrow \infty}}  \psi_n = \psi(x), 
  \quad \lim_{\substack{a \rightarrow 0 \\ N\rightarrow \infty}}  \sum_{n=1}^N a = \int dx, \\
 & \lim_{\substack{a \rightarrow 0 \\ N\rightarrow \infty}} \deriv{{F}(\psi,\dot{\psi})}{\psi_n}{} 
 = \frac{\delta \tilde{F}[\psi,\dot{\psi}]}{\delta \psi(x)}, \\
\end{aligned}
\end{equation}
such that the Poisson bracket becomes
\begin{equation}
 \begin{aligned}
 \lim_{\substack{a \rightarrow 0 \\ N\rightarrow \infty}} \big\{ F, G\big\}_N  
 = \frac{\omega_0 }{\hbar} \int dx \bigg(
 & \frac{\delta {F}[\psi,\dot{\psi}]}{\delta \psi(x)}  \frac{\delta {G}[\psi,\dot{\psi}]}{\delta \dot{\psi}(x)} \\
 &- \frac{\delta {F}[\psi,\dot{\psi}]}{\delta \dot{\psi}(x)}  \frac{\delta {G}[\psi,\dot{\psi}]}{\delta {\psi}(x)}\bigg). 
 \end{aligned}
\end{equation}
Similarly, the associated flow transformation generates the following change in the fields 
\begin{equation}\label{eq:flow_final}
 \begin{aligned}
 &\deriv{\psi(x,s)}{s}{} = \frac{\omega_0}{\hbar}\derfunc{G[\psi,\dot{\psi}]}{\dot{\psi}(x,s)} ,\\
 &\deriv{\dot{\psi}(x,s)}{s}{} = -\frac{\omega_0}{\hbar}\derfunc{G[\psi,\dot{\psi}]}{{\psi}(x,s)} , 
 \end{aligned}
\end{equation}
where once more, the prefactor ${\omega_0}/{\hbar}$ comes from the rescaling units.

\section{The general structure of Poisson brackets in the continuous case}\label{ann:pb_general}
In the present section, we consider two quadratic functionals $F[\psi,\dot{\psi}]$ and $G[\psi,\dot{\psi}]$ 
from which we derive the general expression of the Poisson bracket. 
We follow the notations of Sec. \ref{sec:two_comp} for the functional $F$ and $G$, \textit{i.e.} 
\begin{equation}
 \begin{aligned}
 F[\psi,\dot{\psi}] = & \int dx \Psi^\dagger(x) \hat{\mathcal{F}} \Psi(x),  
 \end{aligned}
\end{equation}
with the matrix $\hat{\mathcal{F}} $ and vector $\Psi(x)$ being defined as 
\begin{equation}
 \hat{\mathcal{F}} = 
 \begin{pmatrix}
  \hat{F}_{11} &  \hat{F}_{12} \\
  \hat{F}_{21} &  \hat{F}_{22}
 \end{pmatrix}, \quad 
 \Psi(x) = 
 \begin{pmatrix}
  \psi(x) \\
  \dot{\psi}(x) \\
 \end{pmatrix},
\end{equation}
and with hermitian diagonal operators, \textit{i.e.} $\hat{F}_{11} = \hat{F}_{11}^\dagger$ 
and $\hat{F}_{22} = \hat{F}_{22}^\dagger$. 
The functional derivative are then written as 
\begin{equation}
 \begin{aligned}
 \frac{\delta F[\psi,\dot{\psi}]}{\delta \psi(x)} &= 
  2 \hat{F}_{11}  \psi(x) + \big(\hat{F}_{12}+ \hat{F}_{21}^\dagger \big) \dot{\psi}(x) \\
 \frac{\delta F[\psi,\dot{\psi}]}{\delta \dot{\psi}(x)} &= 
  2 \hat{F}_{22}  \dot{\psi}(x) + \big(\hat{F}_{21}+ \hat{F}_{12}^\dagger \big) {\psi}(x).\\
 \end{aligned}
\end{equation}
For the sake of compaction of notation, we introduce the following operators 
\begin{equation}\label{eq:def_alpha}
 \begin{aligned}
 \hat{f}_{12} &= \hat{F}_{12}+ \hat{F}_{21}^\dagger, \quad \hat{f}_{21} = \hat{F}_{21}+ \hat{F}_{12}^\dagger,
 \end{aligned}
\end{equation}
such that the functional derivative become 
\begin{equation}
 \begin{aligned}
 \frac{\delta F[\psi,\dot{\psi}]}{\delta \psi(x)} &= 
 2 \hat{F}_{11} \psi(x) + \hat{f}_{12} \dot{\psi}(x) \\
 \frac{\delta F[\psi,\dot{\psi}]}{\delta \dot{\psi}(x)} &= 
 2 \hat{F}_{22}\dot{\psi}(x) + \hat{f}_{21}{\psi}(x).\\
 \end{aligned}
\end{equation}
Then let us consider the Poisson bracket $\{F,G \}$ of Eq. \eqref{eq:pb_limit} which we write as 
\begin{equation}
 \{F,G \} = \frac{\omega_0}{\hbar}\big(\text{I} - \text{II}\big),
\end{equation}
with 
\begin{equation}
 \text{I} =  \int dx \frac{\delta F[\psi,\dot{\psi}]}{\delta \psi(x)} \frac{\delta G[\psi,\dot{\psi}]}{\delta \dot{\psi}(x)}  , \quad 
 \text{II} = \int dx  \frac{\delta G[\psi,\dot{\psi}]}{\delta \psi(x)} \frac{\delta F[\psi,\dot{\psi}]}{\delta \dot{\psi}(x)} . 
\end{equation}
We can develop the first term of the Poisson bracket  as 
\begin{equation}
 \begin{aligned}
 \text{I} & =\int dx  \big( 2\hat{F}_{11} \psi(x) + \hat{f}_{12} \dot{\psi}(x) \big) 
                             \big( 2\hat{G}_{22}\dot{\psi}(x) + \hat{g}_{21}{\psi}(x) \big) \\
                 & = 4 \int dx \psi(x)  \hat{F}_{11}\hat{G}_{22}\dot{\psi}(x) 
                            +2\psi(x) \hat{F}_{11} \hat{g}_{21}{\psi}(x) \\
           & + 2\int dx \dot{\psi}(x)  \hat{f}_{12}^\dagger \hat{G}_{22}\dot{\psi}(x) 
                      +\dot{\psi}(x)  \hat{f}_{12}^\dagger \hat{g}_{21}{\psi}(x),\\
 \end{aligned}
\end{equation}
which can be put into a $2\times 2$ formalism as 
\begin{equation}
 \begin{aligned}
 \text{I}        & = \int dx \Psi^\dagger(x) 
    \begin{pmatrix}
           2\hat{F}_{11} \hat{g}_{21}            & 4\hat{F}_{11}\hat{G}_{22} \\
           \hat{f}_{12}^\dagger \hat{g}_{21} & 2\hat{f}_{12}^\dagger \hat{G}_{22} \\
    \end{pmatrix}
   \Psi(x) .
 \end{aligned}
\end{equation}
One can then compute the second part of the Poisson bracket 
\begin{equation}
 \begin{aligned}
 \text{II}  & = \int dx  
 \big( 2\hat{G}_{11} \psi(x) + \hat{g}_{12} \dot{\psi}(x) \big) 
 \big( 2\hat{F}_{22}\dot{\psi}(x) + \hat{f}_{21}{\psi}(x) \big) \\
            & = 2\int dx     \psi(x) \hat{G}_{11} \hat{f}_{21}{\psi}(x)
  +            4\int dx    \psi(x) \hat{G}_{11}\hat{F}_{22}\dot{\psi}(x) \\
& + \int dx   \dot{\psi}(x) \hat{g}_{12}^\dagger\hat{f}_{21}{\psi}(x)
            +  2\int dx   \dot{\psi}(x) \hat{g}_{12}^\dagger\hat{F}_{22}\dot{\psi}(x),
 \end{aligned}
\end{equation}
or in a matrix form as 
\begin{equation}
 \begin{aligned}
 \text{II}        & = \int dx \Psi^\dagger(x) 
    \begin{pmatrix}
           2\hat{G}_{11} \hat{f}_{21} & 4\hat{G}_{11}\hat{F}_{22} \\
           \hat{g}_{12}^\dagger\hat{f}_{21} &2\hat{g}_{12}^\dagger\hat{F}_{22} \\
    \end{pmatrix}
   \Psi(x) .
 \end{aligned}
\end{equation}
Therefore, the Poisson bracket is expressed as a functional written in a two-components framework 
\begin{equation}\label{eq:poisson_final}
 \begin{aligned}
 \{F,G \} &= \frac{\omega_0}{\hbar}\big(\text{I} - \text{II}\big) \\
 & = \int dx \Psi^\dagger(x) \{\hat{\mathcal{FG}}\}
   \Psi(x) ,
 \end{aligned}
\end{equation}
where the matrix $\{\hat{\mathcal{FG}}\}$ is then given by 
\begin{equation}\label{eq:poisson_final_2}
 \begin{aligned}
 \{\hat{\mathcal{FG}}\}  = 2\frac{\omega_0}{\hbar} 
    \begin{pmatrix}
 \hat{F}_{11} \hat{g}_{21}-\hat{G}_{11}\hat{f}_{21} & 2 \big(\hat{F}_{11}\hat{G}_{22}-\hat{G}_{11}\hat{F}_{22}\big)\\
\frac{1}{2}\big(\hat{f}_{12}^\dagger\hat{g}_{21}-\hat{g}_{12}^\dagger\hat{f}_{21}\big) 
 & \hat{f}_{12}^\dagger \hat{G}_{22} - \hat{g}_{12}^\dagger\hat{F}_{22}\\
    \end{pmatrix}
 \end{aligned}
\end{equation}
which is the result of Eq. \eqref{eq:poisson_final_main}.  

\section{Derivation of the constraints for constants of motion}\label{ann:der_cst}
A generic functional $F[\psi,\dot{\psi}]$ is a constant of motion if $\{F,H_v\}=0$, 
which, using Eq. \eqref{eq:poisson_final_2}, can be explicitly written as 
\begin{equation}
 \begin{aligned}
  &\frac{\hbar}{\omega_0}\int dx  \dot{\psi}(x)\hat{f}_{12}^\dagger \dot{\psi}(x) 
  -mc^2\int dx \psi(x) \hatom_v^2\hat{f}_{21}\psi(x) \\
 & + 2\!\int\! dx \psi(x) \big( \frac{\hbar}{\omega_0}\hat{F}_{11}- mc^2\hatom_v^2\hat{F}_{22} \big) \dot{\psi}(x)
 \! = 0 \,\,\forall \,\,\psi(x),\dot{\psi}(x),
 \end{aligned}
\end{equation}
and where we recall that $mc^2\hatom_v^2 = mc^2 + \hat{h}_v$ where $\hat{h}_v$ is the usual non relativistic Hamiltonian 
with a potential $v$.  
Using the definitions of the $\hat{f}_{12}$ and $\hat{f}_{21}$ operators of Eq. \eqref{eq:def_alpha}, it yields 
\begin{equation}\label{eq:h_pb_zero_main}
 \begin{aligned}
&  + \frac{\hbar}{\omega_0} \int dx \dot{\psi}(x)(\hat{F}_{12}+ \hat{F}_{21}^\dagger)^\dagger \dot{\psi}(x) \\
&- mc^2\int dx \psi(x)\hatom_v^2 ( \hat{F}_{21}+ \hat{F}_{12}^\dagger ) \psi(x) \\
 & + 2\!\int\! dx \psi(x) \big( \frac{\hbar}{\omega_0}\hat{F}_{11}- mc^2\hatom_v^2\hat{F}_{22} \big) \dot{\psi}(x)
 \! = 0 \,\,\forall \,\,\psi(x),\dot{\psi}(x).
 \end{aligned}
\end{equation}
As the variables $\big(\psi(x),\dot{\psi}(x)\big)$ are independent, we cannot rely on cancellation of terms 
in order to fulfil Eq. \eqref{eq:h_pb_zero_main}. 
Therefore, a sufficient condition is to require the nullity of each of the three integrals 
composing Eq. \eqref{eq:h_pb_zero_main}, 
which can be decomposed in terms involving the "diagonal" operators $\hat{F}_{11}$ and $\hat{F}_{22}$, 
and that involving the "off-diagonal" operators $\hat{F}_{21}$ and $\hat{F}_{12}$. 

Let us begin by analyzing the condition for the nullity involving the diagonal operators $\hat{F}_{11}$ 
and $\hat{F}_{22}$ in Eq. \eqref{eq:h_pb_zero_main}, \textit{i.e.} 
\begin{equation}
 \begin{aligned}
 &\int dx \psi(x) \big( \frac{\hbar}{\omega_0}\hat{F}_{11}- mc^2\hatom_v^2\hat{F}_{22} \big) \dot{\psi}(x)= 0 \quad \forall \psi(x),\dot{\psi}(x),
 \end{aligned}
\end{equation}
which translates into 
\begin{equation}\label{eq:diag_term_0_main}
 \frac{\hbar}{\omega_0}\hat{F}_{11} - mc^2\hatom_v^2\hat{F}_{22}  =0,
\end{equation}
or equivalently to 
\begin{equation}
 \begin{aligned}
\hat{F}_{11} \hat{F}_{22}^{-1} = \frac{\omega_0}{\hbar} mc^2\hatom_v^2,
 \end{aligned}
\end{equation}
which can be fulfilled by choosing 
\begin{equation}
 \hat{F}_{11} = \hatom_v^{2\zeta}, \quad \hat{F}_{22} = \frac{1}{\omega_0^2}\hatom_v^{2(\zeta-1)},\quad \zeta\in\mathbb{R} .
\end{equation}

If we now focus on the nullity of the term involving the off-diagonal operators $\hat{F}_{21}$ and $\hat{F}_{12}$ 
in Eq. \eqref{eq:h_pb_zero_main}, we obtain the following condition 
\begin{equation}\label{eq:off_diag_1_main}
 \begin{aligned}
&- \int dx \psi(x)mc^2\hatom_v^2 ( \hat{F}_{21}+ \hat{F}_{12}^\dagger ) \psi(x) \\
&+ \frac{\hbar}{\omega_0} \int dx \dot{\psi}(x)(\hat{F}_{12}^\dagger+ \hat{F}_{21}) \dot{\psi}(x) =0 
\quad \forall \psi,\dot{\psi},
 \end{aligned}
\end{equation}
and as these are integrals involving unrelated functions $\psi(x)$ and $\dot{\psi}(x)$, 
we obtain two separated conditions, namely  
\begin{equation}\label{sec:psi_2_condition}
 \int dx \psi(x)\hatom_v^2 ( \hat{F}_{21}+ \hat{F}_{12}^\dagger ) \psi(x) = 0, \,\,\forall\,\, \psi,
\end{equation}
and 
\begin{equation}\label{sec:dot_psi_2_condition}
  \int dx \dot{\psi}(x)(\hat{F}_{12}^\dagger+ \hat{F}_{21}) \dot{\psi}(x) =0, \,\,\forall\,\, \dot{\psi}. 
\end{equation}
From Eq. \eqref{sec:dot_psi_2_condition} we can guess the condition $\hat{F}_{21}=- \hat{F}_{12}^\dagger$,
but it leads to a trivial case as the corresponding contribution to the functional $F[\psi,\dot{\psi}]$ vanishes. 
One can nevertheless remark that the two conditions of Eq. \eqref{sec:psi_2_condition} and Eq. \eqref{sec:dot_psi_2_condition} 
can be put into a similar form 
\begin{equation}\label{eq:anti_hermit}
 \int dx f(x) \hat{O} f(x) = 0, \,\,\forall\,\, f \Leftrightarrow \hat{O}^\dagger = -\hat{O},
\end{equation}
which means that $\hat{O}$ has to be anti hermitian. 
Therefore from Eq. \eqref{sec:dot_psi_2_condition} and Eq. \eqref{eq:anti_hermit} we obtain that 
\begin{equation}\label{eq:anti_hermit_2}
 \big( \hat{F}_{21} + \hat{F}_{12}^\dagger\big)^\dagger = - \big(\hat{F}_{21} + \hat{F}_{12}^\dagger\big),
\end{equation}
and from Eq. \eqref{sec:psi_2_condition} and Eq. \eqref{eq:anti_hermit} we obtain that 
\begin{equation}\label{eq:commut_0}
 \begin{aligned}
&\bigg( \hatom_v^2 \big( \hat{F}_{21}+ \hat{F}_{12}^\dagger \big) \bigg)^\dagger = - \hatom_v^2 ( \hat{F}_{21}+ \hat{F}_{12}^\dagger ) \\
 \Leftrightarrow & 
\big(  \hat{F}_{21}+ \hat{F}_{12}^\dagger \big)^\dagger\big(\hatom_v^2)^\dagger  = - \hatom_v^2 ( \hat{F}_{21}+ \hat{F}_{12}^\dagger ),
 \end{aligned}
\end{equation}
where we used that $\hatom_v$ is hermitian. 
Also, as $\hatom_v^2=mc^2\mathbb{1}+\hat{h}_v$ is hermitian, using Eq. \eqref{eq:anti_hermit_2} 
we can rewrite the condition of Eq. \eqref{eq:commut_0} as 
\begin{equation}\label{eq:commut_1_annex}
 \begin{aligned}
 - & \big(\hat{F}_{21} + \hat{F}_{12}^\dagger\big)\hat{h}_v   = - \hat{h}_v ( \hat{F}_{21}+ \hat{F}_{12}^\dagger ) \\
\Leftrightarrow &\big[\hat{F}_{21} + \hat{F}_{12}^\dagger,\hat{h}_v \big]= 0.
 \end{aligned}
\end{equation}
We can therefore conclude on the two types of conditions to obtain a constant of motion for a functional $F$ 
\begin{equation}\label{eq:cst_mot}
 \{F,H_v\} = 0 \Rightarrow 
 \hat{F}_{11} = \hatom_v^{2\zeta}  \text{ and } \hat{F}_{22} = \hatom_v^{2(\zeta-1)}{\omega_0}^{-2}, \zeta \in \mathbb{R}\\
\end{equation}
\begin{equation}\label{eq:cst_mot}
 \begin{aligned}
 \{F,H_v\} = 0 \Rightarrow 
&\big( \hat{F}_{21} + \hat{F}_{12}^\dagger\big)^\dagger = - \big(\hat{F}_{21} + \hat{F}_{12}^\dagger\big) \\
 &\text{ and } \big[\hat{F}_{21} + \hat{F}_{12}^\dagger,\hat{h}_v \big]= 0. 
 \end{aligned}
\end{equation}

\section{Rewriting classical functionals as quantum expectation values through a non local change of variables}
\label{sec:func_alpha_annex}
We recall the class of complex-valued non local change of variable 
\begin{equation}\label{eq:phialpha_annex}
 \phialpha(x,t) = \frac{1}{\sqrt{2}}\big(\hatom_v^{1-\alpha} \psi(x,t) + \frac{i}{\omega_0} \hatom_v^{-\alpha} \dot{\psi}(x,t)\big).
\end{equation}
The vector $\Psi=(\psi,\dot{\psi})$ can be written in terms of the vector $\Phialpha=(\phialpha,\phialpha^*)$ as follows 
\begin{equation}
 \Psi(x) = \mathcal{U}_\alpha \,\Phialpha(x),
\end{equation}
where the transformation matrix $\mathcal{U}$ is written as 
\begin{equation}
 \mathcal{U}_\alpha = \frac{1}{\sqrt{2}} 
 \begin{pmatrix}
 \hatom_v^{\alpha-1}& \hatom_v^{\alpha-1}\\
 -i\omega_0\hatom_v^{\alpha} & +i\omega_0\hatom_v^{\alpha}\\
 \end{pmatrix}.
\end{equation}
As a consequence, any functional $F[\psi,\dot{\psi}]$ written as 
\begin{equation}
 \begin{aligned}
  F[\psi,\dot{\psi}] = \int dx \Psi^\dagger(x) \mathcal{F} \Psi(x), 
 \end{aligned}
\end{equation}
where 
\begin{equation}
 \mathcal{F} = 
 \begin{pmatrix}
 \hat{F}_{11} & \hat{F}_{12} \\
 \hat{F}_{21} & \hat{F}_{22} \\
 \end{pmatrix},
\end{equation}
can be written as 
\begin{equation}
 \begin{aligned}
  F[\phialpha,\phialpha^*] = \int dx \Phialpha^\dagger(x) \tilde{\mathcal{F}_\alpha}  \Phialpha(x) ,
 \end{aligned}
\end{equation}
where 
\begin{equation}\label{eq:f_alpha}
 \begin{aligned}
 \tilde{\mathcal{F}_\alpha}& =  \mathcal{U}_\alpha^\dagger \mathcal{F} \mathcal{U}_\alpha.
 \end{aligned}
\end{equation}
We derive here the restrictions on the operators of the functional $F$ such that it is written as an expectation value of the type 
\begin{equation}
 F[\phialpha,\phialpha^*] = \int dx \phialpha^*(x)\hat{F}_\alpha \phialpha(x), 
\end{equation}
which is equivalent to a diagonal matrix $\tilde{\mathcal{F}_\alpha}$. 
Also, we impose that 
\begin{equation}
 \hat{F}_{11} = \hat{F}_{11}^\dagger, \quad \hat{F}_{22} = \hat{F}_{22}^\dagger, \quad \hat{F}_{21} = -\hat{F}_{12}, \quad \hat{F}_{12}^\dagger = -\hat{F}_{12}.
\end{equation}
Because of the linearity of the change of basis, we can treat these operators in a separate way: the diagonal operators $\hat{F}_{11}$ and $\hat{F}_{22}$ on one hand, and the $\hat{F}_{12}$ operator on the other hand. 

\subsection{Diagonal operators}
The first class of functionals are those which are diagonal in the $(\psi,\dot{\psi})$ representation 
\begin{equation}
 \begin{aligned}
  F[\psi,\dot{\psi}] = \int dx \Psi^\dagger(x) 
 \begin{pmatrix}
 \hat{F}_{11} & 0\\
 0 & \hat{F}_{22} \\
 \end{pmatrix}\Psi(x),
 \end{aligned}
\end{equation}
and where both operators $\hat{F}_{11}$ and $\hat{F}_{22}$ are necessarily hermitian. 
The corresponding $\tilde{\mathcal{F}}_\alpha$ matrix as defined in Eq. \eqref{eq:f_alpha} can then be written as 
\begin{equation}\label{eq:def_f_alpha_diag_0}
 \begin{aligned}
 \tilde{\mathcal{F}}_\alpha = \frac{1}{2}
 \begin{pmatrix}
 \tilde{F}_{11} + \tilde{F}_{22} & \tilde{F}_{11} - \tilde{F}_{22}\\
 \tilde{F}_{11} - \tilde{F}_{22} & \tilde{F}_{11} + \tilde{F}_{22}\\
 \end{pmatrix},
 \end{aligned}
\end{equation}
where we introduced 
\begin{equation}\label{eq:def_f_alpha_diag_1}
 \tilde{F}_{11} = \hatom_v^{\alpha-1} \hat{F}_{11} \hatom_v^{\alpha-1} , 
  \quad \tilde{F}_{22} = \omega_0^2\hatom_v^{\alpha}\hat{F}_{22} \hatom_v^{\alpha}.
\end{equation}
As the operators $\hat{F}_{11}$, $\hat{F}_{22}$ and $\hatom_v$ are hermitian, the corresponding functional can then be written as 
\begin{equation}\label{eq:func_phi}
 \begin{aligned}
 F[\phialpha,\phialpha^*] = & \int dx \phialpha^*(x) \big(\tilde{F}_{11} + \tilde{F}_{22}\big) \phialpha(x) \\ 
 +  &\frac{1}{2} \int dx \phialpha(x) \big(\tilde{F}_{11} - \tilde{F}_{22}\big) \phialpha(x) \\
 + &\frac{1}{2} \int dx \phialpha^*(x) \big(\tilde{F}_{11} - \tilde{F}_{22}\big) \phialpha^*(x).
 \end{aligned}
\end{equation}
If we want the functional to have the usual form of quantum mechanics, 
we then want that the second and third terms in the right-handside of Eq. \eqref{eq:func_phi} to vanish, 
which imposes that 
\begin{equation}\label{eq:f_diag_func}
 \tilde{F}_{11}=\tilde{F}_{22} \Leftrightarrow \hat{F}_{22} = \frac{1}{\omega_0^2}\hatom_v^{-1}\hat{F}_{11}\hatom_v^{-1}. 
\end{equation}
The quantum hermitian operator associated to the functional is then 
\begin{equation}\label{eq:diag_op_qm}
 \hat{F}  = 2 \hatom_v^{\alpha-1} \hat{F}_{11} \hatom_v^{\alpha-1},
\end{equation}
and the functional written in the $\big(\phialpha,\phialpha^*\big)$ basis simply becomes 
\begin{equation}\label{eq:generic_f_alpha}
 F[\phialpha,\phialpha^*] = \int dx \phialpha^*(x) \hat{F} \phialpha(x),
\end{equation}
which has the generic form of a quantum expectation value. 

As an application of the result of Eq. \eqref{eq:f_diag_func}, we consider the case of the functionals expressed as averages 
of a function $f(x)$ over the density $\rho_\zeta$ 
\begin{equation}\label{eq:f_general_annex}
 F[\psi,\dot{\psi}] = \int dx f(x)\rho_\zeta(\psi,\dot{\psi},x),
\end{equation}
where we recall that $\rho_\zeta$ is defined in Eq. \eqref{eq:def_measure_zeta}. 
The normalized position functional of Baros and Gomes as defined in Eq. \eqref{eq:def_x_baros} together with the 
unnormalized energy barycentre of Eq. \eqref{eq:x_no_norm} fulfill the requirements of Eq. \eqref{eq:f_general_annex}.  
Because $\hatom_v$ is hermitian, these functionals can be written in the two-components formalism with purely diagonal operators as follows 
\begin{equation}
 \begin{aligned}
 \hat{F}_{11} = \frac{1}{2}\hatom_v^{\zeta}f(x)\hatom_v^{\zeta}, \quad \hat{F}_{22} = \frac{1}{2\omega_0^2}\hatom_v^{\zeta-1}f(x)\hatom_v^{\zeta-1},
 \end{aligned}
\end{equation}
which fulfill therefore the condition of Eq. \eqref{eq:f_diag_func} to be written as a quantum expectation value. 
As a consequence, the associated operator is obtained from Eq. \eqref{eq:diag_op_qm} as follows 
\begin{equation}\label{eq:func_meas}
 \begin{aligned}
 \hat{F}_\alpha & = \hatom_v^{\alpha+\zeta-1} f(x) \hatom_v^{\alpha+\zeta-1}.
 \end{aligned}
\end{equation}
As a consequence, the functionals of Eq. \eqref{eq:f_general_annex} are written as quantum expectation values, 
whatever the potential $v(x)$. 
Also, applied to the case where $\zeta=1/2$ and $\alpha=1/2$, 
which corresponds to the density of Barros and Gomes (see Eq. \eqref{eq:measure_baros_expl}) 
and the Foldy representation, respectively, then the hermitian operator of Eq. \eqref{eq:func_meas} is simply $f(x)$, 
such that the functional of Eq. \eqref{eq:f_general_annex} is written as
\begin{equation}
F[\psilad,\psilad] = \int dx |\psilad(x)|^2 f(x),
\end{equation}
while when choosing the energy density corresponding to $\zeta=1$, the functionals of Eq. \eqref{eq:func_meas} reads 
\begin{equation}
F[\psilad,\psilad] = \int dx \psilad^*(x) \hatom_v^{\frac{1}{2}} f(x)\hatom_v^{\frac{1}{2}}\psilad(x).
\end{equation}

\subsection{Off diagonal functionals}
The second class of functionals considered are those involving only off diagonal terms with anti hermitian operators 
\begin{equation}
 \begin{aligned}
  F[\psi,\dot{\psi}] = \int dx \Psi^\dagger(x) 
 \begin{pmatrix}
 0 & \hat{F}_{12}\\
 -\hat{F}_{12} & 0 \\
 \end{pmatrix}
\Psi(x), 
 \end{aligned}
\end{equation}
such as the linear or angular momentum.  
We can introduce the two operators 
\begin{equation}\label{eq:def_f_alpha_off_diag_0}
 \tilde{F}_{1} = \hatom_v^{\alpha} \hat{F}_{12} \hatom_v^{\alpha-1} 
 , \quad  \tilde{F}_{2} = \hatom_v^{\alpha-1} \hat{F}_{12} \hatom_v^{\alpha} ,
\end{equation}
which are related by the relation $\tilde{F}_{1}^\dagger = -\tilde{F}_{2}$, and the matrix of 
Eq. \eqref{eq:f_alpha} associated to $F$ is then 
\begin{equation}\label{eq:def_f_alpha_off_diag_1}
 \tilde{\mathcal{F}}_\alpha = \frac{i\omega_0}{2}
 \begin{pmatrix}
 -\big(\tilde{F}_{1} + \tilde{F}_{2} \big) & -\big(\tilde{F}_{1} - \tilde{F}_{2} \big) \\
  \tilde{F}_{1} - \tilde{F}_{2}  &  \tilde{F}_{1} + \tilde{F}_{2}  \\
 \end{pmatrix},
\end{equation}
which, using the fact that $\big(\tilde{F}_{1} + \tilde{F}_{2} \big)^\dagger=-\big(\tilde{F}_{1} + \tilde{F}_{2} \big)$, 
corresponds then to the following functional 
\begin{equation}
 \begin{aligned}
 F[\phialpha,\phialpha^*] = & -i\omega_0\int dx \phialpha^*(x) \big(\tilde{F}_{1} + \tilde{F}_{2} \big) \phialpha(x) \\
 &-\frac{i\omega_0}{2}\int dx \phialpha^*(x) \big(\tilde{F}_{1} - \tilde{F}_{2} \big) \phialpha^*(x) \\
& +\frac{i\omega_0}{2}\int dx \phialpha(x) \big(\tilde{F}_{1} - \tilde{F}_{2} \big) \phialpha(x) .
 \end{aligned}
\end{equation}
If we want the functional to have the form of standard quantum mechanics, we need to impose that 
\begin{equation}
 \int dx \phialpha(x) \big(\tilde{F}_{1} - \tilde{F}_{2} \big) \phialpha(x)  = 0\quad \forall \,\,\phialpha(x),
\end{equation}
but as $\big(\tilde{F}_{1} - \tilde{F}_{2} \big)$ is hermitian, it implies that the operator $\big(\tilde{F}_{1} - \tilde{F}_{2} \big)$ itself should vanish, \textit{i.e.} 
\begin{equation}
 \tilde{F}_{1} = \tilde{F}_{2} \Leftrightarrow \hatom_v^\alpha \hat{F}_{12} \hatom_v^{\alpha-1} = \hatom_v^{\alpha-1}\hat{F}_{12} \hatom_v^\alpha  ,
\end{equation}
which is valid when $\hat{F}_{12}$ commutes with $\hatom_v$. As $\hatom_v = \sqrt{1+\hat{h}_v/(2mc^2)}$, 
a sufficient condition is that $\hat{F}_{12}$ commutes with the non relativistic Hamiltonian $\hat{h}_v$,  
such that we conclude that 
\begin{equation}\label{eq:off_diag}
 \big[\hat{F}_{12},\hat{h}_v\big]=0 \Rightarrow 
\int dx \phialpha(x) \big(\tilde{F}_{1} - \tilde{F}_{2} \big) \phialpha(x)  = 0\quad \forall \,\,\phialpha(x),
\end{equation}
Under the condition of Eq. \eqref{eq:off_diag}, the corresponding quantum hermitian operator is then 
\begin{equation}\label{eq:hat_f_off_diag}
 \hat{F}_\alpha = -2i\omega_0 \hatom_v^{2\alpha-1}\hat{F}_{12} ,
\end{equation}
functional takes then simply the following form 
\begin{equation}
 F[\phialpha,\phialpha^*] = \int dx \phialpha^*(x) \hat{F}_\alpha \phialpha(x), 
\end{equation}
which is of the form a quantum expectation value, and as $\hat{F}_{12}$ is anti hermitian, 
the operator $\hat{F}$ of Eq. \eqref{eq:hat_f_off_diag} is hermitian. 
One should nevertheless pay attention to the fact that the condition $[\hat{F}_{12},\hat{h}_v] = 0$ 
therefore depends on the external potential $v(x)$ of the system. 
Therefore, when $v(x)=cst$, the two various definitions of linear momentum of 
Eq. \eqref{eq:momentum_classic} and Eq. \eqref{eq:momentum_relat} can then be written as quantum expectation values, 
while the angular momentum of Eq. \eqref{eq:ang_mom_tot} can also be written in such form 
only when $v(\bfr{r})=v(|\bfr{r}|)$.

\section{Linking the continuous Poisson bracket in the $(\psi,\dot{\psi})$ and $(\psilad,\psilad^*)$ representations}
\label{sec:pb_link}
In the present section we give the details of the derivations of the link between the Poisson bracket expressions 
in the $(\psi,\dot{\psi})$ representation and the corresponding quantum algebra in $(\psilad,\psilad^*)$ basis. 

\subsection{Flow transformations in the $(\psilad,\psilad^*)$ basis}\label{sec:flow_link}
We consider here only functionals $G[\Psi]$ that can be written as quantum expectation values 
in the $(\psilad,\psilad^*)$ basis, such that we know that they are written in the $(\psi,\dot{\psi})$ basis as follows 
\begin{equation}
 \begin{aligned}
 &G[\Psi] = \int dx \Psi^\dagger(x) 
 \begin{pmatrix}
  \hat{G}_{11} & \hat{G}_{12} \\
 -\hat{G}_{12} & \frac{1}{\omega_0^2}\hatom_v^{-1}\hat{G}_{11}\hatom_v^{-1} \\
 \end{pmatrix}
 \Psi(x), \\ 
 &\big(\hat{G}_{11}\big)^\dagger = \hat{G}_{11},\quad \big(\hat{G}_{12}\big)^\dagger = -\hat{G}_{12} ,
 \quad [\hat{G}_{12},\hatom_v] = 0,
 \end{aligned}
\end{equation}
which can be split in terms of functionals involving diagonal and off-diagonal operators, namely 
\begin{equation}\label{eq:func_general}
 G[\Psi] = G^d[\Psi] + G^{od}[\Psi] ,
\end{equation}
with 
\begin{equation}
 G^d[\Psi] = \int dx \Psi^\dagger(x) 
 \begin{pmatrix}
  \hat{G}_{11} &      0       \\
     0         & \frac{1}{\omega_0^2}\hatom_v^{-1}\hat{G}_{11}\hatom_v^{-1} \\
 \end{pmatrix}
 \Psi(x),
\end{equation}
and 
\begin{equation}
 G^{od}[\Psi] = \int dx \Psi^\dagger(x) 
 \begin{pmatrix}
        0      & \hat{G}_{12} \\
 -\hat{G}_{12} &      0       \\
 \end{pmatrix}
 \Psi(x).
\end{equation}
Such functionals can also be written in the $(\phialpha,\phialpha^*)$ basis as 
\begin{equation}
 G[\phialpha,\phialpha^*] = \int dx \phialpha^*(x) \,\hat{G}_\alpha \,\phialpha(x),
\end{equation}
with the operator $\hat{G}$ being also split in terms of "diagonal" and "off-diagonal operators" 
\begin{equation}\label{eq:f_specific}
 \hat{G}_\alpha = \hat{G}^d_\alpha + \hat{G}^{od}_\alpha,
\end{equation}
and where the corresponding operators are 
\begin{equation}\label{eq:recal_def_op}
 \hat{G}^d_\alpha = 2 \hatom_v^{\alpha-1}\hat{G}_{11} \hatom_v^{\alpha-1}, \quad \hat{G}^{od} = -2 i\omega_0\hatom_v^{2\alpha-1} \hat{G}_{12}.
\end{equation}
An important relation coming from the bilinear structure of the functionals in the $(\phialpha,\phialpha^*)$ basis 
is the following 
\begin{equation}\label{eq:flow_deriv_alpha}
 \derfunc{G[\phialpha(s),\phialpha^*(s)]}{\phialpha^*(x,s)} = \hat{G}_\alpha\phialpha(x), 
\end{equation}
and it will be used thoroughly in the present derivations. 
We can now use a generic functional $G[\Psi]$ to generate a flow transformation in phase space as usual 
\begin{equation}
 \begin{aligned}
 \deriv{}{s}{}\psi(x,s) & = \frac{\omega_0}{\hbar} \derfunc{G}{\dot{\psi}(x,s)},\\
 \deriv{}{s}{}\dot{\psi}(x,s) & = -\frac{\omega_0}{\hbar} \derfunc{G}{{\psi}(x,s)},\\
 \end{aligned}
\end{equation}
which, by linearity of the functional derivative, can be decomposed in terms of diagonal and off-diagonal functionals as follows 
\begin{equation}
 \begin{aligned}
 \deriv{}{s}{}\psi(x,s) & = \frac{\omega_0}{\hbar}\big( \derfunc{G^d_\alpha}{\dot{\psi}(x,s)}  + \derfunc{G^{od}_\alpha}{\dot{\psi}(x,s)}\big),\\
 \deriv{}{s}{}\dot{\psi}(x,s) & = -\frac{\omega_0}{\hbar} \big(\derfunc{G^d_\alpha}{{\psi}(x,s)}+ \derfunc{G^{od}_\alpha}{{\psi}(x,s)}\big),\\
 \end{aligned}
\end{equation}
where 
\begin{equation}
 \begin{aligned}
 & \derfunc{G^{d}}{{\psi}(x)} = 2 \hat{G}_{11} \psi(x),
  \quad \derfunc{G^d}{\dot{\psi}(x)} =  \frac{2}{\omega_0^2}\hatom_v^{-1}\hat{G}_{11}\hatom_v^{-1} \dot{\psi}(x), \\
 & \derfunc{G^{od}}{{\psi}(x)} = -2 \hat{G}_{12} \dot{\psi}(x),
  \quad \derfunc{G^{od}}{\dot{\psi}(x)} =  -2 \hat{G}_{12} {\psi}(x). \\
 \end{aligned}
\end{equation}
We start with the off-diagonal functional whose flow equation reads 
\begin{equation}
 \begin{aligned}
&\deriv{\psi(x,s)}{s}{} = -2\frac{\omega_0}{\hbar} \hat{G}_{12} \psi(x,s), \\
&\deriv{\dot{\psi}(x,s)}{s}{} = -2\frac{\omega_0}{\hbar} \hat{G}_{12} \dot{\psi}(x,s),  
 \end{aligned}
\end{equation}
which, as we assumed that $[\hat{G}_{12},\hatom_v]=0$, can be rewritten as 
\begin{equation}\label{eq:flow_tmp2}
 \begin{aligned}
&\hatom_v^{1-\alpha}\deriv{\psi(x,s)}{s}{} = -2\frac{\omega_0}{\hbar} \hat{G}_{12} \hatom_v^{1-\alpha}\psi(x,s), \\
&\hatom_v^{-\alpha} \deriv{\dot{\psi}(x,s)}{s}{} = -2\frac{\omega_0}{\hbar} \hat{G}_{12} \hatom_v^{-\alpha}\dot{\psi}(x,s),  
 \end{aligned}
\end{equation}
and by summing the two equations composing Eq. \eqref{eq:flow_tmp2} with an $i/\omega_0$ factor for the second line, 
we obtain
\begin{equation}\label{eq:d_ds_psi}
 \begin{aligned}
&\deriv{}{s}{}\big( \hatom_v^{1-\alpha}\deriv{\psi(x,s)}{s}{} +\frac{i}{\omega_0}\hatom_v^{-\alpha} \dot{\psi}(x,s)\big) \\
 =& -2\frac{\omega_0}{\hbar} \hat{G}_{12} \big( \hatom_v^{1-\alpha}\psi(x,s) +\frac{i}{\omega_0}\hatom_v^{-\alpha}\dot{\psi}(x,s)\big),  
 \end{aligned}
\end{equation}
where we recognize the definition of $\phialpha$, and therefore one can rewrite Eq. \eqref{eq:d_ds_psi} as simply
\begin{equation}\label{eq:d_ds_psi_2}
 \begin{aligned}
&\deriv{}{s}{} \phialpha(x,s)= -2\frac{\omega_0}{\hbar} \hat{G}_{12} \phialpha(x,s).
 \end{aligned}
\end{equation}
By remembering the definition of the off-diagonal operator $\hat{G}^{od}$ 
in terms of $\hat{G}_{12}$ (\textit{i.e.} Eq. \eqref{eq:recal_def_op}), we know that 
\begin{equation}
 -2 \hat{G}_{12} = \frac{1}{i\omega_0}\hatom_v^{1-2\alpha}\hat{G}^{od}_\alpha,
\end{equation}
such that inserted into Eq. \eqref{eq:d_ds_psi_2} 
we obtain 
\begin{equation}\label{eq:flow_phi_od_2}
 \deriv{}{s}{} \phialpha(x,s) = -\frac{i}{\hbar} \hatom_v^{1-2\alpha}\derfunc{G^{od}_\alpha}{\phialpha^*}.
\end{equation}
We can notice that in the case of $\alpha=1/2$, Eq. \eqref{eq:flow_phi_od_2} reduces to 
\begin{equation}\label{eq:flow_od_phi_half}
 \deriv{}{s}{} \psiladhalf(x,s) = -\frac{i}{\hbar} \derfunc{G^{od}}{\psiladhalf^*}.
\end{equation}
Then, the contribution of the diagonal functional generates the following flow equations 
\begin{equation}
 \begin{aligned}
&\deriv{\psi(x,s)}{s}{} = \frac{2\omega_0}{\hbar\omega_0^2}\hatom_v^{-1}\hat{G}_{11}\hatom_v^{-1} \dot{\psi}(x), \\
&\deriv{\dot{\psi}(x,s)}{s}{} = -\frac{2\omega_0}{\hbar} \hat{G}_{11} \psi(x),
 \end{aligned}
\end{equation}
which can be rewritten as 
\begin{equation}
 \begin{aligned}
&\hatom_v^{1-\alpha}\deriv{\psi(x,s)}{s}{} = \frac{2}{\hbar\omega_0}\hatom_v^{-\alpha}\hat{G}_{11}\hatom_v^{-1} \dot{\psi}(x), \\
&\hatom_v^{-\alpha} \deriv{\dot{\psi}(x,s)}{s}{} = -\frac{2\omega_0}{\hbar} \hatom_v^{-\alpha}\hat{G}_{11} \psi(x),
 \end{aligned}
\end{equation}
and further refined as 
\begin{equation}\label{eq:flow_tmp}
 \begin{aligned}
&\deriv{ \hatom_v^{1-\alpha} \psi(x,s)}{s}{} = \frac{2}{\hbar\omega_0}
\hatom_v^{1-2\alpha}\big(\hatom_v^{\alpha-1}\hat{G}_{11}\hatom_v^{\alpha-1}\big)\hatom_v^{-\alpha} \dot{\psi}(x), \\
&\deriv{ \hatom_v^{-\alpha}  \dot{\psi}(x,s)}{s}{} = -\frac{2\omega_0}{\hbar}  \hatom_v^{1-2\alpha}\big(\hatom_v^{\alpha-1}\hat{G}_{11}\hatom_v^{\alpha-1}\big) \hatom_v^{1-\alpha} \psi(x).
 \end{aligned}
\end{equation}
We recognize in Eq. \eqref{eq:flow_tmp} the hermitian operator associated to the diagonal part of the functional of Eq. \eqref{eq:recal_def_op}, 
\textit{i.e.} $\hat{G}^d_\alpha=2\hatom_v^{\alpha-1}\hat{G}_{11}\hatom_v^{\alpha-1}$, such Eq. \eqref{eq:flow_tmp} 
can then be written as 
\begin{equation}\label{eq:two_eq}
 \begin{aligned}
&\deriv{ \hatom_v^{1-\alpha} \psi(x,s)}{s}{} = \frac{1}{\hbar\omega_0}\hatom_v^{1-2\alpha}\hat{G}^d_\alpha \hatom_v^{-\alpha} \dot{\psi}(x), \\
&\deriv{ \hatom_v^{-\alpha}  \dot{\psi}(x,s)}{s}{} = -\frac{\omega_0}{\hbar}\hatom_v^{1-2\alpha}\hat{G}^d_\alpha \hatom_v^{1-\alpha} \psi(x).
 \end{aligned}
\end{equation}
We can then sum the two equations of Eq. \eqref{eq:two_eq} as follows 
\begin{equation}\label{eq:diag_flow}
 \begin{aligned}
& \deriv{}{s}{}\big(\hatom_v^{1-\alpha}  \psi(x,s) + \frac{i}{\omega_0} \hatom_v^{-\alpha} \dot{\psi}(x,s)\big) \\
 & = -\frac{i}{\hbar}\hatom_v^{1-2\alpha}\hat{G}^d_\alpha\big(\hatom_v^{1-\alpha}\psi(x) + \frac{i}{\omega_0}\hatom_v^{-\alpha} \dot{\psi}(x) \big),
 \end{aligned}
\end{equation}
and remembering the definition of $\phialpha$, Eq. \eqref{eq:diag_flow} is then written as 
\begin{equation}\label{eq:flow_phi_d}
 \deriv{}{s}{}\phialpha(x,s) = -\frac{i}{\hbar}\hatom_v^{1-2\alpha}\hat{G}^d_\alpha\phialpha(x,s).
\end{equation}
Remembering that 
\begin{equation}
 \derfunc{G^d}{\phialpha^*} = \hat{G}^d_\alpha\phialpha,
\end{equation}
Eq. \eqref{eq:flow_phi_d} can be written as 
\begin{equation}\label{eq:flow_phi_d_2}
 \deriv{}{s}{}\phialpha(x,s) = -\frac{i}{\hbar}\hatom_v^{1-2\alpha}\derfunc{G^d}{\phialpha^*}.
\end{equation}
Therefore, summing the contribution of the diagonal and off-diagonal operators, 
we obtain the general case 
\begin{equation}\label{eq:flow_phi_general_1}
 \begin{aligned}
 \deriv{}{s}{}\phialpha(x,s) & = -\frac{i}{\hbar} \hatom_v^{1-2\alpha} \hat{G}_\alpha\phialpha(x,s).\\
 \end{aligned}
\end{equation}
As a consequence, the flow transformation generated by the functional $G$ does not couple 
$\phialpha(s)$ and $\phialpha^*(s)$, while it does couple $\psi(s)$ and $\dot{\psi}(s)$ in the general case. 
Also, using Eq. \eqref{eq:flow_deriv_alpha}, the flow equation of Eq. \eqref{eq:flow_phi_general_1} 
can be expressed as follows 
\begin{equation}\label{eq:flow_phi_general_1_bis}
 \begin{aligned}
 \deriv{}{s}{}\phialpha(x,s) & = -\frac{i}{\hbar} \hatom_v^{1-2\alpha} \derfunc{G}{\phialpha^*},\\
 \end{aligned}
\end{equation}
which resembles the usual form of the flow transformations in terms of functional derivatives.  

One can also notice that in the specific case of $\alpha=1/2$,  
the flow equations of \eqref{eq:flow_phi_general_1} simplies 
\begin{equation}\label{eq:flow_phi_half_1}
 \deriv{}{s}{}\psiladhalf(x,s) = -\frac{i}{\hbar}\hat{G}\psiladhalf(x,s),
\end{equation}
where $\hat{G}$ is the operator 
\begin{equation}
 \hat{G} = -2i\omega_0 \hat{F}_{12} + 2 \hatom_v^{-\frac{1}{2}} \hat{F}_{11} \hatom_v^{-\frac{1}{2}}, 
\end{equation}
such that Eq. \eqref{eq:flow_phi_half_1} can therefore be written as 
\begin{equation}\label{eq:flow_phi_half_2}
 \deriv{}{s}{}\psiladhalf(x,s) = -\frac{i}{\hbar}\derfunc{G}{\psiladhalfb}.
\end{equation}

\subsection{Poisson brackets in the $(\phialpha,\phialpha^*)$ and $(\psiladhalf,\psiladhalf^*)$ representations}
One can then compute how a generic functional is $F[\phialpha,\phialpha^*]$ is changed by 
the action of the flow transformation generated by the functional $G[\phialpha,\phialpha^*]$. 
Using the chain rule in the $(\phialpha,\phialpha^*)$ basis, the derivative is written as 
\begin{equation}\label{eq:d_ds_phi_alpha}
 \begin{aligned}
 \frac{d}{ds}F[\phialpha(s),\phialpha^*(s)] = 
 \int dx \bigg( &\derfunc{F[\phialpha(s),\phialpha^*(s)]}{\phialpha(x,s)}\deriv{}{s}{}{\phialpha(x,s)} \\
 + &\derfunc{F[\phialpha(s),\phialpha^*(s)]}{\phialpha^*(x,s)}\deriv{}{s}{}{\phialpha^*(x,s)}\bigg),
 \end{aligned}
\end{equation}
and using Eq. \eqref{eq:flow_deriv_alpha} together with Eq. \eqref{eq:flow_phi_general_1}, 
the flow equation of Eq. \eqref{eq:d_ds_phi_alpha} can be written as 
\begin{equation}
\begin{aligned}\label{eq:deriv_ds}
 & \frac{d}{ds}F[\phialpha(s),\phialpha^*(s)] = \\ 
 &\frac{1}{i\hbar}\int dx  \big(\hat{F}_\alpha \phialpha^*(x,s)\big)
 \big(\hatom_v^{1-2\alpha}\hat{G}_\alpha\phialpha(x,s)\big) \\
 -&\frac{1}{i\hbar}\int dx \big(\hat{F}_\alpha \phialpha(x,s)\big)\big(\hatom_v^{1-2\alpha} \hat{G}_\alpha\phialpha^*(x,s)\big),
\end{aligned}
\end{equation}
which, using the fact that all operators in Eq. \eqref{eq:deriv_ds} are hermitian, is then written as 
\begin{equation}\label{eq:flow_fg}
\begin{aligned}
 \frac{d}{ds}F[\phialpha(s),\phialpha^*(s)] = 
\frac{1}{i\hbar} \int dx & \phialpha^*(x,s)
 \hat{F}_\alpha\hatom_v^{1-2\alpha} \hat{G}_\alpha\phialpha(x,s) \\
 -\frac{1}{i\hbar} \int dx &\phialpha^*(x,s)\hat{G}_\alpha \hatom_v^{1-2\alpha}\hat{F}_\alpha \phialpha(x,s),
\end{aligned}
\end{equation}
where we used $\big(\hatom_v^{1-2\alpha} \hat{G}_\alpha\big)^\dagger = \hat{G}_\alpha \hatom_v^{1-2\alpha}$ as both $\hat{G}_\alpha$ and $ \hatom_v$ 
are hermitian.
One can then notice that, in the specific case of $\alpha=1/2$, the terms $\hatom_v^{1-2\alpha}=1$, such that Eq. \eqref{eq:flow_fg} simplifies to 
\begin{equation}\label{eq:flow_fg_half_1}
\begin{aligned}
 \frac{d}{ds}F[\psiladhalf(s),\psiladhalf^*(s)] = 
\frac{1}{i\hbar} \int dx & \psiladhalf^*(x,s)
 \hat{F}\hat{G}\psiladhalf(x,s) \\
 -\frac{1}{i\hbar}\int dx  &\psiladhalf^*(x,s)\hat{G}\hat{F} \psiladhalf(x,s),
\end{aligned}
\end{equation}
where we recall that $\hat{G}$ and $\hat{F}$ are the operator corresponding to $\alpha=1/2$ which are defined in Eq. \eqref{eq:f_qm_general_bis}. 
Therefore, using the notation of Eq. \eqref{eq:phifoldy} for the Foldy representation (\textit{i.e.} $\psilad\equiv \psiladhalf$), 
Eq. \eqref{eq:flow_fg_half_1} can then be rewritten as  
\begin{equation}\label{eq:flow_fg_half_2}
\begin{aligned}
 \frac{d}{ds}F[\psilad(s),\psilad^*(s)] = -\frac{i}{\hbar}
 \int dx & \psilad^*(x,s)\big[ \hat{F} , \hat{G} \big] \psilad(x,s).
\end{aligned}
\end{equation}
Using the functional derivatives of Eq. \eqref{eq:flow_deriv_alpha}, Eq. \eqref{eq:flow_fg_half_2} can 
be alternatively written as 
\begin{equation}\label{eq:flow_fg_half_4}
\begin{aligned}
 \frac{d}{ds}F(s) = -\frac{i}{\hbar}\poisson{F(s)}{G(s)}.
\end{aligned}
\end{equation}
where $\poisson{F}{G}$ is the new Poisson bracket defined as  
\begin{equation}\label{eq:poisson_bis}
\begin{aligned}
 \poisson{F}{G} = 
 \int dx \bigg(&\derfunc{F[\psilad,\psilad^*]}{\psilad(x)}
 \derfunc{G[\psilad,\psilad^*]}{\psilad^*(x)}\\
 - &\derfunc{F[\psilad,\psilad^*]}{\psilad^*(x)}
 \derfunc{G[\psilad,\psilad^*]}{\psilad(x)}\bigg).
\end{aligned}
\end{equation}
We therefore conclude that we have the following identity 
\begin{equation}
 \begin{aligned}
 \{F,G\} & = -\frac{i}{\hbar} \poisson{F}{G} \\
         & = -\frac{i}{\hbar} \langle \psilad | \big[ \hat{F} , \hat{G} \big] |\psilad \rangle,
 \end{aligned}
\end{equation}
or equivalently 
\begin{equation}
  \langle \psilad | \big[ \hat{F} , \hat{G} \big] |\psilad \rangle = i\hbar\{F,G\},
\end{equation}
which is similar to the Dirac canonical quantization scheme. 

\section{Non relativistic limit of the non local change of variable in the case of $v(x)\ne0$}\label{sec:nr_v_x}
In the present section we examine the case of the non local change of variable for the generic case where $v(x)=0$, 
and its non relativistic limit. We begin in Sec. \ref{sec:hatom_taylor} by analyzing the generic form of the operator $\hatom_v^\alpha$ 
in the non relativistic limit. Based on the latter results, we then focus in Sec. \ref{sec:non_r_lim_func} on the expressions of the functionals the non relativistic limit 
in the $(\phialpha,\phialpha^*)$ basis and show that they coincide with the usual non relativistic expectation values. 
\subsection{Non relativistic limit of $\hatom_v^\alpha$ and related quantities}\label{sec:hatom_taylor}
The non local change of variable in the case of a system with an external potential $v(x)\ne 0$ implies the 
following operator 
\begin{equation}
 \hatom_v^\alpha = \big( 1 + 2 \frac{\hat{h}_v}{mc^2}\big)^\frac{\alpha}{2},
\end{equation}
which can be expanded using 
\begin{equation}
 (1+x)^a = \sum_{k=0}^\infty x^k \binom{a}{k},
\end{equation}
such that it can be formally written as
\begin{equation}
 \hatom_v^\alpha = \sum_{k=0}^\infty 2^k \bigg(\frac{\hat{h}_v}{mc^2}\bigg)^k \binom{\frac{\alpha}{2}}{k}.
\end{equation}
We write the non relativistic Hamiltonian $\hat{h}_v$ as 
\begin{equation}
 \hat{h}_v = \hat{h} + v(x),
\end{equation}
where $\hat{h}$ is the usual quantum non relativistic kinetic energy operator, such that we can write 
\begin{equation}
 \big(\hat{h}_v\big)^k = \big(\hat{h}\big)^k  + \big(v(x)\big)^k  + \delta_v^{(k)},
\end{equation}
where $ \delta_v^{(k)}$ is an operator containing all mixed terms between $v(x)$ and $\hat{h}$, 
which satisfies 
\begin{equation}
  \delta_v^{(k)} = 0 \text{ for }k\le1, \quad \delta_v^{(2)} = \hat{h} v(x) + v(x)\hat{h}.
\end{equation}
We can then rewrite the operator $\hatom_v^\alpha$ as follows 
\begin{equation}\label{eq:hatom_v}
 \begin{aligned}
 \hatom_v^\alpha = \hatom^{\alpha} + \frac{\alpha v(x)}{2mc^2} + \frac{1}{m^2c^4}\hat{\delta}^{(\alpha)}_v.  
 \end{aligned}
\end{equation}
where 
\begin{equation}
 \hatom^{\alpha} = \sum_{k=0}^\infty 2^k \bigg(\frac{\hat{h}}{mc^2}\bigg)^k \binom{\frac{\alpha}{2}}{k} ,
\end{equation}
and 
\begin{equation}
 \hat{\delta}_v^{(\alpha)} = 
  \sum_{k=2}^\infty 2^k {mc^2}^{k-2}\bigg(v(x)+\delta_v^{(k)}\bigg)^k \binom{\frac{\alpha}{2}}{k} .
\end{equation}
We give here the first few terms expansion in powers of $c^2$ of the operator $\hatom_v^{\alpha}$ 
\begin{equation}\label{eq:hatom_alpha_taylor}
 \hatom_v^{\alpha} = 1 + \frac{\alpha}{2mc^2}\hat{h}_{v} 
  + o(c^{-4}), 
\end{equation}
such that in the non relativistic limit, we obtain 
\begin{equation}\label{eq:lim_hatom}
 \begin{aligned}
 \lim_{c\rightarrow \infty}\hatom_v^{\alpha} & = 1, 
 \end{aligned}
\end{equation}
and 
\begin{equation}
 \lim_{c\rightarrow \infty}mc^2(\hatom^{\alpha}-1) = \frac{\alpha}{2}\hat{h}_v .  
\end{equation}
As a consequence of Eq. \eqref{eq:lim_hatom}, 
the non relativistic limit of the normalized and unnormalized momentum 
of Eq. \eqref{eq:p_phihalf} and Eq. \eqref{eq:momentum_relat_phihalf} both coincide with the usual 
definition of the non relativistic quantum momentum. 
Similarly, the $c\rightarrow \infty$ limit of the unnormalized energy barycentre of Eq. \eqref{eq:x_no_norm_half} and the normalized position 
functional of Eq. \eqref{eq:x_half} both coincide with the definition of the position in non relativistic quantum mechanics. 

\subsection{Non relativistic limit of functionals}\label{sec:non_r_lim_func}
We now focus our attention on the non relativistic limit of functionals not necessarily fulfilling the properties 
enabling them to be written as quantum expectation values 
in the $(\psilad,\psilad^*)$. More precisely, we consider the following functionals 
\begin{equation}
 F[\psi,\dot{\psi}] = \int dx \Psi^\dagger(x) \hat{\mathcal{F}} \Psi(x),
\end{equation}
where the matrix $\hat{\mathcal{F}}$ is written as 
\begin{equation}
 \hat{\mathcal{F}} = \frac{1}{2}
 \begin{pmatrix}
 \hat{F}_{11} & \lambda_c^2\hat{F}_{12} \\
 -\lambda_c^2\hat{F}_{12} & \omega_0^{-2}\hat{F}_{22} \\
 \end{pmatrix},
\end{equation}
and where the operators $\hat{F}_{11}$, $\hat{F}_{22}$ and $\hat{F}_{12}$ are of zeroth-order in $c$ and 
do not fulfill the conditions of Eq. \eqref{eq:f_diag_func} and Eq. \eqref{eq:off_diag}. 
Among important examples falling in that category are the momentum functionals of 
Eq. \eqref{eq:momentum_classic} and Eq. \eqref{eq:momentum_relat} in the case of a generic potential $v(x)$, 
the angular momentum of Eq. \eqref{eq:ang_mom_tot} in the case of a non central potential, 
and the average forces of Eq. \eqref{eq:ex_force}. 

The functionals are then written in the $(\psilad,\psilad^*)$ basis as follows 
\begin{equation}\label{eq:func_psilad}
 \begin{aligned}
 F[\psilad,\psilad^*] = & \int dx \psilad^*(x) \hat{F}_d \psilad(x) \\
 +  &\int dx \psilad(x) \hat{F}_{od} \psilad(x) \\
 +  &\int dx \psilad^*(x) \hat{F}_{od}^\dagger \psilad^*(x),\\
 \end{aligned}
\end{equation}
where the operators $\hat{F}_d$ and $\hat{F}_{od}$ are given by setting $\alpha=1/2$ in 
Eqs. \eqref{eq:def_f_alpha_diag_0}, \eqref{eq:def_f_alpha_diag_1}, 
\eqref{eq:def_f_alpha_off_diag_0} and \eqref{eq:def_f_alpha_off_diag_1}, which then reads 
\begin{equation}\label{eq:f_d}
 \begin{aligned}
  \hat{F}_d = &\frac{1}{2} \big(
  \hatom_v^{-\frac{1}{2}} \hat{F}_{11} \hatom_v^{-\frac{1}{2}} + 
 \hatom_v^{\frac{1}{2}} \hat{F}_{22} \hatom_v^{\frac{1}{2}} \big)\\
& -i\frac{\hbar}{2}\big(\hatom_v^{\frac{1}{2}}\hat{F}_{12}\hatom_v^{-\frac{1}{2}} + 
 \hatom_v^{-\frac{1}{2}}\hat{F}_{12}\hatom_v^{\frac{1}{2}}\big),
 \end{aligned}
\end{equation}
\begin{equation}\label{eq:f_od}
 \begin{aligned}
  \hat{F}_{od} = &\frac{1}{2} \big(
  \hatom_v^{-\frac{1}{2}} \hat{F}_{11} \hatom_v^{-\frac{1}{2}} - 
 \hatom_v^{\frac{1}{2}} \hat{F}_{22} \hatom_v^{\frac{1}{2}} \big)\\
& -i\frac{\hbar}{2}\big(\hatom_v^{\frac{1}{2}}\hat{F}_{12}\hatom_v^{-\frac{1}{2}} - 
 \hatom_v^{-\frac{1}{2}}\hat{F}_{12}\hatom_v^{\frac{1}{2}}\big).
 \end{aligned}
\end{equation}
We emphasize that, because the operators do not fulfill the conditions of 
Eq. \eqref{eq:f_diag_func} and Eq. \eqref{eq:off_diag}, the functional as written in Eq. \eqref{eq:func_psilad} 
contains the terms involving $\psilad\hat{F}_{od}\psilad$ and $\psilad^*\hat{F}_{od}^\dagger\psilad^*$, which 
therefore are not present in the usual expectation values of non relativistic quantum mechanics. 
Nevertheless, in the non relativistic limit, these terms will vanish as we shall see here. 

To perform the non relativistic limit, we need to take the $c\rightarrow \infty$ limit 
of the $\hatom_v$ operator obtained in Sec. \ref{sec:hatom_taylor} and rewrite the functional in terms of the 
slowly-varying variables $(\lad,\lad^*)$ of Eq. \eqref{eq:lad_foldy}. 
As we assumed that the operators are of order $0$ in $c$, the $c\rightarrow \infty$  limit of the operators 
$\hat{F}_{d}$ and $\hat{F}_{od}$ can be obtained trivially by substituting $\hatom_v$ by the identity operators 
in Eq. \eqref{eq:f_d} and Eq. \eqref{eq:f_od}, 
such that it leads to 
\begin{equation}\label{eq:f_d_2}
 \begin{aligned}
  \lim_{c\rightarrow \infty}\hat{F}_d = &\frac{1}{2} \big(
  \hat{F}_{11} + 
 \hat{F}_{22} \big) -i\hbar\hat{F}_{12},  
 \end{aligned}
\end{equation}
\begin{equation}\label{eq:f_od_2}
 \begin{aligned}
  \lim_{c\rightarrow \infty}\hat{F}_{od} = &\frac{1}{2} \big(
  \hat{F}_{11} - 
 \hat{F}_{22} \big).  
 \end{aligned}
\end{equation}
We therefore notice from Eq. \eqref{eq:f_od_2} that for functionals of the type of momentum fulfilling 
$\hat{F}_{11} = \hat{F}_{22} = 0$, we obtain $\lim_{c\rightarrow \infty}\hat{F}_{od}=0$, which implies 
that they are in form akin to a quantum expectation value. 

Then, when writing the functionals of Eq. \eqref{eq:func_psilad} in the slowly-varying variables $(\lad,\lad^*)$ 
it leads to 
\begin{equation}\label{eq:func_psilad_1}
 \begin{aligned}
 F[\lad,\lad^*] = & \int dx \lad^*(x) \hat{F}_d \lad(x) \\
 +  &e^{-2i\omega_0t}\int dx \lad(x) \hat{F}_{od} \lad(x) \\
 +  &e^{+2i\omega_0t}\int dx \lad^*(x) \hat{F}_{od}^\dagger \lad^*(x).\\
 \end{aligned}
\end{equation}
The non relativistic limit is equivalent to  $\omega_0\rightarrow\infty$ and therefore 
the two terms involving $\hat{F}_{od}$ oscillates at an infinite frequency. 
As a consequence, if we assume that 
\begin{equation}\label{eq:slow}
 \bigg|\frac{\derivb{t}{}\lad(x,t)}{\lad(x,t)}\bigg| \ll \omega_0,
\end{equation}
which corresponds to the well known slowly varying envelope approximation, 
the fastly oscillating terms vanish over a time average of the functional 
over a typical time scale of $\delta t= 2\pi\omega_0^{-1}$ and we recover the usual form of the quantum expectation 
value 
\begin{equation}
 \begin{aligned}
 &\frac{1}{\delta t} \int_{t}^{t+\delta t} ds F[\lad,\lad^*] \approx 
\int dx \lad^*(x,t) \hat{F}_d \lad(x,t) . 
 \end{aligned}
\end{equation}

\section{The 1+1 Poincar\'e group through the classical Poisson brackets}\label{eq:annex_group}
In this section we show that the Lie algebra of the 1+1 Poincar\'e group can be represented 
by the  $H$, $X$ and  $P$ functionals together with their Poisson brackets.  
We begin in Sec. \ref{sec:pc_11} by recalling the basic properties of the Poincar\'e group 
and then show how some of the classical functionals and their Poisson brackets 
allows to represent its algebra. Based on this, we present the adjoint representation 
of the 1+1 Poincar\'e group in Sec. \ref{sec:adjoint}.  

\subsection{Summary of Poincar\'e transformations and algebra in 1+1 dimension}
\label{sec:pc_11}
The Poincar\'e group in 1+1 dimension consists in all symmetry operations of special relativity (\textit{i.e.} of the Minkowski metric $g_{\mu\nu}$) in one spatial dimension, 
which consists in spatial translation,  time translations and Lorentz boost. 
These symmetry operations can be generated by three operations which are $\tilde{P}_0$ (for time translations), $\tilde{P}_1$ (for spatial translations) 
and $K$ (for Lorentz boosts), which fulfill the following Lie algebra 
\begin{equation}\label{eq:lie_pc}
 [K,\tilde{P}_1] = \tilde{P}_0, \quad [K,\tilde{P}_0] = \tilde{P}_1, \quad [\tilde{P}_1,\tilde{P}_0] = 0,
\end{equation}
and where the notation $"[,]"$ denotes the Lie bracket associated with the Lie group. 
The relations of the Lie algebra of Eq. \eqref{eq:lie_pc}  can be recovered from 
the continuous Poisson bracket together with the functional forms of $X$, $P$ and $H$ provided some rescaling units. 

To do so, one defines the following dimension-less Poisson bracket $\pbdim{.}{.} $ as follows 
\begin{equation}
 \pbdim{.}{.} = \hbar\{.,. \},  
\end{equation}
together with the rescaled variables 
\begin{equation}
 X = \lambda_c \scalex, \quad H = mc^2 \scaleh, \quad P = mc \scalep. 
\end{equation}
Then, as shown in Sec. II-B of the supplementary material, 
the Poisson brackets between the $H$, $X$ and $P$ functionals are given by 
\begin{equation}
 \{X,P\} = \frac{H}{mc^2}, \quad \{X,H\} = \frac{P}{m}, \quad \{P,H\} = 0,
\end{equation}
and expressed in terms of the scaled variables it becomes 
\begin{equation}\label{eq:pb_poincarre_11_bis}
 \pbdim{\scalex}{\scalep} = \scaleh, \quad \pbdim{\scalex}{\scaleh} = \scalep, \quad \pbdim{\scalep}{\scaleh} = 0,
\end{equation}
which is exactly the Lie algebra of the 1+1 Poincar\'e group of Eq. \eqref{eq:lie_pc} provided that $K=\scalex$, $\tilde{P}_1=\scalep$ 
and $\tilde{P}_0=\scaleh$. 
We will now build a representation of the $\liep$ Lie algebra and its Casimir invariant. 

\subsection{Adjoint representation of $\liep$}\label{sec:adjoint}
For the sake of compactness of notations, let $X_0=\scaleh$, $X_1=\scalex$ and $X_2=\scalep$, such that the 
Lie algebra relations of Eq. \eqref{eq:pb_poincarre_11_bis} is then written as 
\begin{equation}
 \pbdim{X_1}{X_2} = X_0, \quad \pbdim{X_1}{X_0} = X_2, \quad\pbdim{X_2}{X_0} = 0.
\end{equation}
For now, we simply use the standard $[,]$ notations for the Lie bracket. 
The Lie algebra $\liep$ is a vector space of dimension $3$, and we choose as a basis 
$\mathcal{B}=\{X_0,X_1,X_2\}$. 
As in any vector space, a vector $V\in\liep$ of the Lie algebra can be associated to its components 
$(v_0,v_1,v_2)$, namely  $ V=\sum_{m=0}^2 v_m X_m$. 

What differentiates a Lie algebra from a simple vector space is the Lie bracket which gives  
a correspondence between any two elements in $\liep$ and another element in $\liep$. 
In our basis $\mathcal{B}=\{X_0,X_1,X_2\}$, we can then compute the so-called structure constant tensor 
$C_{kj}^i$ defined by 
\begin{equation}\label{eq:def_struct}
 [ X_i,X_j ] = \sum_k C_{kj}^i X_k,
\end{equation}
which lists all such connexions (we acknowledge that the notation in Eq. \eqref{eq:def_struct} is unusual but it is handy for the matrix representation). 
Therefore, for each $X_i$ one can associate a matrix $\bfr{C}^i$ whose matrix elements are the $C_{kj}^i$, \textit{i.e.}  
\begin{equation}
 \big(\bfr{C}^i\big)_{kj} = C_{kj}^i,
\end{equation}
which explicitly reads 
\begin{equation}\label{eq:c_matrices}
 \bfr{C}^0 = 
 \begin{pmatrix}
  0 &0 &0\\
  0 &0 &0\\
  0 &-1&0\\
 \end{pmatrix}, 
 \bfr{C}^1 = 
 \begin{pmatrix}
  0 &0 &1 \\
  0 &0 &0 \\
  1 &0 &0 \\
 \end{pmatrix}, 
 \bfr{C}^2 = 
 \begin{pmatrix}
  0 &-1 &0\\
  0 &0 &0\\
  0 &0 &0\\
 \end{pmatrix}. 
\end{equation}
In a differential equation form, the use of the matrix $\bfr{C}^i$ can be understood as how an element $X(s)$ 
is changed by some generator $X_i$ through the Lie bracket, \textit{i.e.} 
\begin{equation}
 \begin{aligned}
 \frac{d}{ds} 
 \begin{pmatrix}
 X_0(s) \\
 X_1(s) \\
 X_2(s) \\
 \end{pmatrix}
 &=  -\big(\bfr{C}^i\big)^T
 \begin{pmatrix}
 X_0(s) \\
 X_1(s) \\
 X_2(s) \\
 \end{pmatrix}.
 \end{aligned}
\end{equation}
More generally, to each vector $\bfr{\vecpp}=(\vecpp_0,\vecpp_1,\vecpp_2) \in\liep$ we can associate a transformation matrix 
$\bfr{\Vecpp}=\vecpp_0 \bfr{C}^0 + \vecpp_1 \bfr{C}^1 + \vecpp_2 \bfr{C}^2 $ which generates the flow transformation of a vector $X(s)\in\liep$  
as follows
\begin{equation}
 \frac{d}{ds} X(s) = -\bfr{\Vecpp}^T X(s) ,
\end{equation}
and whose formal solution is then of course
\begin{equation}\label{eq:x_s}
 X(s) = e^{-s\bfr{\Vecpp}^T}X(0).
\end{equation}
This is the usual exponential map at the identity which allows to go from any vector in the algebra $\liep$ to an element of the group. 
In terms of physical transformations, the vector $\bfr{\Vecpp}$ corresponds to three physical parameters $(t,l,\mu)$ 
which characterizes any transformation of the Poincar\'e group. 
Therefore, the matrix representation associated to $\bfr{\Vecpp}$ in the adjoint representation is given by 
\begin{equation}
 \Gamma^{\text{ad}}(\bfr{\Vecpp}) \equiv e^{- \omega_0 t (\bfr{C}^0)^T - \frac{\mu}{mc}(\bfr{C}^1)^T - \frac{l}{\lambda_c}(\bfr{C}^2)^T}.
\end{equation}

\bibliography{biblio}
 \end{document}